%% file: main.tex
\pdfoutput=1
\documentclass[12pt,a4paper]{article}

\usepackage{ifthen} 
\newboolean{pdflatex}
\setboolean{pdflatex}{true} 

\newboolean{articletitles}
\setboolean{articletitles}{true} 

\newboolean{uprightparticles}
\setboolean{uprightparticles}{false} 
\usepackage{array}
\usepackage{booktabs}          
    
\def\paperauthors{LHCb collaboration} 
\def\paperasciititle{Observation of a Lb-Lbbar production asymmetry in proton-proton collisions at sqrt(s) = 7 and 8 TeV} 
\def\papertitle{Observation of a $\Lb-\Lbbar$ production asymmetry in proton--proton collisions at $\sqrt{s} = 7 \text{ and } 8\tev$ } 
\def\paperkeywords{{High Energy Physics}, {LHCb}} 
\def\papercopyright{\the\year\ CERN for the benefit of the LHCb collaboration} 
\def\paperlicence{CC BY 4.0 licence}
\def\paperlicenceurl{https://creativecommons.org/licenses/by/4.0/}

\input{preamble}

\input{macros}

\usepackage{longtable} 
\usepackage[capitalise]{cleveref}
\creflabelformat{equation}{#2\textup{#1}#3}

\begin{document}

\renewcommand{\thefootnote}{\fnsymbol{footnote}}
\setcounter{footnote}{1}

\input{title-LHCb-PAPER}


\renewcommand{\thefootnote}{\arabic{footnote}}
\setcounter{footnote}{0}

\cleardoublepage


\pagestyle{plain} 
\setcounter{page}{1}
\pagenumbering{arabic}
\clearpage

\input{body}

\input{acknowledgements}

\input{appendix}
\clearpage

\addcontentsline{toc}{section}{References}
\bibliographystyle{LHCb}
\bibliography{main,standard,LHCb-PAPER,LHCb-CONF,LHCb-DP,LHCb-TDR}

\newpage
\input{Authorship_LHCb-PAPER-2021-016.tex}

\end{document}

%% file: preamble.tex
\usepackage[top=1in, bottom=1.25in, left=1in, right=1in]{geometry}

\usepackage{microtype}
\usepackage{lineno}  
\usepackage{xspace} 
\usepackage{caption} 

\usepackage{graphicx}  
\usepackage{color}
\usepackage{colortbl}
\graphicspath{{./figs/}} 

\usepackage{amsmath} 
\usepackage{amssymb}
\usepackage{amsfonts}
\usepackage{upgreek} 

\newcommand*\patchAmsMathEnvironmentForLineno[1]{%
\expandafter\let\csname old#1\expandafter\endcsname\csname #1\endcsname
\expandafter\let\csname oldend#1\expandafter\endcsname\csname
end#1\endcsname
 \renewenvironment{#1}%
   {\linenomath\csname old#1\endcsname}%
   {\csname oldend#1\endcsname\endlinenomath}%
}
\newcommand*\patchBothAmsMathEnvironmentsForLineno[1]{%
  \patchAmsMathEnvironmentForLineno{#1}%
  \patchAmsMathEnvironmentForLineno{#1*}%
}
\AtBeginDocument{%
\patchBothAmsMathEnvironmentsForLineno{equation}%
\patchBothAmsMathEnvironmentsForLineno{align}%
\patchBothAmsMathEnvironmentsForLineno{flalign}%
\patchBothAmsMathEnvironmentsForLineno{alignat}%
\patchBothAmsMathEnvironmentsForLineno{gather}%
\patchBothAmsMathEnvironmentsForLineno{multline}%
\patchBothAmsMathEnvironmentsForLineno{eqnarray}%
}

\usepackage{hyperxmp}

\usepackage[pdftex,
            pdfauthor={\paperauthors},
            pdftitle={\paperasciititle},
            pdfkeywords={\paperkeywords},
            pdfcopyright={Copyright (C) \papercopyright},
            pdflicenseurl={\paperlicenceurl}]{hyperref}

\usepackage[bottom,flushmargin,hang,multiple]{footmisc}

\usepackage[all]{hypcap} 

\input{lhcb-symbols-def} 

\usepackage{cite} 
\usepackage{mciteplus}

%% file: lhcb-symbols-def.tex
\usepackage{xspace} 
\usepackage{upgreek}

\def\lhcb   {\mbox{LHCb}\xspace}

\def\lhc    {\mbox{LHC}\xspace}

\def\velo   {VELO\xspace}

\def\ttracker {TT\xspace}

\def\MagUp {\mbox{\em Mag\kern -0.05em Up}\xspace}

\ifthenelse{\boolean{uprightparticles}}%
{

 \def\Pmu         {\ensuremath{\upmu}\xspace}                 
 \def\Pnu         {\ensuremath{\upnu}\xspace}                 
                  
 \def\Ppi         {\ensuremath{\uppi}\xspace}

 \def\Ppsi        {\ensuremath{\uppsi}\xspace}

 \def\PDelta      {\ensuremath{\Delta}\xspace}                 
 \def\PXi         {\ensuremath{\Xi}\xspace}                 
 \def\PLambda     {\ensuremath{\Lambda}\xspace}                 
 \def\PSigma      {\ensuremath{\Sigma}\xspace}                 
 \def\POmega      {\ensuremath{\Omega}\xspace}                 
 \def\PUpsilon    {\ensuremath{\Upsilon}\xspace}

 \def\PB      {\ensuremath{\mathrm{B}}\xspace}                 
                  
 \def\PD      {\ensuremath{\mathrm{D}}\xspace}

 \def\PJ      {\ensuremath{\mathrm{J}}\xspace}                 
 \def\PK      {\ensuremath{\mathrm{K}}\xspace}

 \def\PX      {\ensuremath{\mathrm{X}}\xspace}

 \def\Pb      {\ensuremath{\mathrm{b}}\xspace}                 
 \def\Pc      {\ensuremath{\mathrm{c}}\xspace}                 
 \def\Pd      {\ensuremath{\mathrm{d}}\xspace}

 \def\Ph      {\ensuremath{\mathrm{h}}\xspace}                 
 \def\Pi      {\ensuremath{\mathrm{i}}\xspace}

 \def\Pp      {\ensuremath{\mathrm{p}}\xspace}

 \def\Ps      {\ensuremath{\mathrm{s}}\xspace}                 
                  
 \def\Pu      {\ensuremath{\mathrm{u}}\xspace}

 \def\thebaroffset{0.0em}
}
{

 \def\Pmu         {\ensuremath{\mu}\xspace}                 
 \def\Pnu         {\ensuremath{\nu}\xspace}                 
                  
 \def\Ppi         {\ensuremath{\pi}\xspace}

 \def\Ppsi        {\ensuremath{\psi}\xspace}                 
                  
 \mathchardef\PDelta="7101
 \mathchardef\PXi="7104
 \mathchardef\PLambda="7103
 \mathchardef\PSigma="7106
 \mathchardef\POmega="710A
 \mathchardef\PUpsilon="7107
                  
 \def\PB      {\ensuremath{B}\xspace}                 
                  
 \def\PD      {\ensuremath{D}\xspace}

 \def\PJ      {\ensuremath{J}\xspace}                 
 \def\PK      {\ensuremath{K}\xspace}

 \def\PX      {\ensuremath{X}\xspace}

 \def\Pb      {\ensuremath{b}\xspace}                 
 \def\Pc      {\ensuremath{c}\xspace}                 
 \def\Pd      {\ensuremath{d}\xspace}

 \def\Ph      {\ensuremath{h}\xspace}                 
 \def\Pi      {\ensuremath{i}\xspace}

 \def\Pp      {\ensuremath{p}\xspace}

 \def\Ps      {\ensuremath{s}\xspace}                 
                  
 \def\Pu      {\ensuremath{u}\xspace}

 \def\thebaroffset{0.18em}
}
\newcommand{\offsetoverline}[2][\thebaroffset]{\kern #1\overline{\kern -#1 #2}}%

\makeatletter
\ifcase \@ptsize \relax
  \newcommand{\miniscule}{\@setfontsize\miniscule{4}{5}}
\or
  \newcommand{\miniscule}{\@setfontsize\miniscule{5}{6}}
\or
  \newcommand{\miniscule}{\@setfontsize\miniscule{5}{6}}
\fi
\makeatother

\DeclareRobustCommand{\optbar}[1]{\shortstack{{\miniscule (\rule[.5ex]{1.25em}{.18mm})}
  \\ [-.7ex] $#1$}}

\def\mup        {{\ensuremath{\Pmu^+}}\xspace}
\def\mun        {{\ensuremath{\Pmu^-}}\xspace} 

\def\neu        {{\ensuremath{\Pnu}}\xspace}
\def\neub       {{\ensuremath{\overline{\Pnu}}}\xspace}
\def\neum       {{\ensuremath{\neu_\mu}}\xspace}
\def\neumb      {{\ensuremath{\neub_\mu}}\xspace}

\def\uquark    {{\ensuremath{\Pu}}\xspace}

\def\dquark    {{\ensuremath{\Pd}}\xspace}

\def\squark    {{\ensuremath{\Ps}}\xspace}

\def\cquark    {{\ensuremath{\Pc}}\xspace}

\def\bquark    {{\ensuremath{\Pb}}\xspace}
\def\bquarkbar {{\ensuremath{\overline \bquark}}\xspace}

\def\pion   {{\ensuremath{\Ppi}}\xspace}

\def\pip    {{\ensuremath{\pion^+}}\xspace}
\def\pim    {{\ensuremath{\pion^-}}\xspace}

\def\kaon    {{\ensuremath{\PK}}\xspace}
\def\Kbar    {{\ensuremath{\offsetoverline{\PK}}}\xspace}

\def\KorKbar {\kern \thebaroffset\optbar{\kern -\thebaroffset \PK}{}\xspace}
\def\Kz      {{\ensuremath{\kaon^0}}\xspace}
\def\Kzb     {{\ensuremath{\Kbar{}^0}}\xspace}
\def\Kp      {{\ensuremath{\kaon^+}}\xspace}
\def\Km      {{\ensuremath{\kaon^-}}\xspace}

\def\KS      {{\ensuremath{\kaon^0_{\mathrm{S}}}}\xspace}

\def\D       {{\ensuremath{\PD}}\xspace}

\def\DorDbar {\kern \thebaroffset\optbar{\kern -\thebaroffset \PD}\xspace}

\def\Dp      {{\ensuremath{\D^+}}\xspace}
\def\Dm      {{\ensuremath{\D^-}}\xspace}

\def\DpDm    {\ensuremath{\Dp {\kern -0.16em \Dm}}\xspace}

\def\Ds      {{\ensuremath{\D^+_\squark}}\xspace}

\def\B       {{\ensuremath{\PB}}\xspace}

\def\BorBbar {\kern \thebaroffset\optbar{\kern -\thebaroffset \PB}\xspace}

\def\Bd      {{\ensuremath{\B^0}}\xspace}

\def\BdorBdbar {\kern \thebaroffset\optbar{\kern -\thebaroffset \Bd}\xspace}
\def\Bu      {{\ensuremath{\B^+}}\xspace}

\def\Bs      {{\ensuremath{\B^0_\squark}}\xspace}

\def\BsorBsbar {\kern \thebaroffset\optbar{\kern -\thebaroffset \Bs}\xspace}

\def\jpsi     {{\ensuremath{{\PJ\mskip -3mu/\mskip -2mu\Ppsi}}}\xspace}

\def\Y#1S{\ensuremath{\PUpsilon{(#1S)}}\xspace}

\def\proton      {{\ensuremath{\Pp}}\xspace}
\def\antiproton  {{\ensuremath{\overline \proton}}\xspace}

\def\Lz          {{\ensuremath{\PLambda}}\xspace}
\def\Lbar        {{\ensuremath{\offsetoverline{\PLambda}}}\xspace}
\def\LorLbar     {\kern \thebaroffset\optbar{\kern -\thebaroffset \PLambda}\xspace}

\def\Lc          {{\ensuremath{\Lz^+_\cquark}}\xspace}
\def\Lcbar       {{\ensuremath{\Lbar{}^-_\cquark}}\xspace}

\def\Lb           {{\ensuremath{\Lz^0_\bquark}}\xspace}
\def\Lbbar        {{\ensuremath{\Lbar{}^0_\bquark}}\xspace}

\newcommand{\decay}[2]{\ensuremath{#1\!\to #2}\xspace} 

\def\to                 {\ensuremath{\rightarrow}\xspace}

\def\CP                {{\ensuremath{C\!P}}\xspace}

\def\AT#1     {\ensuremath{A_{\mathrm{T}}^{#1}}\xspace}           

\def\C#1      {\ensuremath{\mathcal{C}_{#1}}\xspace}                       
\def\Cp#1     {\ensuremath{\mathcal{C}_{#1}^{'}}\xspace}                    
\def\Ceff#1   {\ensuremath{\mathcal{C}_{#1}^{\mathrm{(eff)}}}\xspace}        
\def\Cpeff#1  {\ensuremath{\mathcal{C}_{#1}^{'\mathrm{(eff)}}}\xspace}       
\def\Ope#1    {\ensuremath{\mathcal{O}_{#1}}\xspace}                       
\def\Opep#1   {\ensuremath{\mathcal{O}_{#1}^{'}}\xspace}                    

\newcommand{\nospaceunit}[1]{\ensuremath{\text{#1}}}       
\newcommand{\aunit}[1]{\ensuremath{\text{\,#1}}}       

\newcommand{\tev}{\aunit{Te\kern -0.1em V}\xspace}
\newcommand{\gev}{\aunit{Ge\kern -0.1em V}\xspace}
\newcommand{\mev}{\aunit{Me\kern -0.1em V}\xspace}
\newcommand{\kev}{\aunit{ke\kern -0.1em V}\xspace}
\newcommand{\ev}{\aunit{e\kern -0.1em V}\xspace}
 
\newcommand{\mevc}{\ensuremath{\aunit{Me\kern -0.1em V\!/}c}\xspace}
\newcommand{\gevc}{\ensuremath{\aunit{Ge\kern -0.1em V\!/}c}\xspace}
\newcommand{\mevcc}{\ensuremath{\aunit{Me\kern -0.1em V\!/}c^2}\xspace}
\newcommand{\gevcc}{\ensuremath{\aunit{Ge\kern -0.1em V\!/}c^2}\xspace}

\def\mum  {\ensuremath{\,\upmu\nospaceunit{m}}\xspace}

\def\mbarn{\aunit{mb}\xspace}

\def\fb   {\ensuremath{\aunit{fb}}\xspace}
\def\invfb   {\ensuremath{\fb^{-1}}\xspace}

\def\gsim{{~\raise.15em\hbox{$>$}\kern-.85em
          \lower.35em\hbox{$\sim$}~}\xspace}
\def\lsim{{~\raise.15em\hbox{$<$}\kern-.85em
          \lower.35em\hbox{$\sim$}~}\xspace}

\def\sqs   {\ensuremath{\protect\sqrt{s}}\xspace}

\def\pt         {\ensuremath{p_{\mathrm{T}}}\xspace}

\def\evtgen     {\mbox{\textsc{EvtGen}}\xspace}

\def\geant      {\mbox{\textsc{Geant4}}\xspace}

\def\photos     {\mbox{\textsc{Photos}}\xspace}

\def\pythia     {\mbox{\textsc{Pythia}}\xspace}

\def\tell1  {TELL1\xspace}
\def\ukl1   {UKL1\xspace}



%% file: macros.tex
\newcommand{\Araw}{\ensuremath{A_\mathrm{raw}}\xspace}
\newcommand{\Adet}{\ensuremath{A_D}}
\newcommand{\Aprod}{\ensuremath{A_{\text{P}}}}

\newcommand{\recLb}{\ensuremath{\Lc(\rightarrow \proton\Km\pip)\mun}}
\newcommand{\recLbbar}{\ensuremath{\Lcbar(\rightarrow \antiproton\Kp\pim)\mup}}
\newcommand{\aintpi}{\ensuremath{A_\text{int}^i(\proton)}}
\newcommand{\aintp}{\ensuremath{A_\text{int}(\proton)}}
\newcommand{\apidpi}{\ensuremath{A_{\text{PID}}^i(\proton)}}
\newcommand{\apidp}{\ensuremath{A_{\text{PID}}(\proton)}}
\newcommand{\adetmui}{\ensuremath{A_{\text{trigger}+\text{PID}}^i(\mun)}}
\newcommand{\adetmu}{\ensuremath{A_{\text{trigger}+\text{PID}}(\mun)}}
\newcommand{\atrackpmui}{\ensuremath{A_{\text{track}}^i(\proton\mun)}}
\newcommand{\atrackpmu}{\ensuremath{A_{\text{track}}(\proton\mun)}}
\newcommand{\adetkpii}{\ensuremath{A_D^i\!\left(\Km\pip\right)}}
\newcommand{\adetkpi}{\ensuremath{A_D\!\left(\Km\pip\right)}}
\newcommand{\kpi}{\ensuremath{\Km\pip}}
\newcommand{\y}{\ensuremath{y}}
\newcommand{\yrec}{\ensuremath{y}}

\newcommand{\ptrec}{\pt}
\newcommand{\NA}{\ensuremath{\mathcal{N}_\mathrm{A}}} 
\newcommand{\pvec}{\ensuremath{\vec{p}}} 

\DeclareMathOperator\arctanh{artanh}

%% file: title-LHCb-PAPER.tex

\begin{titlepage}
\pagenumbering{roman}

\vspace*{-1.5cm}
\centerline{\large EUROPEAN ORGANIZATION FOR NUCLEAR RESEARCH (CERN)}
\vspace*{1.5cm}
\noindent
\begin{tabular*}{\linewidth}{lc@{\extracolsep{\fill}}r@{\extracolsep{0pt}}}
\ifthenelse{\boolean{pdflatex}}
{\vspace*{-1.5cm}\mbox{\!\!\!\includegraphics[width=.14\textwidth]{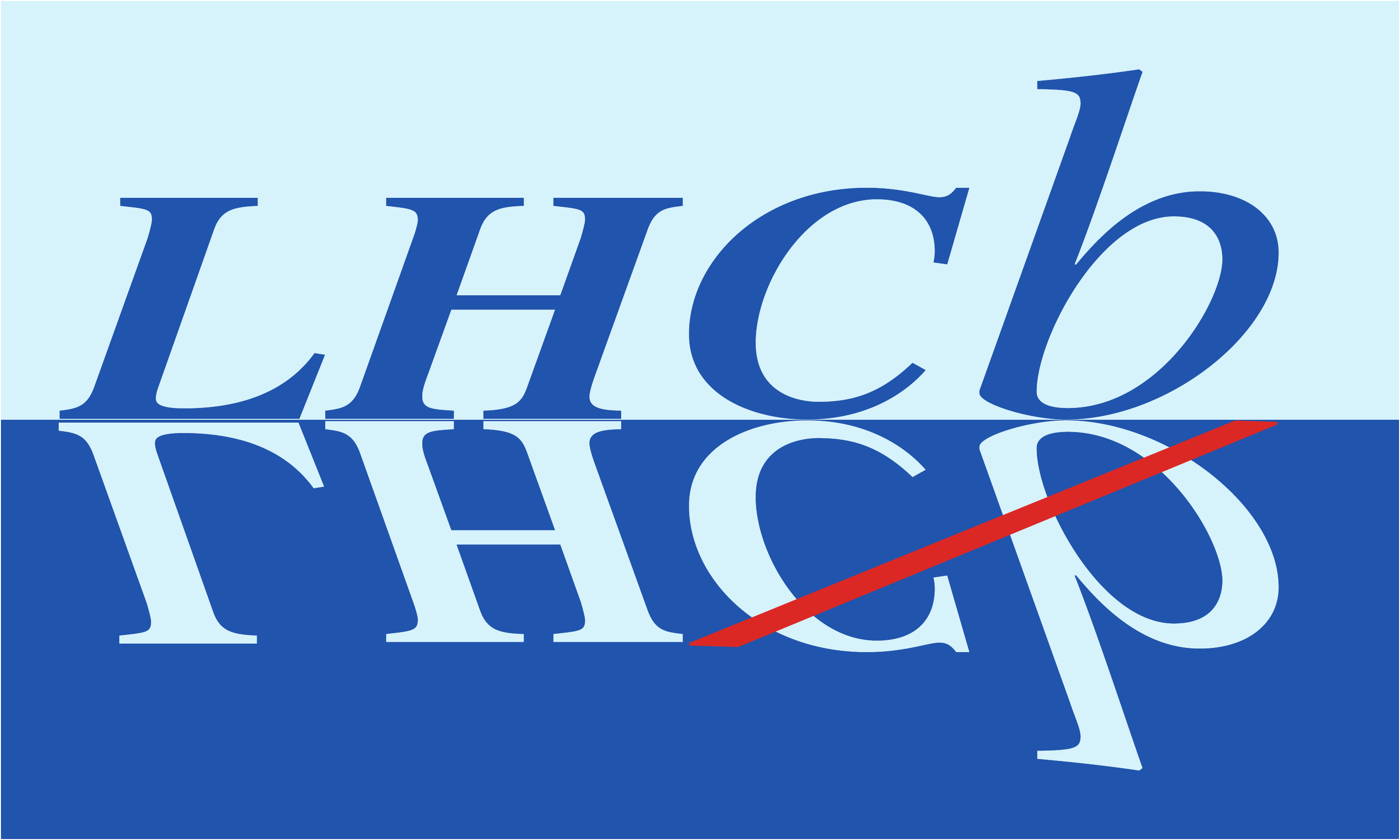}} & &}%
{\vspace*{-1.2cm}\mbox{\!\!\!\includegraphics[width=.12\textwidth]{figs/lhcb-logo.eps}} & &}%
\\
 & & CERN-EP-2021-121 \\  
 & & LHCb-PAPER-2021-016 \\  
 & & October 7, 2021 \\ 
 & & \\
\end{tabular*}

\vspace*{4.0cm}

{\normalfont\bfseries\boldmath\huge
\begin{center}
  \papertitle 
\end{center}
}

\vspace*{1.0cm}

\begin{center}
\paperauthors\footnote{Authors are listed at the end of this paper.}
\end{center}

\vspace{\fill}

\begin{abstract}
  \noindent
  This article presents differential measurements of the asymmetry between $\Lb$ and $\Lbbar$ baryon production rates in proton-proton collisions at centre-of-mass energies of $\sqs=7$ and $8\tev$ collected with the \lhcb experiment, corresponding to an integrated luminosity of 3\invfb. The $\Lb$ baryons are reconstructed through the inclusive semileptonic decay $\decay{\Lb}{\Lc\mun\neumb\PX}$. The production asymmetry is measured both in intervals of rapidity in the range $2.15<y<4.10$ and transverse momentum in \mbox{$2<\pt<27 \gevc$}. The results are found to be incompatible with symmetric production with a significance of 5.8 standard deviations for both $\sqs=7$ and $8\tev$ data, assuming no \CP violation in the decay. There is evidence for a trend as a function of rapidity with a significance of 4 standard deviations. Comparisons to predictions from hadronisation models in \pythia and heavy-quark recombination are provided. This result constitutes the first observation of a particle-antiparticle asymmetry in $\bquark$-hadron production at LHC energies.

\end{abstract}

\vspace*{0.5cm}

\begin{center}
  Published in JHEP 10 (2021) 060.
\end{center}

\vspace{\fill}

{\footnotesize 
\centerline{\copyright~\papercopyright. \href{\paperlicenceurl}{\paperlicence}.}}
\vspace*{2mm}

\end{titlepage}


\newpage
\setcounter{page}{2}
\mbox{~}
%
%
%
%

%% file: body.tex
\section{Introduction}
Measurements of production asymmetries allow the dynamics of quarks and gluons in high-energy particle collisions to be studied and, consequently, are a tool to test effective descriptions of the strong interaction.  Beauty quark and antiquark (\bquark\bquarkbar{}) pairs are produced in inelastic proton-proton ($pp$) collisions at the \lhc~\cite{Norrbin:2000zc}.
In the following hadronisation process, the interaction with the proton remnants can lead to a different production rate of \Lb{} baryons, with \uquark\dquark\bquark as their valence quarks,
compared to \Lbbar{} baryons, consisting of \(\overline{\uquark\dquark\bquark}\). The production asymmetry is defined as the relative difference between the production rates, \(\sigma(\decay{\proton\proton}{\Lb} Y)\) and \(\sigma(\decay{\proton\proton}{\Lbbar} Y)\), where $Y$ represents other produced particles,
\begin{equation}
 \label{eq:def_aprod}
 \Aprod \equiv \frac{\sigma(\decay{\proton\proton}{\Lb}Y)-\sigma(\decay{\proton\proton}{\Lbbar}Y)}{\sigma(\decay{\proton\proton}{\Lb}Y)+\sigma(\decay{\proton\proton}{\Lbbar}Y)}.
 \end{equation}
Models of the hadronisation process predict increased production-rate asymmetries at small angles with respect to the beam direction. In addition, the production asymmetry is expected to decrease with increasing centre-of-mass energy~\cite{Norrbin:2000jy, Norrbin:1999by, Braaten:2001bf, LbprodLai}.  

A precise measurement of the \Lb production asymmetry is crucial to improve the precision of \CP-violation measurements in the decays of \bquark baryons at the \lhcb experiment. The production asymmetry appears as background asymmetry in such measurements, and is the limiting systematic uncertainty for some of them~\cite{LHCB-PAPER-2018-025}. 
There have been several measurements of the production asymmetry of \Lb{} baryons at the \lhc{} \cite{LHCb-PAPER-2015-032,LHCb-PAPER-2016-062,CMS-2012136}; in none of them any significant effect was observed. In Ref.~\cite{LHCb-PAPER-2016-062} the \Lb{} production asymmetry is estimated indirectly from the production asymmetries measured in \Bd, \Bu and \Bs decays by assuming that \bquark-hadrons are produced in particle-antiparticle pairs.

This paper reports the measurement of the production asymmetry using the semileptonic \(\decay{\Lb}{\Lc\mun\neumb\PX}\) and its charge conjugate \(\decay{\Lbbar}{\Lcbar\mup\neum\PX}\) decay, where \PX{} denotes possible additional particles, for example from decays of excited charm baryons to \(\Lc\PX\) and \(\Lcbar\PX\) final states.\footnote{The inclusion of the charge-conjugated decays is implied when not specified otherwise.} The measurement is performed in intervals of the rapidity and the transverse momentum of the \Lb{} hadron. The chosen \Lb{} decay modes benefit from a high branching fraction of about 10\%~\cite{PDG2020}, but have the disadvantage that the momentum of the \Lb and \Lbbar baryons is not fully reconstructed.
The \Lc{} and \Lcbar{} baryons are reconstructed through their hadronic decays to $\proton\Km\pip$ and $\antiproton\Kp\pim$ final states, respectively. In the following it is assumed that no significant \CP-violating asymmetry is present in the reconstructed \Lb{} and \Lc{} decays, as each decay is dominated by one tree-level \bquark- or \cquark-quark transition, respectively.

The asymmetries caused by different reconstruction efficiencies for positively and negatively charged particles are a challenge in any measurement of particle-antiparticle asymmetries. One source of these so-called detection asymmetries is the difference in rates of interaction with the detector material for particles and antiparticles. A new method to precisely determine the interaction asymmetry of protons and antiprotons, which is of general interest, is developed as part of this analysis, and presented in detail.
The other sources of detection asymmetries are corrected using methods applied in previous \lhcb measurements.

\section{Detector and simulation}\label{sec:Detector}
The \lhcb detector~\cite{LHCb-DP-2008-001,LHCb-DP-2014-002} is a single-arm forward spectrometer designed for the study of particles containing \bquark or \cquark
quarks. It covers the pseudorapidity range $2<\eta <5$ where the pseudorapidity, \(\eta\), of a particle with momentum, $\vec{p}$, is defined as \(\arctanh(p_z/|\vec{p}|)\).\footnote{The \lhcb coordinate system is right-handed, with the \(z\) axis pointing along the beam axis, \(y\) the vertical direction, and \(x\) the horizontal direction.} The detector includes a high-precision tracking system consisting of a silicon-strip vertex detector (\velo) surrounding the $pp$ interaction region, a large-area silicon-strip detector (\ttracker) located
upstream of a dipole magnet, and three stations of silicon-strip detectors and straw
drift tubes (T stations) placed downstream of the magnet. The magnetic field of the dipole magnet has a bending power of about $4{\mathrm{\,Tm}}$ and its polarity is regularly reversed during data taking.  The horizontal  plane, \((x, z)\), is the bending plane of the dipole magnet. Its two polarities are referred to as up and down.
The tracking system provides a measurement of the momentum $\vec{p}$ of charged particles with
a relative uncertainty on the magnitude that varies from 0.5\% at low momentum to 1.0\% at 200\gevc.
The minimum distance of a track to a primary $pp$ collision vertex (primary vertex), the impact parameter, is measured with a resolution of $(15+29/\pt)\mum$,
where \pt is the component of the momentum transverse to the beam, in\,\gevc. Different types of charged hadrons are distinguished using information
from two ring-imaging Cherenkov detectors (RICH). Photons, electrons and hadrons are identified by a calorimeter system consisting of
scintillating-pad and preshower detectors, an electromagnetic and a hadronic calorimeter. Muons are identified by a system composed of alternating layers of iron and multiwire
proportional chambers. The online event selection is performed by a trigger, which consists of a hardware stage, based on information from the calorimeter and muon systems, followed by a software stage, based on information from a partial event reconstruction and subsequently a full event reconstruction. The trigger selection of candidates is described in more detail in \cref{sec:Selection}.

Simulation is required to model the effects of the detector response and the imposed selection requirements. In the simulation, $pp$ collisions are generated using \pythia~\cite{Sjostrand:2007gs,*Sjostrand:2006za} with a specific \lhcb configuration~\cite{LHCb-PROC-2010-056}. Decays of unstable particles are described by \evtgen~\cite{Lange:2001uf}, in which final-state radiation is generated using \photos~\cite{davidson2015photos}. The interaction of the generated particles and antiparticles with the detector, and its response, are implemented using the \geant toolkit~\cite{Allison:2006ve, *Agostinelli:2002hh} as described in Ref.~\cite{LHCb-PROC-2011-006}. However, detection asymmetries are corrected using data from control samples and do not rely on a precise description of particle--antiparticle differences in simulation. 

\section{Data samples and selection}\label{sec:Selection}
This analysis uses data samples of $pp$ collisions collected with the \lhcb experiment in 2011 at a centre-of-mass energy of 7\tev and in 2012 at a centre-of-mass energy of 8\tev.
These data correspond to integrated luminosities of 1\invfb and 2\invfb, respectively. The fraction of
data collected with up (down) polarity of the magnetic field is 40\% (60\%) in 2011 and 52\% (48\%) in 2012.

In the online processing, events consistent with $\decay{\Lb}{\Lc(\rightarrow \proton\Km\pip)\mun\neumb\PX}$ decays are first required to pass the hardware trigger, which selects events containing at least one muon with a transverse momentum, estimated from the hits in the muon stations, of $\pt>1.48\gevc$ in the 7\tev data or $\pt>1.76\gevc$ in the 8\tev data. In the subsequent software trigger, where the momentum is measured with high precision using the tracking system, a muon candidate is required to have $\pt>1.0\gevc$ and a significant impact parameter with respect to any primary vertex. In the next trigger stage, the muon and at least one of the \Lc{} decay products are required to be consistent with the topological signature of \(b\)-hadron decays~\cite{BBDT}.

In the offline processing, signal \Lb{} decays are inclusively reconstructed as \mbox{\(\Lc(\rightarrow \proton\Km\pip)\mun\)} candidates, \textit{i.e.} tracks identified as protons, kaons and pions are combined to an intermediate \Lc{} candidate which is subsequently combined with a muon candidate to form a \Lb{} candidate. In order to suppress background, all final-state particles are required to be reconstructed as tracks of high quality, to have a significant impact parameter with respect to any primary vertex and to have a relatively high (transverse) momentum. The muons are required to have $|\vec{p}|>6\gevc$ and $\pt>1.2\gevc$, kaons $|\vec{p}|>2.5\gevc$ and $\pt>250\mevc$, pions $|\vec{p}|>3.3\gevc$ and $\pt>250\mevc$, and protons $|\vec{p}|>10\gevc$ and $\pt>900\mevc$. Information from the RICH, calorimeter and muon systems is used to identify proton, kaon, pion and muon candidates with high confidence and small misidentification probability. To select \Lc{} candidates, protons, kaons and pions are required to form a good-quality vertex displaced from any primary vertex, and to have an invariant mass of \(\left| m(\proton\Km\pip)- 2286.46\mevcc \right|<80\mevcc\). Finally, \Lb{} candidates are selected by requiring a good-quality displaced \Lc\mun{} vertex, and the invariant mass of the \Lc\mun pair, \(m(\Lc\mun)\), is required to be in the range \([3.0, 5.6]\gevcc\). The momentum perpendicular to the \Lb{} flight direction that is missing due to unreconstructed decay products is equal in magnitude and opposite in direction compared to that of \Lc\mun pair, \(p_{\perp}(\Lc\mun)\), where the \Lb{} flight direction is defined by the primary vertex with the smallest impact parameter and the \Lc\mun{} vertex. This missing momentum is considered in the corrected mass\cite{SLD:1997ihw} defined as
\begin{equation}
\sqrt{m(\Lc\mun)^2 +p_{\perp}(\Lc\mun)^2 / c^2}+|p_{\perp}(\Lc\mun)|/ c \ .
\end{equation}
The corrected mass is required to be larger than 4.2\gevcc to further reject random-track (combinatorial) background. In total, about \(1.7\times10^6\) \Lb{} signal candidates are selected, \(0.5\times10^6\) in the \(\sqs=7\tev\) data set and \(1.2\times10^6\) in the \(\sqs=8\tev\) data set.

Additional requirements are applied to reduce detection asymmetries induced by the detector geometry. In some regions of phase space particles of a given charge are swept out of the detector by the dipole magnet. In order to exclude these regions, the horizontal  components of kaons and pions momenta are required to fulfil \(|p_x|<0.317\cdot(|\vec{p}|-3.4\gevc)\) when the rapidity, \(y\),  of the \Lb{} baryon, is smaller than \(2.58\). The rapidity is defined by the energy of the \Lb{} baryon, \(E\), and the component of its momentum along the beam direction, \(p_z\), as \(y = \frac{1}{2} \ln{\frac{E+p_z c}{E-p_z c}}\).
The rapidity region corresponds to the first interval of the measurement as defined in \cref{sec:bins}. The aforementioned momentum requirements on final state particles are more stringent for protons to reduce detection asymmetries from material interactions. In addition, protons are required to be within a pseudorapidity range of $2$ to $4.25$, to exclude regions of phase space where protons have to traverse significant amounts of material. 

\section{Formalism}\label{sec:Formalism}
The measured (raw) asymmetry between the decays \decay{\Lb}{\Lc\mun\neumb\PX} and \decay{\Lbbar}{\Lcbar\mup\neum\PX} is defined as
\begin{equation}
    \Araw = \frac{N\left( \recLb \right) - N\left( \recLbbar \right)}{N\left( \recLb \right) + N\left( \recLbbar \right)},
\end{equation}
where $N$ denotes the number of observed decays. The raw asymmetry can be expressed in terms of the production rates of \Lb and \Lbbar hadrons, $\sigma\left(\decay{\proton\proton}{\Lb Y} \right)$ and $\sigma\left(\decay{\proton\proton}{\Lbbar Y} \right)$, and the reconstruction efficiencies $\epsilon(\Ph^\pm)$ of given particle species $h$ denoting proton, kaon, pion, and muon tracks, as
\begin{equation}
    \Araw = \frac{\sigma(\decay{\proton\proton}{\Lb Y}) \epsilon(\proton)\epsilon(\Km)\epsilon(\pip)\epsilon(\mun) - \sigma(\decay{\proton\proton}{\Lbbar Y}) \epsilon(\antiproton)\epsilon(\Kp)\epsilon(\pim)\epsilon(\mup)}{\sigma(\decay{\proton\proton}{\Lb Y}) \epsilon(\proton)\epsilon(\Km)\epsilon(\pip)\epsilon(\mun) + \sigma(\decay{\proton\proton}{\Lbbar Y}) \epsilon(\antiproton)\epsilon(\Kp)\epsilon(\pim)\epsilon(\mup)}.
\end{equation}
The detection asymmetry of a particle \(\Ph^+\) is defined as
\begin{equation}
    \label{eq:def_adet}
    A_D(\Ph^+) = \frac{\epsilon(\Ph^+)-\epsilon(\Ph^-)}{\epsilon(\Ph^+)+\epsilon(\Ph^-)}\ ,
\end{equation}
and similarly for an antiparticle \(\Ph^-\). 
Using the definitions in \cref{eq:def_aprod}, \cref{eq:def_adet}, and neglecting third-order and higher terms in the asymmetries, the raw asymmetry can be approximated as
\begin{equation}
    \label{eq:meas_aprod}
    \Araw \approx \Aprod + A_D(\proton\Km\pip\mun) =\Aprod + \Adet(\proton) + \Adet(\Km) + \Adet(\pip) + \Adet(\mun)\;.
\end{equation}
Equation~\ref{eq:meas_aprod} is valid up to corrections of order \(10^{-4}\), which is well below the statistical uncertainty of the measurement.

The detection asymmetries of different particle species have different components. All charged particles are affected by the magnetic dipole field which deflects positively and negatively charged particles in opposite directions. Thus, left-right asymmetric imperfections of the detector in the tracking, particle identification or trigger systems produce detection asymmetries for charged particles. These detection asymmetries can be studied and effectively reduced by reversing the polarity of the dipole magnet.
To be detected, a charged particle must traverse the full tracking system without undergoing an inelastic scatter or an elastic scatter with a large deflection angle~\cite{Dufour:2708057}. Therefore, different interaction rates for particles and antiparticles with the detector material are a source of detection asymmetries, referred to as interaction asymmetries. This type of detection asymmetry is mostly independent of the magnet polarity. A sizable interaction asymmetry in the relevant momentum range of 2 to 100\gevc{} is expected for kaons and protons, a negligible interaction asymmetry is expected for pions, and the interaction asymmetry is absent for muons.

The corrections due to detection asymmetries are determined in separate steps, approximating  \( A_D(\proton\Km\pip\mun)\) in \cref{eq:meas_aprod} as
\begin{equation}
    A_D(\proton\Km\pip\mun)\approx \aintp + \apidp + \adetmu + \atrackpmu + \adetkpi,
\end{equation}
where \aintp{} denotes the interaction asymmetry of protons (\cref{sec:proton_int_det}); \apidp{} the asymmetry from identifying protons (\cref{sec:proton_pid_det}); \adetmu{} the asymmetry from identifying muons and triggering on the muon candidate (\cref{sec:muon_det}); \atrackpmu{} is the tracking asymmetry of the muon and proton independent of interaction asymmetries (\cref{sec:pmu_track_det}); and \adetkpi{} the combined detection asymmetry for the \kpi{} pair due to tracking, particle identification and material interaction (\cref{sec:kpi_det}).

The analysis is performed separately for the \(\sqs=7\tev\) and \(8\tev\) data as the production asymmetry is expected to change with  centre-of-mass energy. In addition, detection asymmetries may change due to different operational conditions of the detector. In principle, the method does not rely on any cancellation due to the regular field reversal, since all detection asymmetries are corrected for. This is tested by performing the analysis separately for data samples collected with the two magnet polarities.
To reduce any residual biases, the final results are determined from the arithmetic mean of the results obtained for these two samples.

\subsection{Measurement as a function of \boldmath{\Lb{}} kinematics}\label{sec:bins}
\begin{figure}[t]
	\centering
	\includegraphics[width=0.59\textwidth]{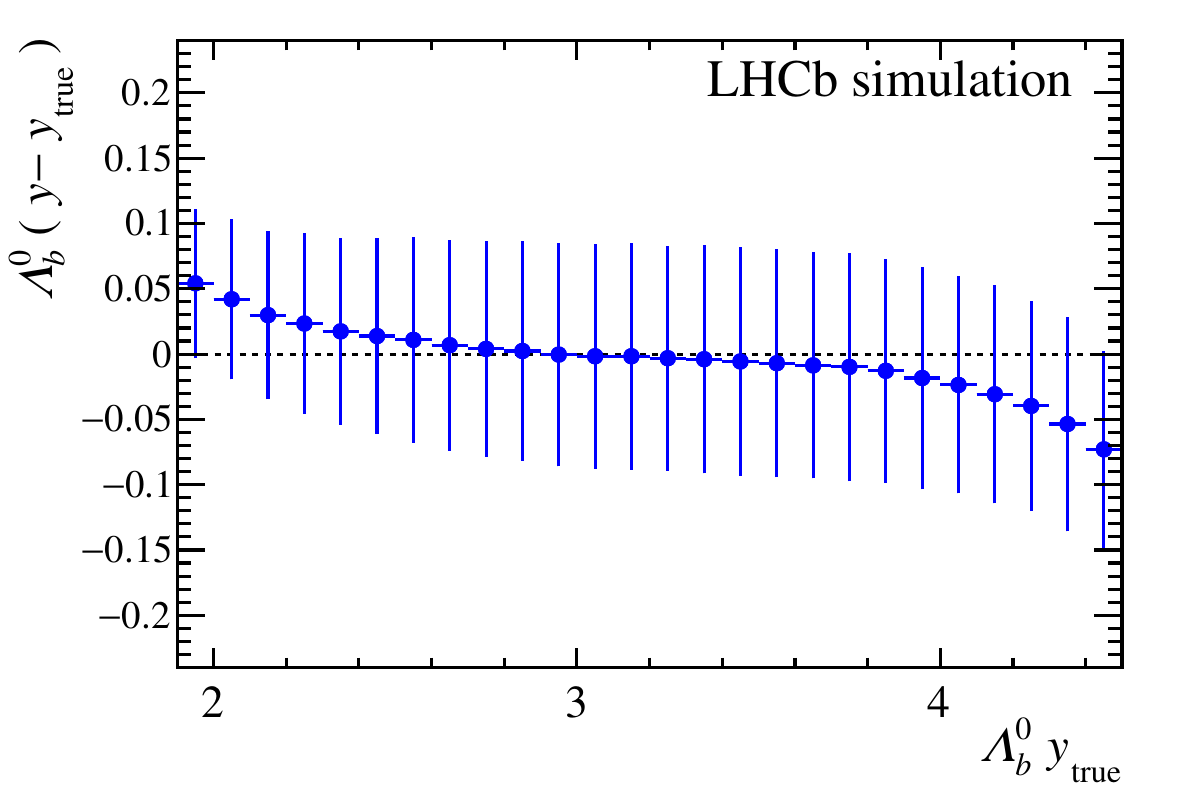}
	\caption{Mean and the standard deviation of the difference of reconstructed and true rapidity of $\Lb$ baryons as a function of the true rapidity obtained from simulation. The reconstructed rapidity $y$ is approximated by the rapidity of the $\Lc\mun$ system. The markers correspond to the means and the error bars correspond to the standard deviations of the distributions in a $y_\text{true}$ interval.}
	\label{fig:rapidity_resolution}
\end{figure}

Since the production asymmetry is expected to depend on the kinematics of the \Lb{} baryon, the measurement is performed in intervals of rapidity \(y\), and alternatively in intervals of transverse momentum \pt.
Although the semileptonic decay is only partially reconstructed, the measured quantities of the \Lc\mun{} pair are a good proxy for the true rapidity of the \Lb{} baryon and no correction is applied.
The bias and dispersion of this approximation are studied using simulated $\decay{\Lb}{\Lc(\rightarrow \proton\Km\pip)\mun\neumb}$ decays, and shown in \cref{fig:rapidity_resolution} as a function of the true rapidity of a \Lb{} baryon. The resolution is less than 0.1 units of rapidity in most regions and small biases are observed at high and low rapidities. No significant difference is observed for simulated events at \(\sqs=7\) and 8\tev. The impact of additional missing particles due to decays of excited charm baryons is studied as part of the systematic uncertainties in \cref{sec:systematic}. 
The seven rapidity intervals used in the analysis are
\begin{equation}
    y : [2.15, 2.58]; [2.58, 2.80]; [2.80, 3.00]; [3.00, 3.20]; [3.20, 3.43];[3.43, 3.70]; [3.70, 4.10].
\end{equation}
The intervals are chosen to be larger than the average resolution and to be roughly equally populated, except for the highest-rapidity interval which is less populated. The effects due to migrations with respect to the true rapidity are found to be small compared to the statistical uncertainty in all intervals except for the last one. Residual biases are included in the systematic uncertainties.

\begin{figure}[t]
	\centering
	\includegraphics[width=0.59\textwidth]{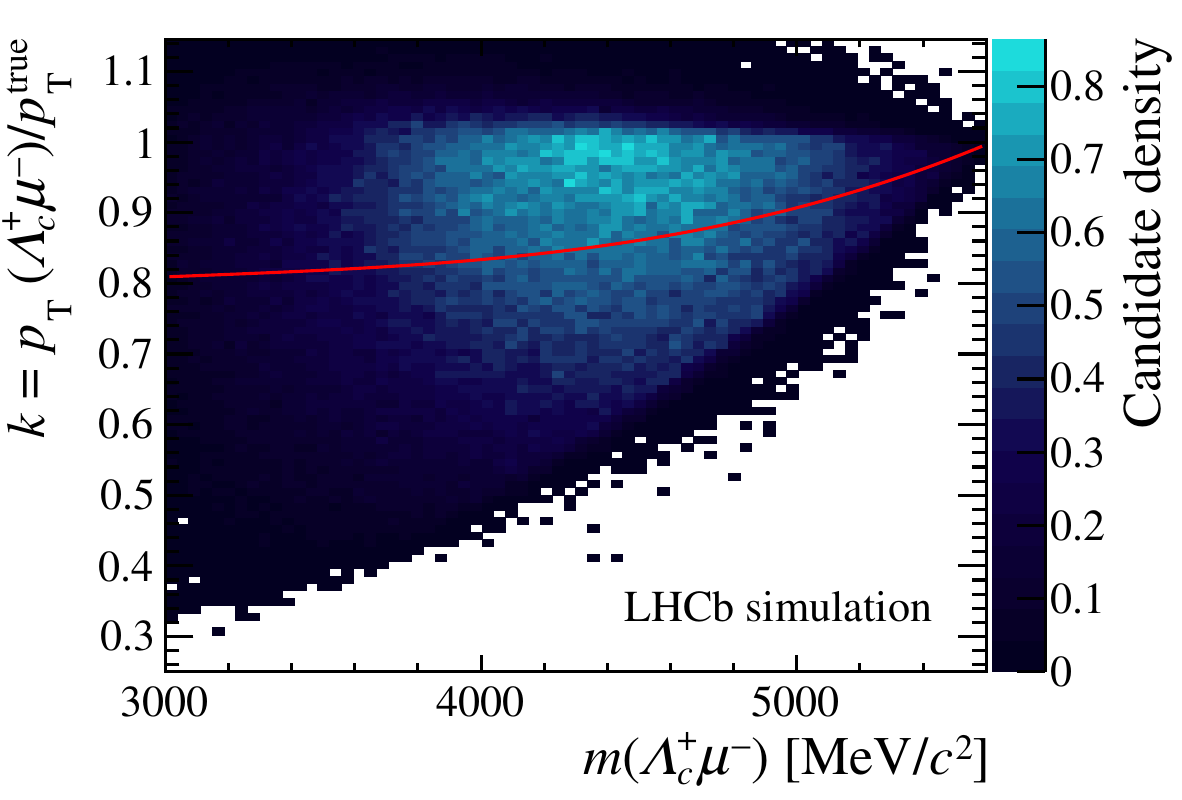}
	\caption{Distributions of the $k$-factors in simulated data as a function of the $\Lc\mun$ invariant mass. The average correction $k(m(\Lc\mun))$  is overlaid in red. The colour code describes the normalised density of the distribution. }
	\label{fig:kfactors_2d_distribution}
\end{figure}

The transverse momenta of the \Lc\mun{} pairs underestimate the transverse momenta of \Lb{} baryons on average. A correction factor, \(k\), defined as the ratio of the true and the reconstructed \pt{}, is determined as a function of the invariant mass of the reconstructed \Lc\mun{} system, \(m(\Lc\mun)\), using simulated events. The distribution of the correction factor is shown in \cref{fig:kfactors_2d_distribution}. A third-order polynomial, \(k(m(\Lc\mun))\), is fitted to the distribution and used to correct the \pt{} of candidates reconstructed in 7 and 8\tev data. The procedure follows other analyses involving semileptonic decays at the \lhcb experiment~\cite{LHCb-PAPER-2015-031, LHCb-PAPER-2019-033}.
Subsequently, the \pt-dependent measurement of the \Lb{} production asymmetry is performed in intervals of \(\ptrec = \pt(\Lc\mun)/k(m)\).
Fewer \pt{} than \y{} intervals are chosen due to the poorer resolution of the reconstructed transverse momentum. Five \pt{} intervals are considered as follows
\begin{equation}
    \pt[\gevc] : [2, 4]; [4, 8]; [8, 12]; [12, 18]; [18, 27]\ .
\end{equation}
Only candidates which fall in the rapidity range \(2.15 < y < 4.10\) are used. Despite the $k$-factor correction, the large spread in the \(\pt(\Lc\mun)/p_T^\text{true}\) distribution results in the migration of events into neighbouring \pt{} intervals.
The impact of this on the measurement is discussed in the context of systematic uncertainties in \cref{sec:systematic}.

\section{Measurement of raw asymmetries}\label{sec:raw_asymmetry_measurement}
\begin{figure}[]
	\begin{center}
		\includegraphics[width=0.48\textwidth]{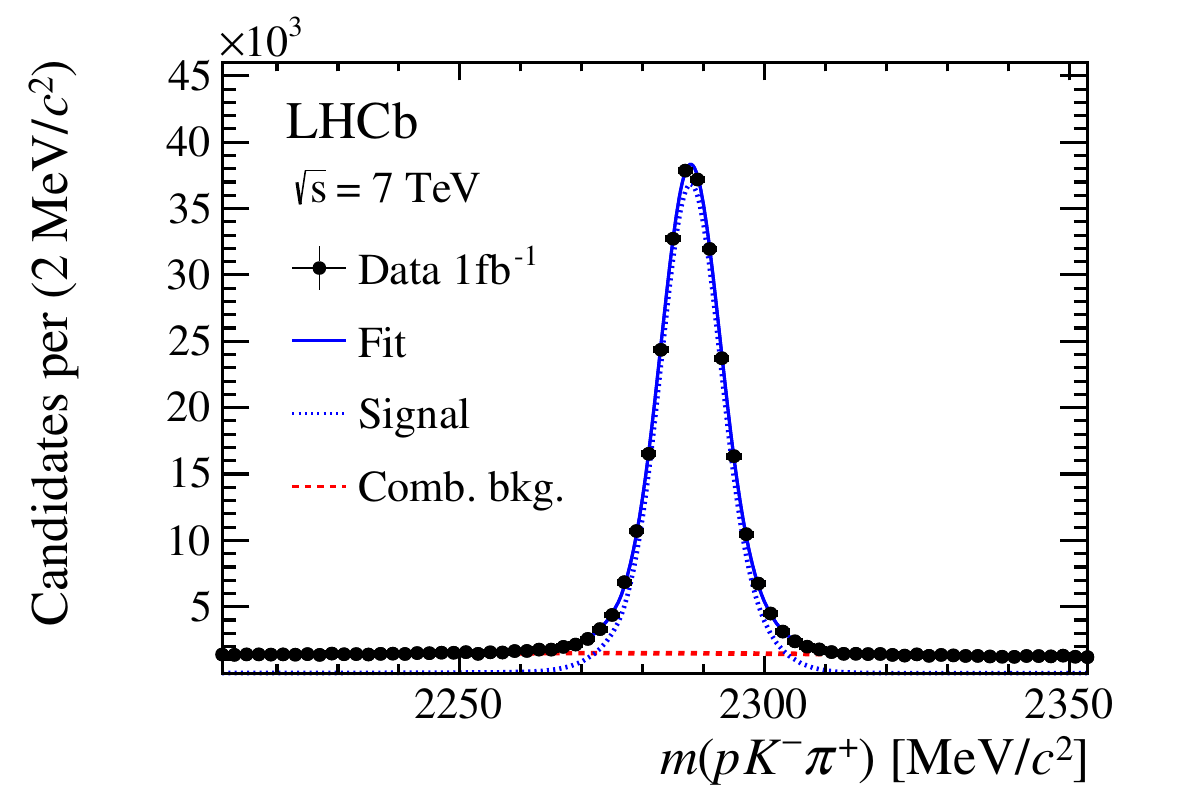}
		\includegraphics[width=0.48\textwidth]{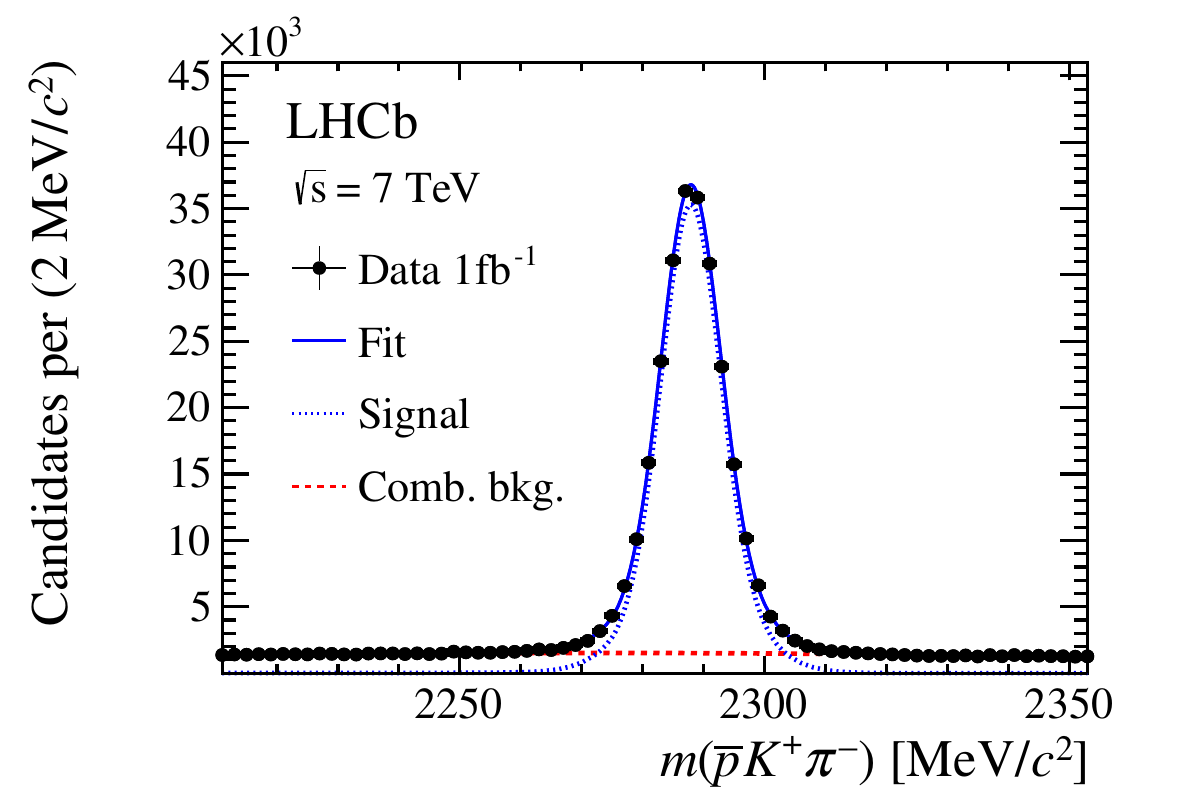} \\
		\includegraphics[width=0.48\textwidth]{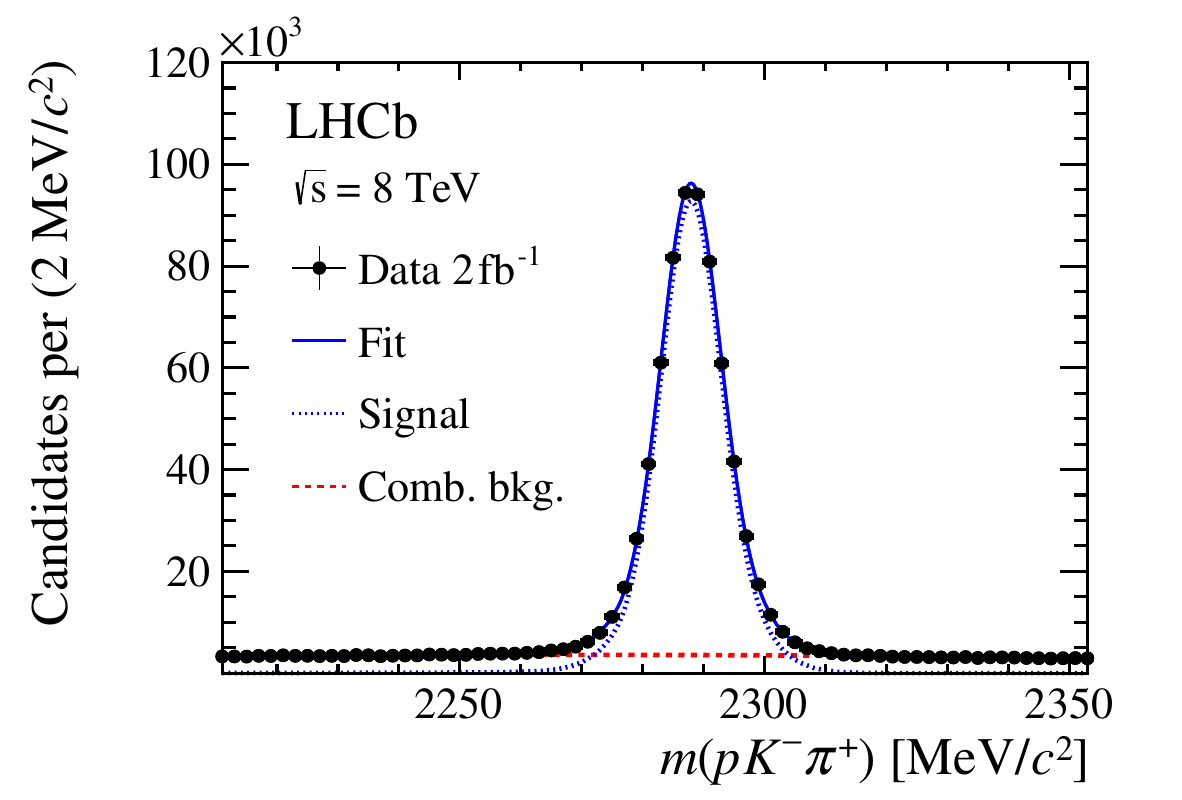}
		\includegraphics[width=0.48\textwidth]{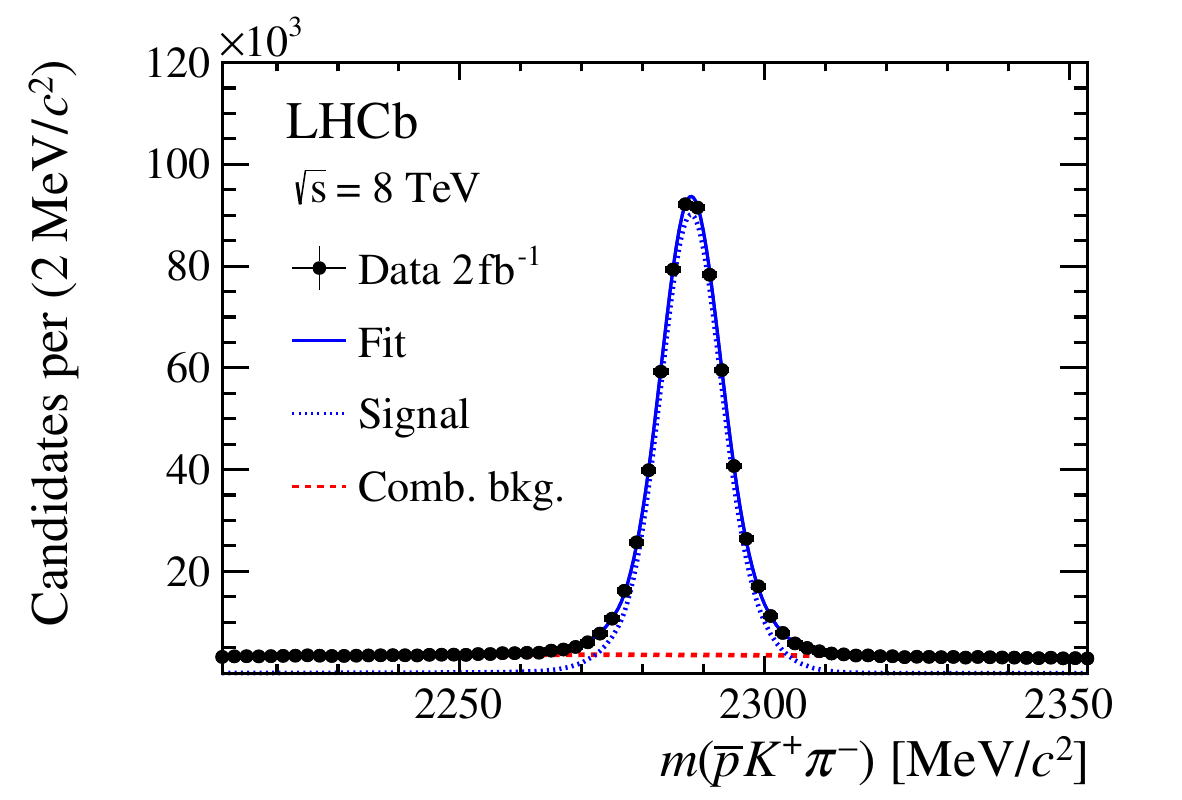} 
		\caption{Distributions of the $\proton\Km\pip$ invariant mass for (left) $\Lc\mun$ and (right) $\Lcbar\mup$ candidates for the (top) $\sqs=7\tev$ and (bottom) $\sqs=8\tev$ data set. Fits to the distributions, as described in the text, are shown as well. }
		\label{fig:raw_asymmetry_example_fit}
	\end{center}
\end{figure}

The signal yields and raw asymmetries are determined through unbinned maximum-likelihood fits to the $\proton\Km\pip$ and $\antiproton\Kp\pim$ invariant-mass distributions in the twelve kinematic bins, separately for the two centre-of-mass energies and the two magnet polarities. The signal components for \Lc and \Lcbar candidates are modelled by the sum of a Gaussian function and a Gaussian function with a power-law tail at low mass, the background components are modelled by second-order Chebyshev polynomials. The \Lc and \Lcbar signal yields are expressed in terms of the total yield and the raw asymmetry, such that the raw asymmetry is a fit parameter. The low-mass tail parameters of the \Lc{} and \Lcbar{} signal components are constrained to be similar, with a maximum relative difference of 20\%, as is the fraction of the second Gaussian function. All other parameters are fit independently for \Lc and \Lcbar signal and background components.
Example fits are shown in \cref{fig:raw_asymmetry_example_fit}.

\begin{figure}[]
	\begin{center}
		\includegraphics[width=0.49\textwidth]{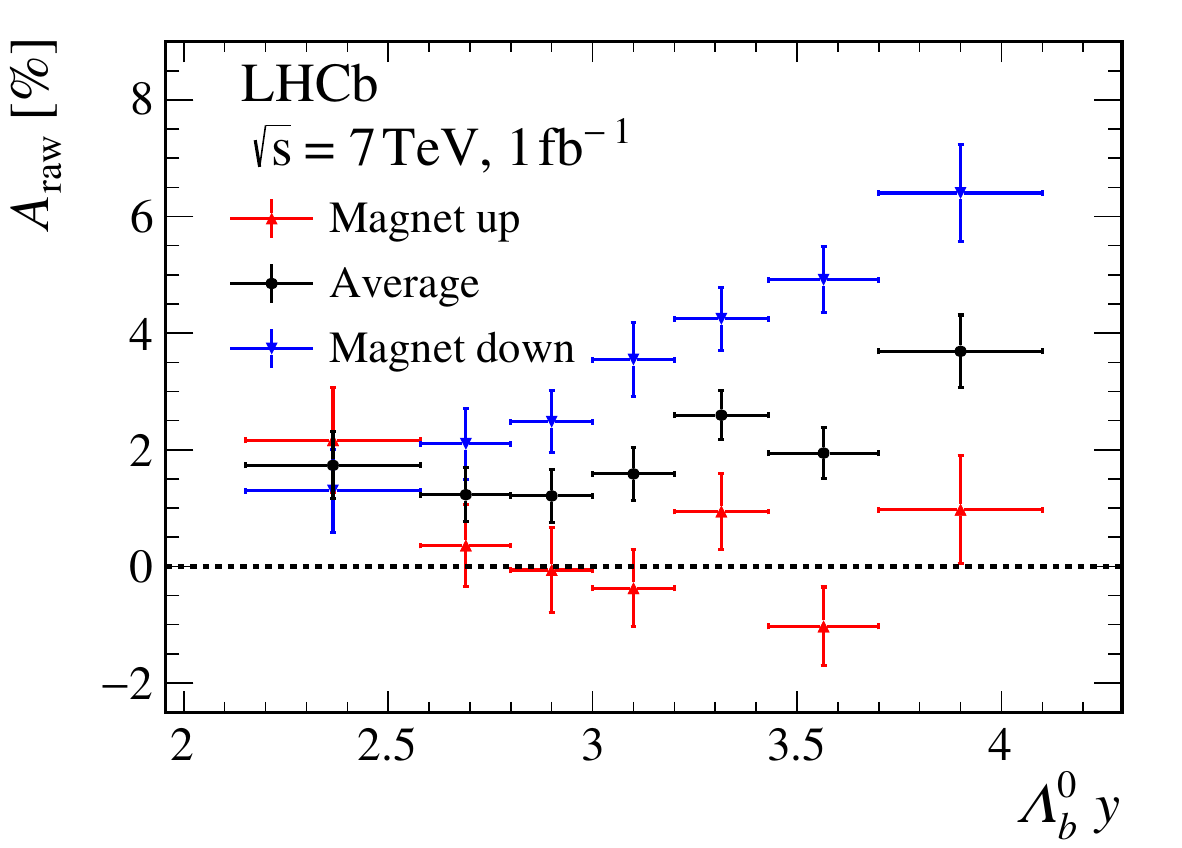}
		\includegraphics[width=0.49\textwidth]{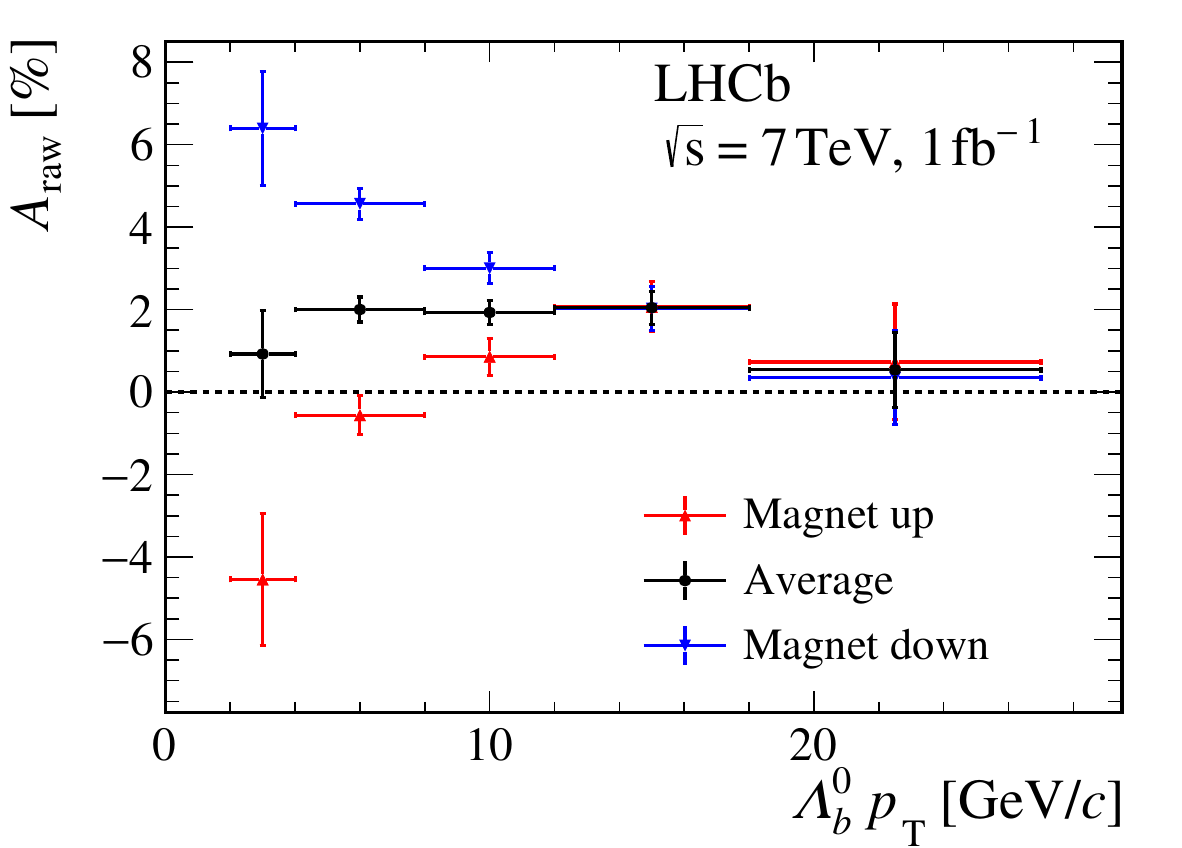}\\
		\includegraphics[width=0.49\textwidth]{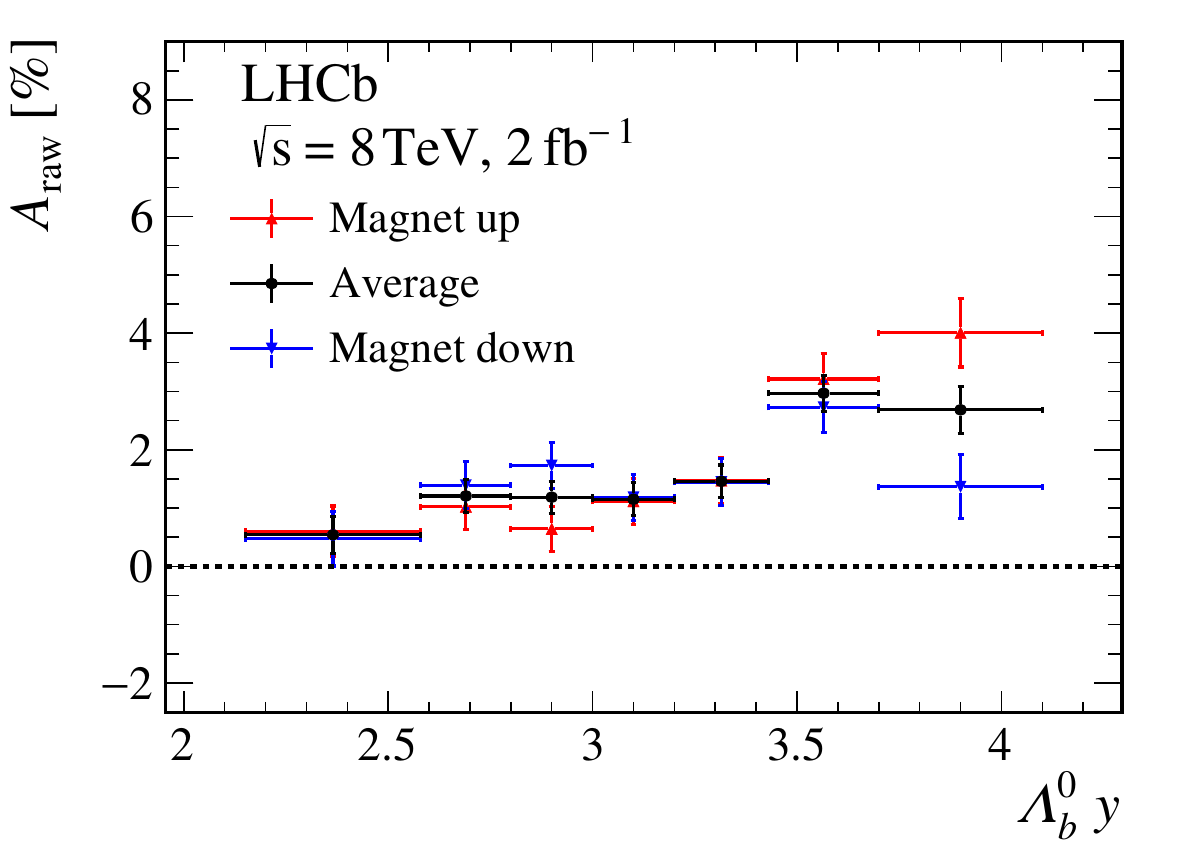}
		\includegraphics[width=0.49\textwidth]{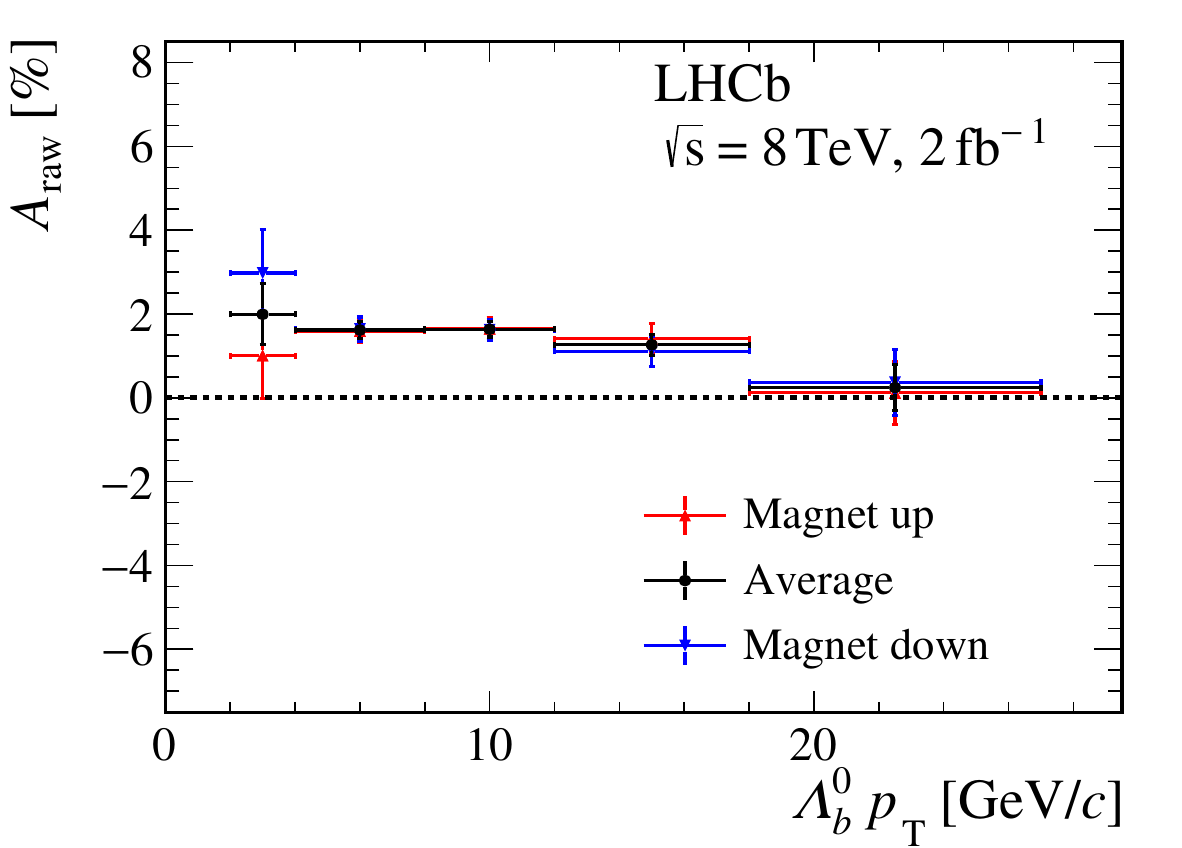}\\
		\caption{Measured raw asymmetry versus (left) rapidity and (right) $\pt$ for $\Lb$ candidates from data taken at centre-of-mass energies of (top) $\sqs=7\tev$ and (bottom) $\sqs=8\tev$. The results are shown separately for (red upward triangles) magnet up, (blue downward triangles) magnet down and (black dots) their average. Uncertainties are statistical only.}
		\label{fig:raw_asymmetry_result_stat}
	\end{center}
\end{figure}

The resulting raw asymmetries in intervals of the \yrec{} and \pt{} of the \Lb{} baryons are shown in \cref{fig:raw_asymmetry_result_stat} for $\sqs=7\tev$ and $\sqs=8\tev$ data, separately for data taken with the two magnet polarities and their arithmetic average. On average, a positive raw asymmetry increasing with \(y\) but independent of \pt{} is measured. The difference between the two magnet polarities, more pronounced for \(\sqs=7\tev\) data, is due to polarity-dependent detection asymmetries discussed in the following. 

\section{Detection asymmetries}\label{sec:detection_asymmetries}
As presented in \cref{sec:Formalism}, the contributions to the measured asymmetry are split into different parts, so that the correction for detection asymmetries per given kinematic interval $i$ can be written as
\begin{eqnarray}
    \label{eq:method_aprod}
    A_D^i(\proton\Km\pip\mun) &=&  \aintpi + \apidpi + \adetkpii\\
                    && + \adetmui + \atrackpmui \ \nonumber . 
\end{eqnarray}
As the kinematic distributions of the final-state particles overlap for the intervals in which the measurement is performed, the corrections are correlated and full correlation matrices are determined. In the following, the measurements of the individual detection asymmetries are presented.

\subsection{Proton-interaction asymmetry}\label{sec:proton_int_det}
A novel technique to calibrate the proton-interaction asymmetry at the \lhcb detector is used and described here. The method is general for the most part and potentially useful for other experiments.
It exploits a combination of external measurements of cross-sections of proton and antiproton scattering on deuterium targets, a detailed \lhcb detector simulation, and large samples of \decay{\Lz}{\proton\pim} decays for a calibration determined from data. 

\subsubsection{Formalism}
Charged hadrons interact strongly with the nuclei inside the detector material. The nuclear collision length, \(\lambda_T^{\pm}(|\pvec|)\), is the typical length that a hadron or an antihadron with momentum \(\pvec\) travels before it undergoes an elastic or inelastic scatter. The inverse of the nuclear collision length is given by
\begin{equation}
    \frac{1}{\lambda_T^\pm(|\pvec|)} = \frac{\sigma_T^\pm(|\pvec|)\;\rho\;\NA}{A}\ ,
\end{equation}
where \(\sigma_T^{\pm}(|\pvec|)\) is the momentum- and material-dependent total hadron or antihadron interaction cross-section, \(\rho\) the material density, \(A\) the number of nucleons in the nuclei of the material and \NA{} Avogadro's number. The efficiencies to reconstruct protons and antiprotons which traverse material with thickness \(d\), are proportional to \(\exp\left({-d/\lambda_{T}^\pm(|\pvec|)}\right)\). As the detector consists of a variety of materials, the total probability for a hadronic interaction is given by the product of the corresponding probabilities.
The interaction asymmetry for protons and antiprotons traversing the detector is then given by 
\begin{equation}
	\aintp(\pvec) = \frac{\exp(- \sum_i d_i/\lambda_{T,i}^{+}(|\pvec|)) - \exp(-\sum_i d_i/\lambda_{T,i}^{-}(|\pvec|))}{\exp(-\sum_i d_i/\lambda_{T,i}^{+}(|\pvec|)) + \exp(-\sum_i d_i/\lambda_{T,i}^{-}(|\pvec)|)} \ ,
    \label{eq:aintp}
\end{equation}
where the sum is taken over all path lengths and materials encountered up to the last tracking station. Critical inputs to this equation are a description of the material and momentum-dependent cross-sections, as well as an accurate simulation of the detector.

\subsubsection{Momentum dependence}
\begin{figure}[tbp]
    \begin{center}
        \includegraphics[width=0.65\textwidth]{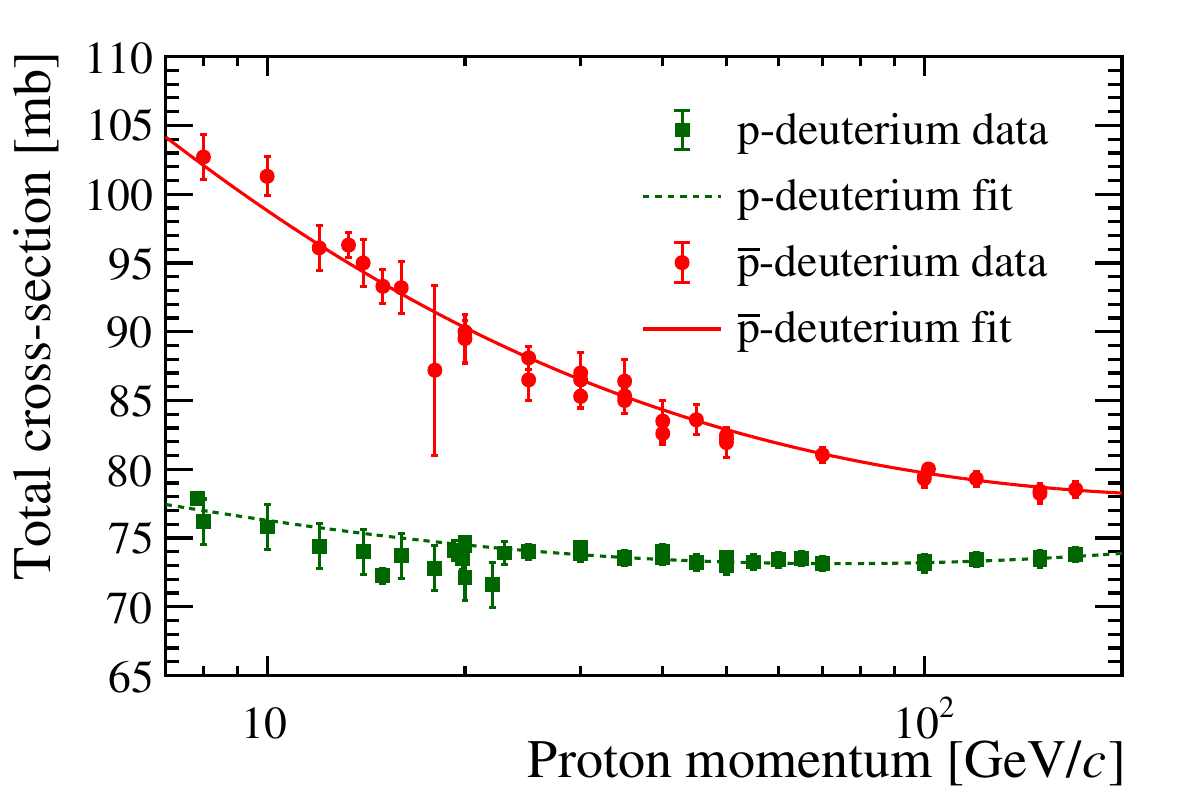}
		\caption{Measured total proton-deuterium and antiproton-deuterium cross-sections as function of proton momentum are shown with green squares and red dots, respectively. Fits according to the model described in the text are overlaid. Data are taken from Ref.~\cite{PDG2016}.}
			\label{fig:proton_asymmetry_momentum_fit}
    \end{center}
\end{figure}

Precise measurements of the total cross-sections of proton and antiproton scattering on isoscalar deuterium targets for momenta up to 270\gevc~\cite{PDG2016} are used to determine the momentum dependence of the interaction asymmetry. The use of these cross-sections is motivated by the fact that the \lhcb detector is almost an isoscalar target, as obtained from simulation. In Ref.~\cite{PDG2016}, the total cross-section data as a function of the centre-of-mass energy squared, \(s\), are fitted with an analytical function. For a more conservative estimate of the model uncertainties compared to Ref.~\cite{PDG2016} additional degrees of freedom per charge, \(Z^\pm\) and \(Y_2^\pm\), are introduced. The function is given by
\begin{equation}
\sigma_d^\pm(\pvec) = Z^\pm + B \log^2\left(s/s_{M}\right) + Y_{1}(s_M/s)^{\eta_1} \mp Y_2^\pm(s_M/s)^{\eta_2}\ ,
\end{equation}
where $s_M$ is defined as $\left(m_p + m_d + M\right)^2$ where the constant $M$ is a model parameter, and $m_{p}$ and $m_{d}$ denoting the masses of proton and deuterium nuclei. The parameters $M$, $B$, $\eta_1$ and $\eta_2$ are taken from Ref.~\cite{PDG2016}, and parameters denoted with \(\pm\) are different in the fits to proton and antiproton deuterium cross-sections. The fits to these data are repeated in the relevant momentum range of 5 to 170\gevc. Data and fit results are shown in \cref{fig:proton_asymmetry_momentum_fit} and the resulting fitted parameters are given in \cref{tab:parameters_cross_section_function}. 

\begin{table}[htbp]
    \caption{Parameters obtained in the description of the (anti)proton-deuterium cross-sections along with their uncertainties.}
    \centering
    \bgroup
    \def\arraystretch{1.25}
    \begin{tabular}{lcc}
    \toprule
     Parameter &  Proton & Antiproton              \\ \midrule
      $Z^\pm [\mbarn]$  & $64.76 \pm 0.32$         & $64.45 \pm 0.29$   \\
      $Y_1 [\mbarn]$ & \multicolumn{2}{c}{$29.66 \pm 0.39$} \\
      $Y_2^\pm [\mbarn]$& $15.97 \pm 0.51$  & $14.80 \pm 0.94$    \\  \bottomrule
      \end{tabular}
      \egroup
    \label{tab:parameters_cross_section_function}
\end{table}

\subsubsection{Material dependence}\label{sec:material_dependence}
For elements with larger nucleon number, an extra nucleon contributes less to the cross-section than in a small atom. 
This so-called screening effect can be described by
\begin{equation}
    \frac{\sigma_A^\pm}{A} \propto A^{-\alpha}\ ,
\end{equation}
where \(\sigma_A^{+(-)}\) is the (anti)proton interaction cross-section for an element with nucleon number \(A\) and \(\alpha\) is the screening factor. In the case of densely-packed hard spheres \(\alpha\) equals \(1/3\), point-like scatterers would give \(\alpha=0\). The momentum dependence is taken from the well known proton- and antiproton-deuterium cross-sections discussed above and, thus, cross-sections used in the interaction asymmetry estimation are approximated as
\begin{equation}
    \frac{\sigma_A^\pm(|\pvec|)}{A} \approx \frac{\sigma_d^\pm(|\pvec|)}{2^{1-\alpha}} A^{-\alpha}\ .
\end{equation}
The same screening factor for protons and antiprotons is used, and it is assumed that the screening factor is independent of momentum. Screening factors are determined from the cross-sections of protons with different materials as implemented in \geant and the LHEP physics list~\cite{Allison:2006ve, *Agostinelli:2002hh}.  
For the inelastic and total cross-sections screening factors of \(\alpha = 0.253\) and \(\alpha = 0.195\), respectively, are used. As not all elastic scatters lead to loss of a particle, the screening factor for the inelastic cross-section is used as default, and that for the total cross-section is considered in systematic studies.

\subsubsection{Modifications to formalism}
Modifications to the model are introduced to better describe the asymmetries observed in simulation and data. First, some elastic hadronic scatters lead to an efficiency loss while some inelastic scatters do not lead to an efficiency loss, \textit{e.g.}\ a proton scattering shortly before the end of the tracking stations can still leave enough hits in the detector for its trajectory to be reconstructed. Momentum-dependent factors, \(C_{|\pvec|}^{\text{sim}}\), are determined from simulation and account for these effects. In addition, these factors depend on the reconstructed final state since the impact of the momentum spread depends on the \(Q\) value of the decay. Therefore, they differ for the \decay{\Lz}{\proton\pim} calibration channel, discussed in more detail later, and the \decay{\Lc}{\proton\Km\pip} signal channel. Second, since the description of the detector material in simulation is not perfect, an overall scaling factor, \(F_{\text{data}}\), is introduced. It is fixed to unity when validating the method with simulation but is allowed to vary freely when calibrating with data. The factor \(F_{\text{data}}\) can also absorb differences in the cross-sections between data and simulation. Consequently, \cref{eq:aintp} is changed to
\begin{equation}
	\aintp(\pvec) = \frac{\exp(-C_{|\pvec|}^{\text{sim}} F_{\text{data}} \sum_i d_i/\lambda_{T,i}^{+}(|\pvec|)) - \exp(-C_{|\pvec|}^{\text{sim}} F_{\text{data}}\sum_i d_i/\lambda_{T,i}^{-}(|\pvec|))}{\exp(-C_{|\pvec|}^{\text{sim}} F_{\text{data}}\sum_i d_i/\lambda_{T,i}^{+}(|\pvec|)) + \exp(-C_{|\pvec|}^{\text{sim}} F_{\text{data}}\sum_i d_i/\lambda_{T,i}^{-}(|\pvec|))} \ .
    \label{eq:aintpfull}
\end{equation}   

\subsubsection{Validation with calibration data}\label{sec:proton_calibration}
\begin{figure}[b]
	\begin{center}
		\includegraphics[width=0.48\textwidth]{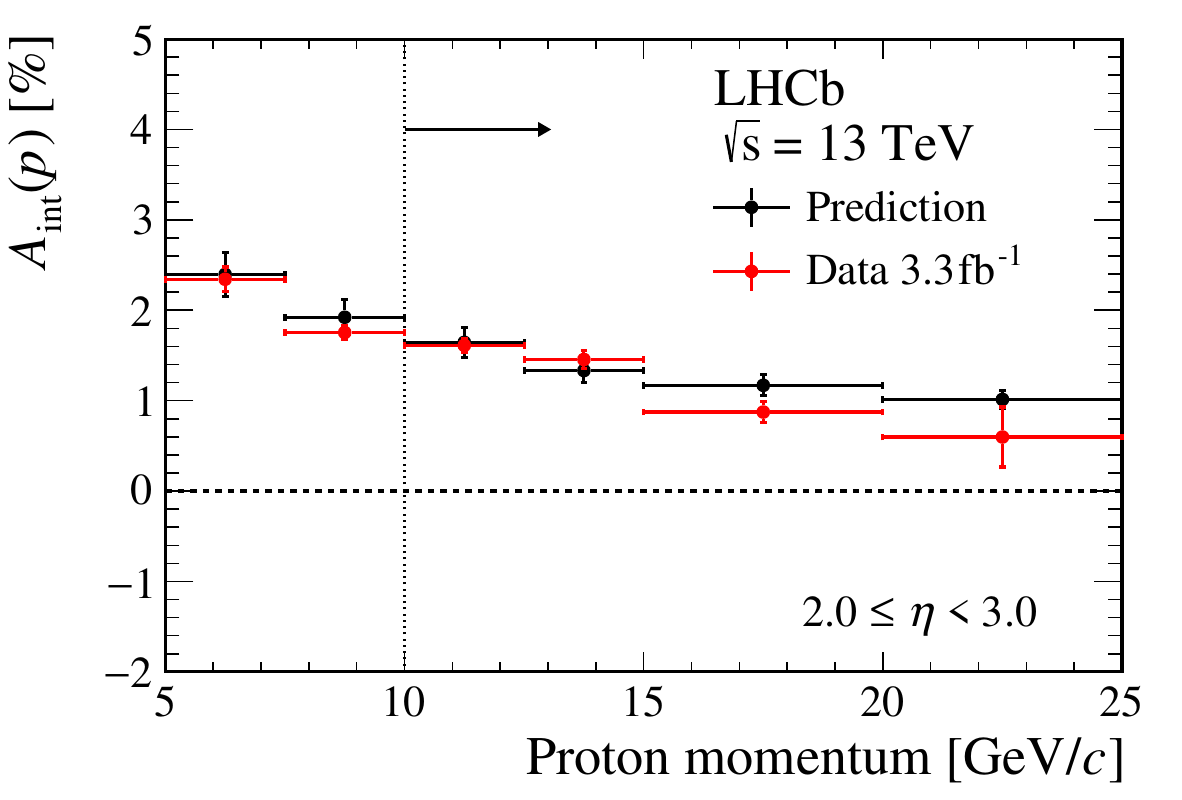}
		\includegraphics[width=0.48\textwidth]{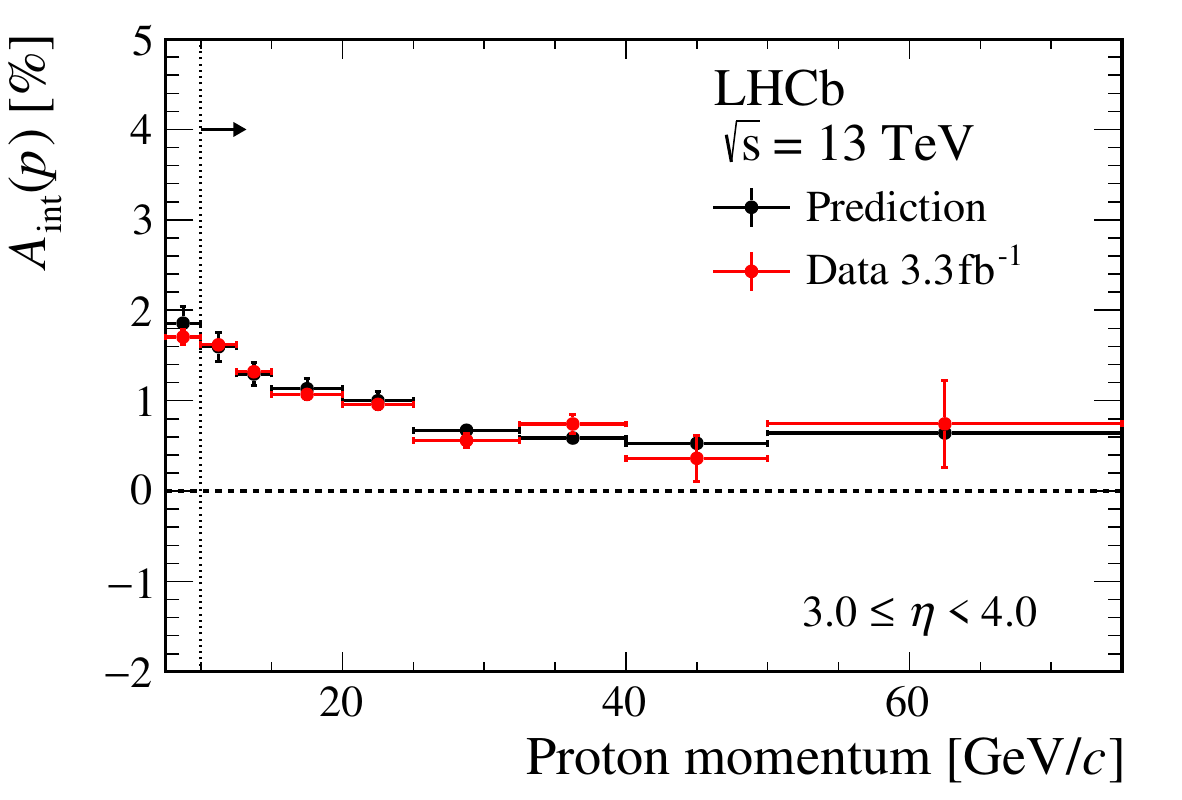}
		\includegraphics[width=0.48\textwidth]{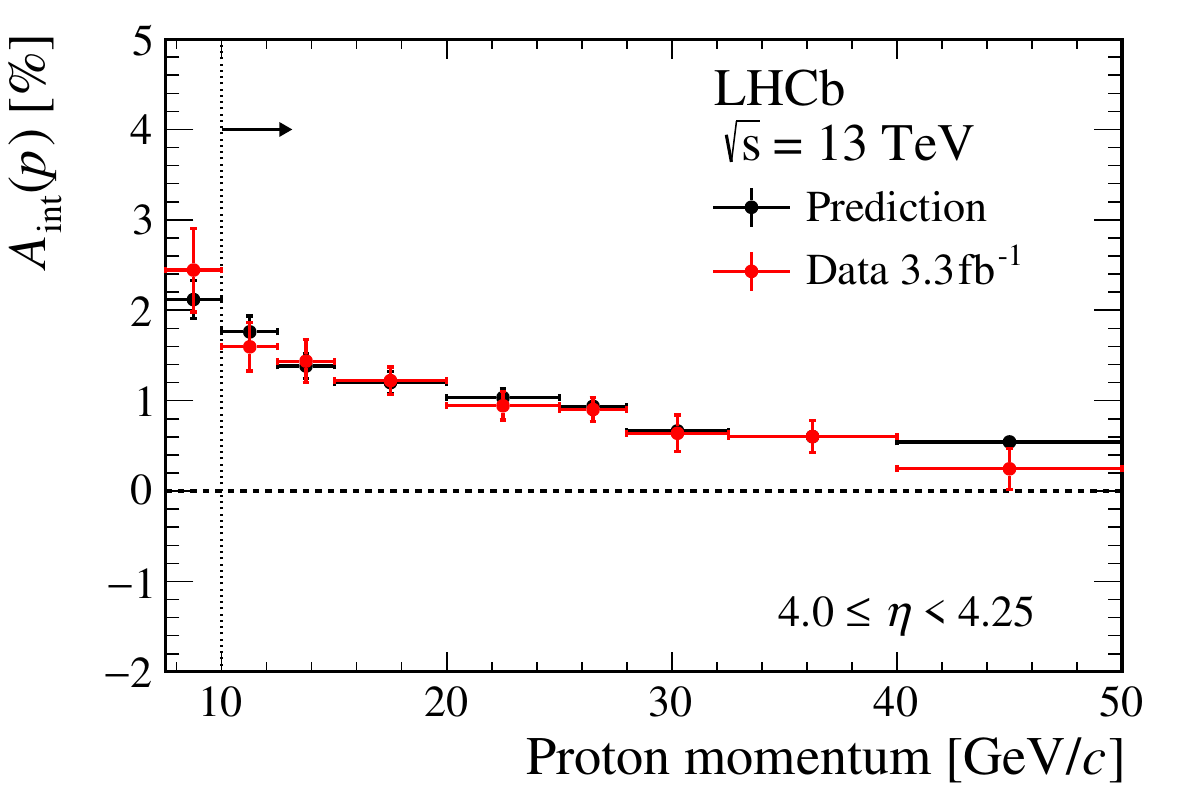}
		\caption{Proton-interaction asymmetry measured as difference in detection asymmetries between protons originating from $\Lz\to\proton\pim$ and pions from $\KS\to\pip\pim$ decays in 2017 and 2018 data, split in three intervals of pseudorapidity. The $\eta$ ranges are $[2, 3]$, $[3, 4]$ and $[4.00, 4.25]$ from top left to bottom. Also shown are the predictions from the proton-deuterium cross-sections. Model uncertainties shown here are a constant $10\%$ uncertainty due to the knowledge on the material budget~\cite{LHCb-DP-2013-002}. These uncertainties are not used in the $\chi^{2}$ values reported in the text. The vertical line shows the lower cutoff in momentum used in the fit.}
			\label{fig:proton_asymmetry_model_data_VELO}
	\end{center}
\end{figure}

The procedure to estimate the proton-interaction asymmetry is validated by measuring the asymmetry with partially reconstructed \decay{\Lz}{\proton\pim} decays exploiting a tag-and-probe method. The method is described in more detail in Ref.~\cite{LHCb-DP-2019-003}, where it is applied to a different decay mode. The proton (probe) is reconstructed using information from the \velo{} only, while the pion (tag) is fully reconstructed using information from \velo, \ttracker and T stations (long track). Due to the large lifetime of \Lz{} baryons, candidates are selected with high purity by requiring a large displacement of the decay vertex with respect to the primary-interaction vertex. Kinematic and geometric constraints allow the mass of the \Lz candidate and the momentum of the proton candidate to be reconstructed with a relative resolution of about 7\%. The efficiency to reconstruct a proton is given by the ratio of matched and selected candidates. A proton track is considered as matched if there is a long track which has at least 65\% of hits in common with the \velo probe (with a minimum of 6 hits), and the correct charge and momentum to form a \Lz{} candidate when combined with the tag pion. The procedure is performed separately for proton and antiproton candidates to determine the detection asymmetry. Contributions to the detection asymmetry other than the interaction cross-sections, \textit{e.g.} left-right asymmetries in the detector efficiency, are controlled by applying the same method to \decay{\KS}{\pip\pim} decays with one pion being the tag and the other the probe. The asymmetry in pion interaction cross-sections is negligible for the relevant momenta above 10\gevc~\cite{PDG2016}. The proton-interaction asymmetry for protons with \(|\vec{p}|>10\gevc\) is then determined as the difference of detection asymmetries obtained for proton and pion probes. The data are split into different intervals of proton momentum and pseudorapidity to test different kinematic regions. Data recorded in 2017 and 2018 at \(\sqs=13\tev\), corresponding to an integrated luminosity of 3.3\invfb, are used to perform the calibration, as no suitable trigger selections existed in 2011 and 2012. However, changes in the detector geometry are minimal and the calibration is valid for all data-taking periods, as it is derived as a function of the proton kinematics. In total, about 3 million \Lz{} and 40 million \KS{} candidates are used. The measured proton-interaction asymmetries as a function of proton momentum and pseudorapidity are given in \cref{fig:proton_asymmetry_model_data_VELO}.

The measured asymmetries are compared to the expected proton-interaction asymmetry from (anti)proton-deuterium cross-sections according to \cref{eq:aintpfull}. The coefficients \(C_{|\pvec|}^{\text{sim}}\) are determined with simulation in momentum ranges of [5, 15], [15, 30] and [30, 75] \gevc to be \(0.91 \pm 0.02\), \(0.85 \pm 0.02\) and \(0.73\pm0.05\). Using the proton momentum and the last hit of the proton track in the \velo, the path of the proton through the detector is determined and the expected proton-interaction asymmetry is calculated according to \cref{eq:aintp}. A test with simulated data shows good agreement between the asymmetry determinations with \decay{\Lz}{\proton\pim} decays and (anti)proton-deuterium cross-sections when the parameter \(F_{\text{data}}\) is fixed to 1. Using \decay{\Lz}{\proton\pim} decays in data, the parameter \(F_{\text{data}}\) is determined from a fit to be \( 0.967 \pm 0.017 \). The fit has a \(\chi^2\) of 20.8 with 19 degrees of freedom, showing a very good statistical agreement of the asymmetry between measurement and estimation from (anti)proton-deuterium cross-sections. The resulting correction factor \(F_{\text{data}}\) is subsequently used in the determination of the proton-interaction asymmetry of \(\decay{\Lb}{\Lc\mun\neumb\PX}\) decays.

As a test, the same procedure is repeated with \Lz{} and \KS{} candidates with the proton reconstructed in the first two tracking detectors, \velo{} and \ttracker, instead of the \velo only. By restricting the method to such tracks, the overall material affecting the inferred efficiency is reduced, but the average composition is different, making it a complementary test of the method. The predicted asymmetries are about two times smaller. Using the previously determined parameter \(F_{\text{data}}\), a very good agreement with a \(\chi^2\) of 18.8 with 19 degrees of freedom is observed between measurement and prediction from (anti)proton-deuterium cross-sections.

\subsubsection{Application to signal decays}
The measurement using \decay{\Lz}{\proton\pim} decays presented in the previous section probes the interaction asymmetry of material downstream of the \velo, which is about two-thirds of the total material budget of the tracking system~\cite{VanTilburg:885750}. Additionally, protons from \Lz decays have a relatively soft momentum spectrum compared to the protons from \Lb{} decays. Therefore, the formalism given in \cref{eq:aintpfull}, which explicitly accounts for the momentum dependence and the full detector geometry, is used to determine the proton-interaction asymmetry for $\Lc(\rightarrow \proton\Km\pip)\mun$ candidates. 
\begin{figure}[b]
	\begin{center}
		\includegraphics[width=0.49\textwidth]{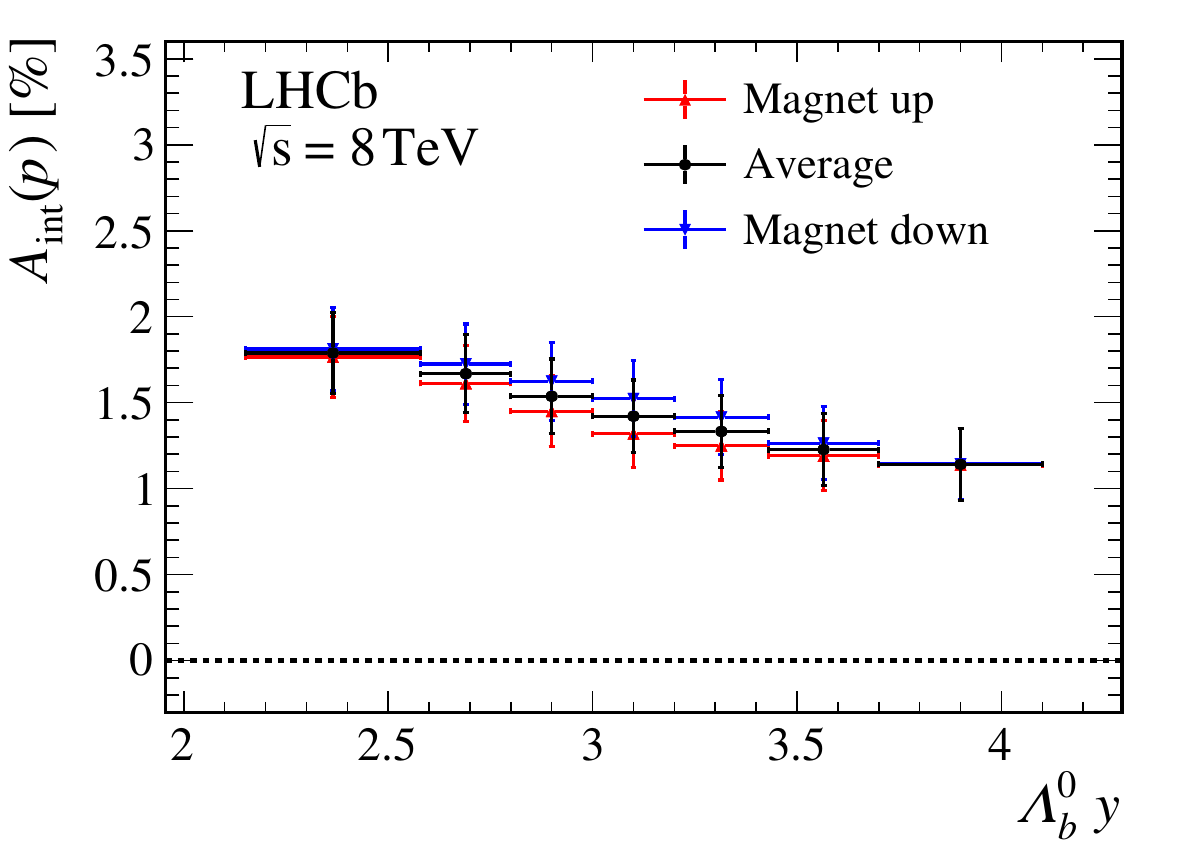}
		\includegraphics[width=0.49\textwidth]{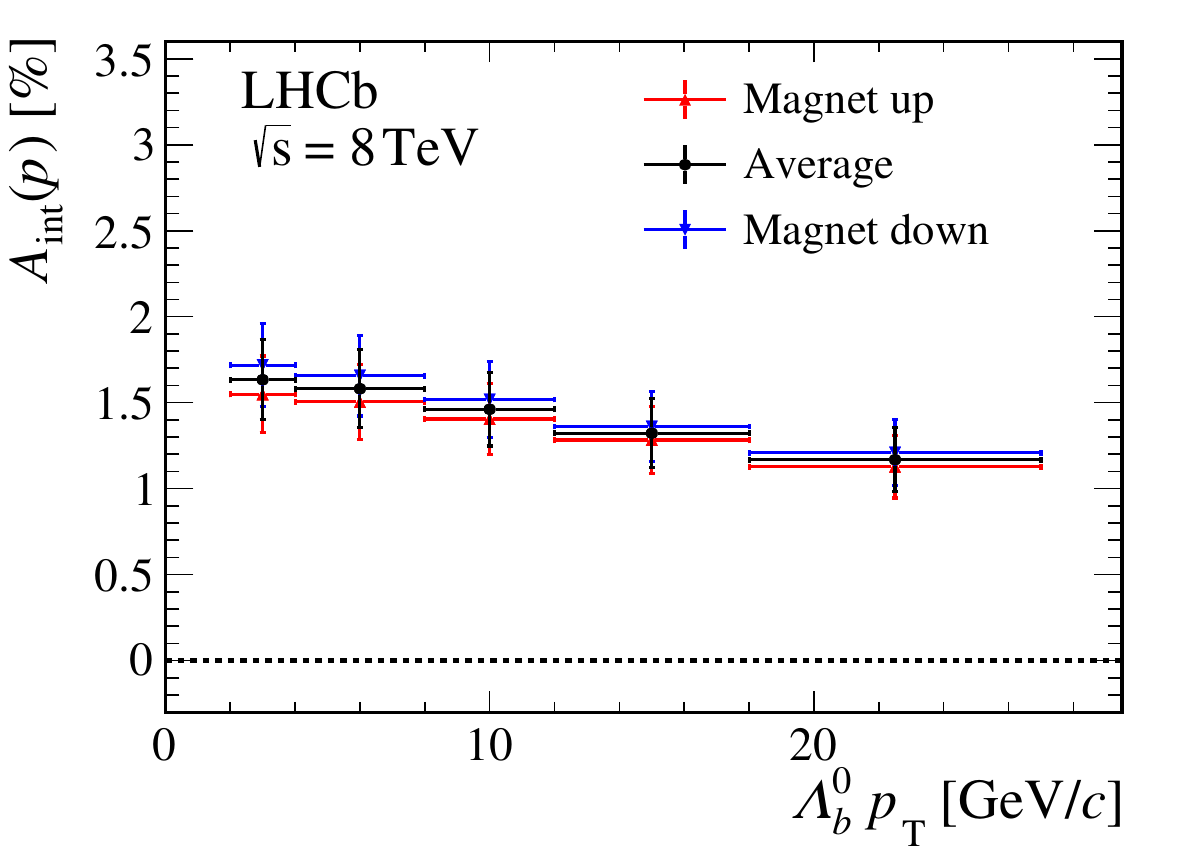}\\
		\caption{Proton-interaction asymmetry versus $\Lb$ (left) rapidity and (right) $\pt$ for data taken at a centre-of-mass energy of $\sqs=8\tev$. The results are shown separately for (red upward triangles) magnet up, (blue downward triangles) magnet down and (black dots) their average.}
		\label{fig:proton_detection_asymmetry}
	\end{center}
\end{figure}

For \decay{\Lc}{\proton\Km\pip} decays, the efficiency correction of hadronic scattering is smaller compared to \Lz{} decays as the \Lc{} decay has a larger \(Q\) value. The correction factors \(C_{|\pvec|}^{\text{sim}}\) vary between \(0.9\) and \(0.95\) for proton momenta from 10 to 150\gevc. The asymmetry due to the proton-interaction asymmetry is then estimated for the protons in the \Lb{} signal decays by averaging \cref{eq:aintpfull} over the hypothetical paths through the detector material given by the \Lc{} decay vertices and the proton momenta. Differences between data and simulation are accounted for by the effective correction of the material map \(F_{\text{data}}\) of \( 0.967 \pm 0.017\). The proton-interaction asymmetry varies between 1\% and 2\% depending on \Lb{} kinematics, and is shown as a function of \Lb{} rapidity and \pt{} in \cref{fig:proton_detection_asymmetry} for $\sqs=8\tev$ data.  The small difference between the results for the two magnet polarities is due to an asymmetric material distribution in the tracking stations. The results for $\sqs=7\tev$ are almost identical, as the material density is the same.

The systematic uncertainties on the proton-interaction asymmetry are summarised in \cref{tab:proton_detection_asymmetry_systematic_table}. They stem from the unknown contribution of elastic scattering, as discussed in \cref{sec:material_dependence}, the uncertainty of the correction of the survival probability of hadronic scattering, the statistical uncertainty on the correction factor due to the material, the uncertainty on the amount of material in the \velo~\cite{LHCb-DP-2014-001}, and the uncertainty of the (anti)proton-deuterium cross-section measurements.
\begin{table}[t]
    \setlength{\tabcolsep}{1em}\centering\caption{Ranges of the systematic uncertainties for the proton-detection asymmetry.}
    \label{tab:proton_detection_asymmetry_systematic_table}\bgroup
    \def\arraystretch{1.25}
    \begin{tabular}{lc}
    \toprule 
     Source of systematic uncertainty & Absolute uncertainty  [$\%$]\\
     \midrule
    Contribution of elastic scattering & $0.13 - 0.15$ \\
    Survival probability correction & $0.04 - 0.05$ \\
    Material map uncertainty & $0.02$  \\
    \velo material uncertainty & $0.03$  \\
    Cross-section uncertainty & $0.05  - 0.06$ \\
    \midrule
    Total systematic uncertainty & $0.21 - 0.24$\\
    \bottomrule
    \end{tabular}
    \egroup
\end{table}

In addition to the interaction asymmetry, a small correction due to the geometric acceptance of protons is determined. It is negligible when averaging the results for the two magnet polarities but has a size of up to 0.2\% in magnitude when considering only one polarity. It is largest at high rapidities where particles are close to the beam pipe.

\subsection{Proton PID asymmetry}\label{sec:proton_pid_det}
The asymmetry of the proton identification is determined with large data samples of \decay{\Lz}{\proton\pim} and \decay{\Lc}{\proton\Km\pip} decays \cite{LHCb-PUB-2016-021}. Only \Lc{} candidates originating from the primary vertex are used to avoid overlap with the \Lb{} signal decay. The decays are selected without any requirement on the proton-identification variables in the trigger and offline processing. The signal yield is determined from a fit to the invariant-mass distributions of \Lz{} or \Lc{} candidates. Tables of efficiencies depending on proton momentum, proton rapidity and per-event track multiplicity are built by determining the fraction of candidates fulfilling the selection requirement of signal candidates in a given interval. The procedure is performed separately for proton and antiproton candidates to calculate the detection asymmetry according to \cref{eq:def_adet}. The average asymmetry correction in each \Lb{} kinematic interval is calculated from the efficiency tables and the proton kinematics. Proton PID asymmetries are measured with an absolute uncertainty of 0.1\% to 0.2\% and are consistent with zero in most regions of the considered \Lb{} decay phase space. Systematic uncertainties are assigned based on differences in the detection asymmetries when slightly increasing the number of intervals per dimension and when using the transverse momentum or the azimuthal angle of the proton to parameterise the efficiencies. The systematic variations are found to be comparable in size to the statistical uncertainty. 

\subsection{Muon trigger and PID asymmetry}\label{sec:muon_det}
The asymmetry due to the muon PID and the muon-based trigger is determined with a tag-and-probe method using \decay{\jpsi}{\mup\mun} decays incompatible with coming from any primary vertex~\cite{LHCb-PAPER-2014-053}.
One of the muon candidates (tag) is required to have a positive trigger decision and to be identified as a muon, while the other muon (probe) is only required to be reconstructed by the tracking system. This selection ensures that the probe is unbiased with respect to trigger and muon identification. Subsequently, it is tested whether the probe muon passes or fails the selection requirements given in \cref{sec:Selection}. The detection asymmetry is determined by a simultaneous fit of the \mup\mun invariant-mass distributions of the samples divided according to the charge of the probe muon and its response to the selection criteria. The procedure is performed in ranges of \pt and \(\eta\) of the probe muon to determine asymmetry tables for different data-taking years and magnet polarities. The chosen intervals have approximately equal signal yields while capturing the variations of the asymmetry. The average asymmetry correction in each \Lb{} kinematic interval is calculated from the asymmetry maps and the muon kinematics in the considered interval. The corrections are measured with a statistical precision of 0.1\% to 0.2\%. While the average asymmetry of the two magnet-polarity samples is consistent with zero in most phase-space regions, a significant difference between the asymmetries in data taken with magnet-polarity up and down is observed in 2011. This feature originates from a charge-dependent bias in the momentum measured by the hardware trigger, which was corrected starting from the 2012 data-taking period. The bias was caused by a combination of misaligned muon stations together with a too simplified momentum determination in the hardware trigger. Systematic uncertainties are obtained from the variations observed in the efficiency maps when adding the azimuthal angle of the muon as an additional dimension and when slightly varying the limits of the \pt and \(\eta\) intervals.

\subsection{Asymmetry of track reconstruction for protons and muons}\label{sec:pmu_track_det}
Imperfections in the tracking system can lead to different reconstruction efficiencies for positively and negatively charged particles. This asymmetry adds to the detection asymmetries due to strong interactions with the detector material and particle identification determined in previous sections. The efficiencies to reconstruct the trajectories of positively and negatively charged particles are measured with a tag-and-probe method using \mbox{\decay{\jpsi}{\mup\mun}} decays where the probe muon is not fully reconstructed by the tracking system \cite{LHCb-DP-2013-002, LHCb-PAPER-2016-013}. Asymmetry maps are extracted as a function of \pt, momentum or \(\eta\) of the probe muon. An average asymmetry correction of each \Lb{} kinematic interval is calculated from the muon and proton kinematics of \Lb{} candidates in the considered interval. The asymmetries obtained from the asymmetry map as a function of \(\eta\) are used as the central values, while the largest differences in the other maps are used as systematic uncertainties. As the proton and muon in the \Lb{} decay have opposite charges, the combined tracking asymmetry is consistent with zero in all $\Lb$ kinematic bins with a precision of about 0.1\%, dominated by the systematic uncertainties.

\subsection{Kaon-pion detection asymmetry}\label{sec:kpi_det}
All detection asymmetries of the \Km\pip pair are determined as a single correction, which includes contributions from track reconstruction, particle identification and material interactions. In particular, kaons have a sizeable interaction asymmetry with the detector material, similar to the one of protons. 

The \Km\pip detection asymmetry is determined with large calibration samples of promptly-produced \mbox{{\decay{\Dp}{\Km\pip\pip}}} decays where the production asymmetry of \Dp mesons and the detection asymmetry of the additional pion are corrected for with promptly-produced \decay{\Dp}{\Kzb\pip} decays~\cite{LHCb-PAPER-2014-013}. This additional pion is required to trigger the event in the selection of \Dp decays. The neutral kaon, reconstructed in the \pip\pim final state, induces a small asymmetry originating from \CP violation, material interaction, and their interference. This asymmetry is corrected for with the \CP-violation parameters of the neutral-kaon system, interaction cross-sections and the material map of the \velo, within which the \Kz{} candidates are required to decay. Asymmetries are determined with fits to the \Dp invariant-mass distributions. As detection asymmetries depend on kinematics, candidates of the \Dp decay are assigned per-candidate weights, chosen such that the weighted kinematic distributions of kaons and pions match the distributions of the ones from the \(\decay{\Lb}{\Lc(\rightarrow\proton\Km\pip)\mun\neumb\PX}\) decay. The weights are determined by considering the distributions of momentum and pseudorapidity of the kaon and the transverse momentum of the pion.

Corrections vary between $-1.5\%$ and $-1.0\%$ depending on the \Lb{} kinematic range. The kaon-pion detection asymmetry is of opposite sign compared to the proton-interaction asymmetry and has a similar magnitude. Statistical uncertainties are about 0.3\% for data taken at $\sqs=7\tev$ and $0.2\%$ for data taken at $\sqs=8\tev$. Systematic uncertainties are determined by varying the bins used in the weighting procedure and by studying the signal model used in the fits to the calibration samples. The systematic uncertainties are significantly smaller than the statistical uncertainties for all \Lb{} kinematic bins.

\subsection{Summary of detection asymmetries}
\begin{figure}[tb]
	\begin{center}
		\includegraphics[width=0.49\textwidth]{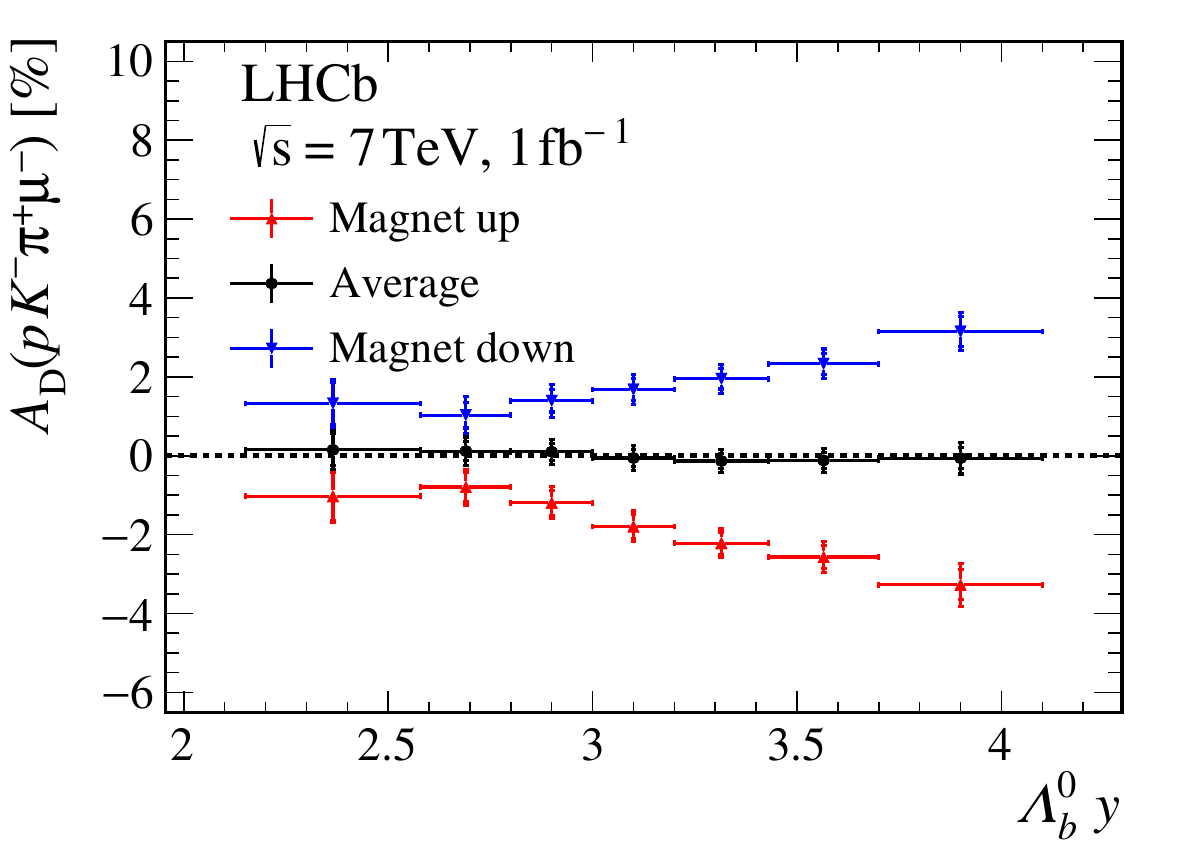}
		\includegraphics[width=0.49\textwidth]{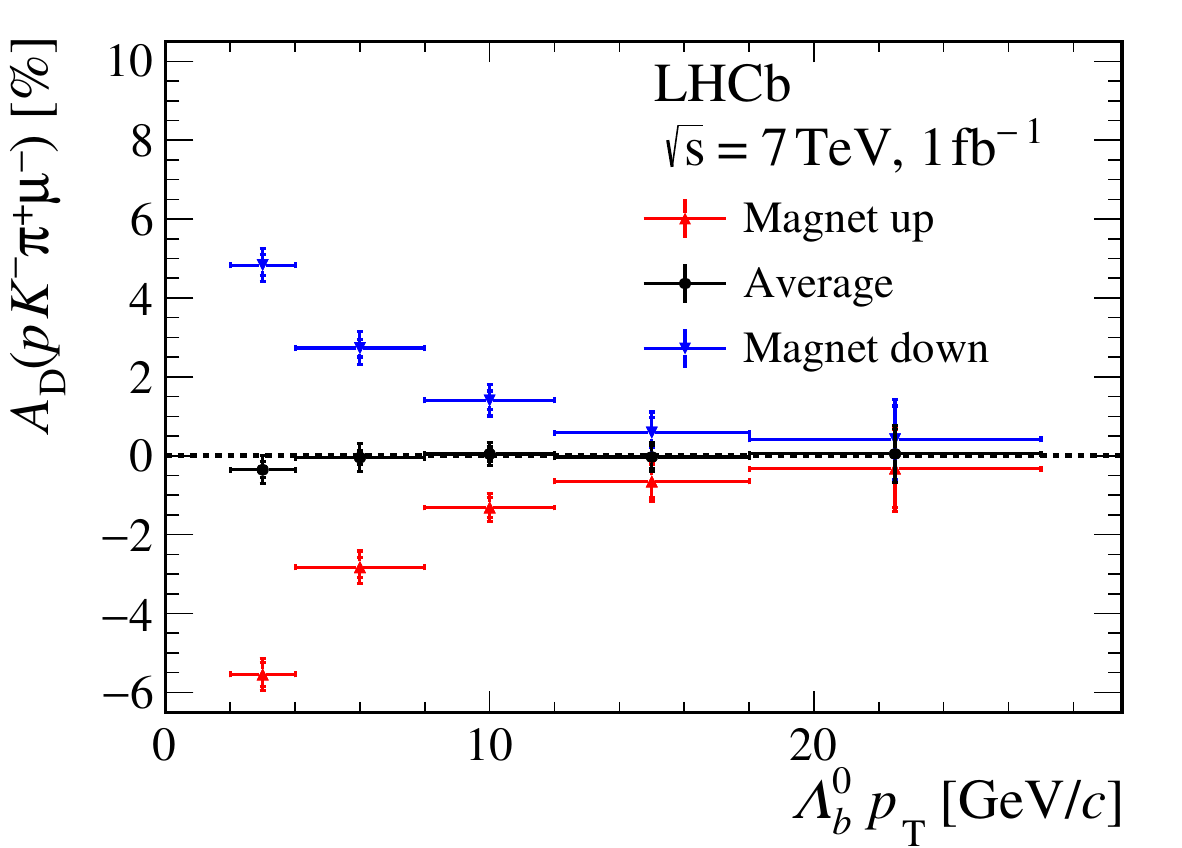}\\
		\includegraphics[width=0.49\textwidth]{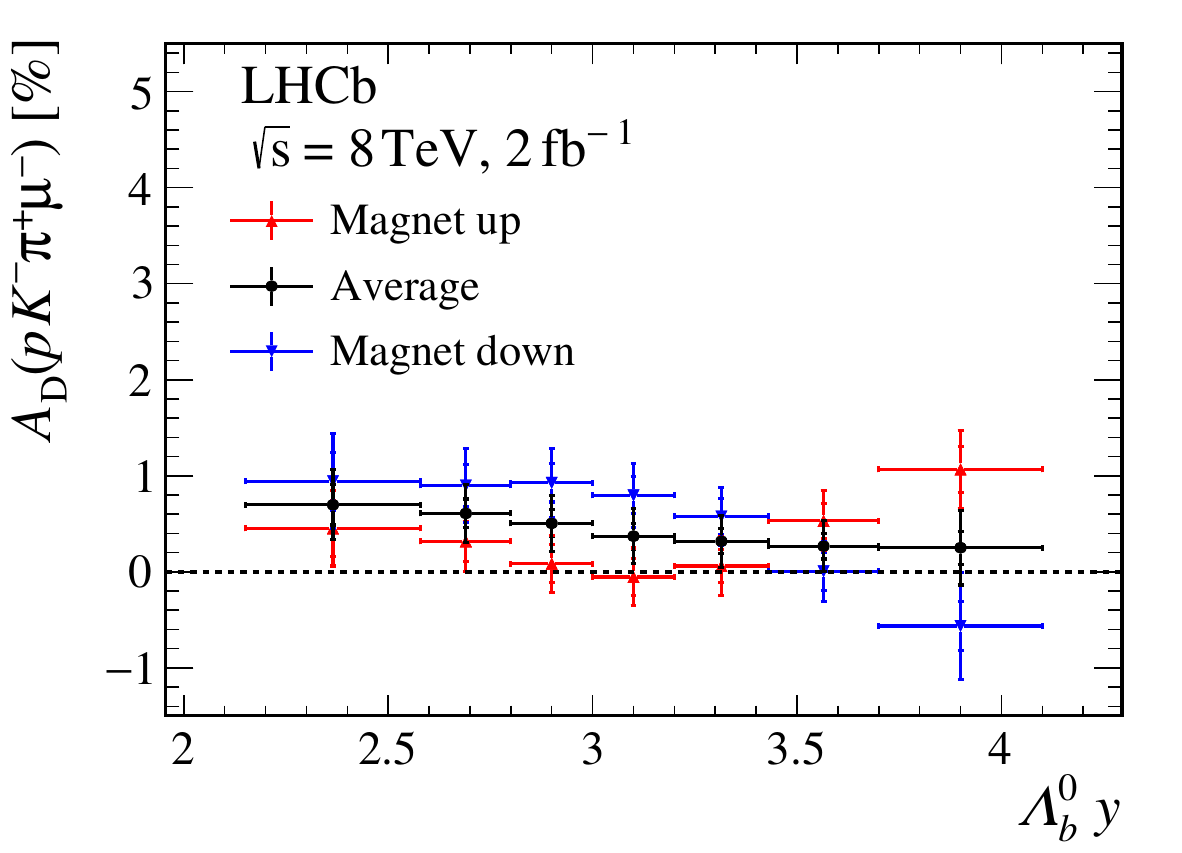}
		\includegraphics[width=0.49\textwidth]{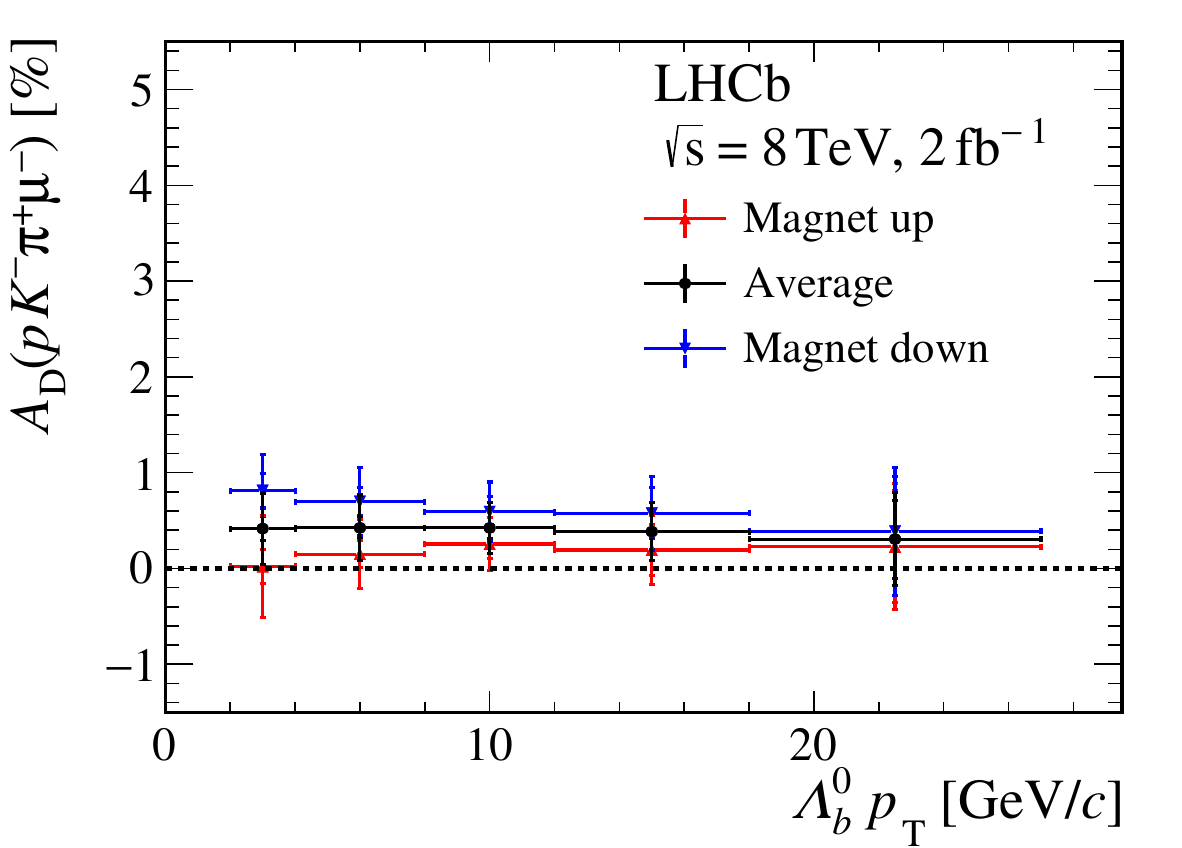}\\
		\caption{Total correction due to detection asymmetries versus $\Lb$ (left) rapidity and (right) $\pt$ for data taken at centre-of-mass energies of (top) $\sqs=7\tev$ and (bottom) $\sqs=8\tev$. The results are shown separately for the (red upward triangles) magnet-up sample, (blue downward triangles) magnet-down sample and (black dots) their average. The first bar indicates the statistical uncertainty and the second bar the total uncertainty. }
		\label{fig:results_sum_of_adet}
	\end{center}
\end{figure}
The total corrections due to detection asymmetries are shown in \cref{fig:results_sum_of_adet} for different magnet-polarity samples and centre-of-mass energies. The largest corrections are the proton-interaction asymmetry and the kaon-pion detection asymmetry. As these are of opposite sign, the total correction is relatively small. The fairly large difference between the total detection asymmetries for data taken with magnet polarities up and down at \(\sqs=7\tev\) is due to the muon trigger and PID asymmetry present in this data-taking period.

\section{Systematic uncertainties}\label{sec:systematic}
Systematic shifts in the measured production asymmetries per \Lb{} kinematic interval can arise from background contributions due to misreconstructed final states, from the finite rapidity and \pt resolution due to non-reconstructed final state particles,
and, as discussed before, from biases in the determination of detection asymmetries. The details of the evaluation of systematic uncertainties in the measurements of the various detection asymmetries are described in \cref{sec:detection_asymmetries}.

Signal yields are determined by a fit to the invariant-mass distribution of reconstructed \Lc candidates (see \cref{sec:raw_asymmetry_measurement}). The combination of a \Lc{} baryon not originating from a \Lb{} baryon decay with a random muon in the event is a potential background with a different production asymmetry. To estimate this background fraction, combinations of \Lc{} candidates with same-sign \mup{} candidates are formed with the otherwise default signal selection and their yield is compared to the total signal yield. This fraction is measured to be less than $0.5\%$. There is no precise measurement of the production asymmetry of \Lc{} baryons produced promptly in \proton\proton collisions to date, but it is not expected to differ by more than a few percent in absolute terms from the production asymmetries of \Lb{} baryons~\cite{LbprodLai}. The \Dp{} and \Ds{} production asymmetries are measured to be less than 1\% in absolute terms with no significant kinematic dependence~\cite{LHCb-PAPER-2012-026, LHCb-PAPER-2018-010}. Thus, associating the full measured asymmetry to the \Lb{} production asymmetry leads to a negligible systematic uncertainty due to the small background contribution.

Systematic uncertainties due to the chosen fit model are evaluated with pseudo-experiments. Invariant-mass distributions are generated with alternative models which empirically describe data and are fitted with the default fit model. The variations include a model with non-Gaussian tails on both sides of the signal peak~\cite{Santos:2013gra} and a model with different tail parameters for \Lc{} and \Lcbar{} candidates. Negligible biases are found and no systematic uncertainty due to the raw-asymmetry determination is assigned.

The rapidity and transverse momentum are determined from the reconstructed momentum of the $\Lc\mun$ system, where the momenta of the neutrino and additional particles in the decay of excited \Lc{} resonances are not considered as discussed in \cref{sec:bins}. The resulting, degraded resolution can lead to a migration of candidates between kinematic intervals. A measure for this migration is the purity per interval, which is defined as the ratio of candidates correctly reconstructed in a given kinematic interval over the total number of candidates in that interval. Studies with simulated events show that the purity in all \Lb{} kinematic intervals is close to 70\% or higher. The rapidity resolution is better for decays proceeding via excited \Lc{} baryons, such as \mbox{$\decay{\Lb}{\Lc(2595)\mun\neumb}$} with \mbox{$\decay{\Lc(2595)}{\Lc\pip\pim}$}, while the transverse momentum resolution degrades due to the additional missing particles. Pseudoexperiments are generated for intervals of $y$ and $\pt$ using efficiency and resolution functions determined from the decays without any intermediate resonance and from those proceeding via the $\Lc(2595)$, $\Lc(2625)$, $\Lc(2765)$, and $\Lc(2880)$ resonances. The relative fraction of these decays is taken from Ref.~\cite{LHCb-PAPER-2017-016}. The generated momentum distribution is taken from the LHCb simulation and the generated production asymmetries are conservatively chosen such that they exceed those observed in data by up to a factor two. The estimated systematic biases vary from 0.03\% at low rapidities to 0.16\% in the largest rapidity interval. For the \pt{} intervals, the systematic biases vary from 0.19\% at low \pt{} to 0.08\% at large \pt.   

All systematic uncertainties are summarized  in \cref{tab:systematic_table} for intervals of \Lb{} rapidity, being similar for those in \pt.
The dominant systematic uncertainty comes from the determination of the proton interaction asymmetry. This systematic uncertainty is around 0.2\% in all \Lb{}~kinematic intervals and is almost fully correlated between all intervals.
\begin{table}[tbp]
	\footnotesize
    \setlength{\tabcolsep}{1em}\centering
    \caption{Absolute statistical and systematic uncertainties affecting the measurement, given in percent. The ranges are taken from the results in intervals of $\Lb$ rapidity.}
    \label{tab:systematic_table}\bgroup
    \def\arraystretch{1.25}
    \begin{tabular}{lcccc}
    \toprule
	& \multicolumn{2}{c}{$\sqrt{s}=7\tev$} & \multicolumn{2}{c}{$\sqrt{s}=8\tev$} \\
	\cmidrule(r{.5em}){2-3} \cmidrule(l{.5em}){4-5}
	 & Stat. [\%] & Syst. [\%] & Stat. [\%] & Syst. [\%]\\
     \midrule
    Raw asymmetry & $0.42-0.62$ & - & $0.27-0.40$  & - \\
    Proton interaction & - & $0.21-0.24$ & - & $0.21-0.24$\\
    Kaon-pion detection & $0.16 - 0.31 $ & $0.05-0.10$ & $0.11 - 0.20 $ & $0.04-0.08$\\
    Muon trigger \& PID & $0.07-0.19$ & $0.00-0.08$ & $0.04-0.11$ & $0.08-0.15$\\
    Tracking & $0.02-0.05$ & $0.05 - 0.10$ & $0.01-0.04$ & $0.02 - 0.19 $\\
    Proton PID & $0.01 - 0.16$ & $0.02-0.05$ & $0.01-0.09$ & $0.01-0.05$\\ 
	Interval migration & - & $0.02-0.17$ & - & $0.02 - 0.17$\\ 
	\midrule
    Total uncertainty & $0.46 - 0.70$ & $0.23-0.30$ & $0.31 - 0.44$ & $0.25 - 0.35$\\
    \bottomrule
    \end{tabular}
    \egroup
\end{table}

\section{Results}\label{sec:results}
\begin{figure}[ht]
	\begin{center}
		\includegraphics[width=0.49\textwidth]{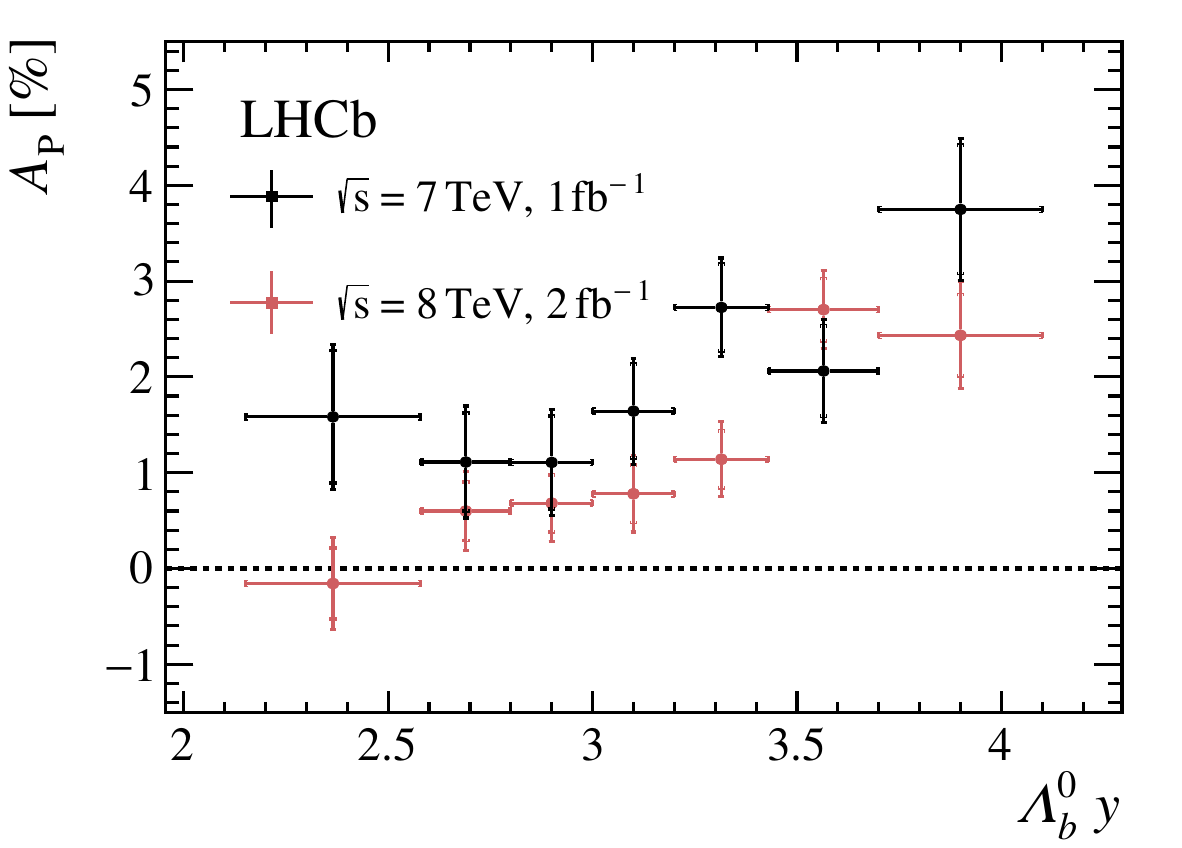}
		\includegraphics[width=0.49\textwidth]{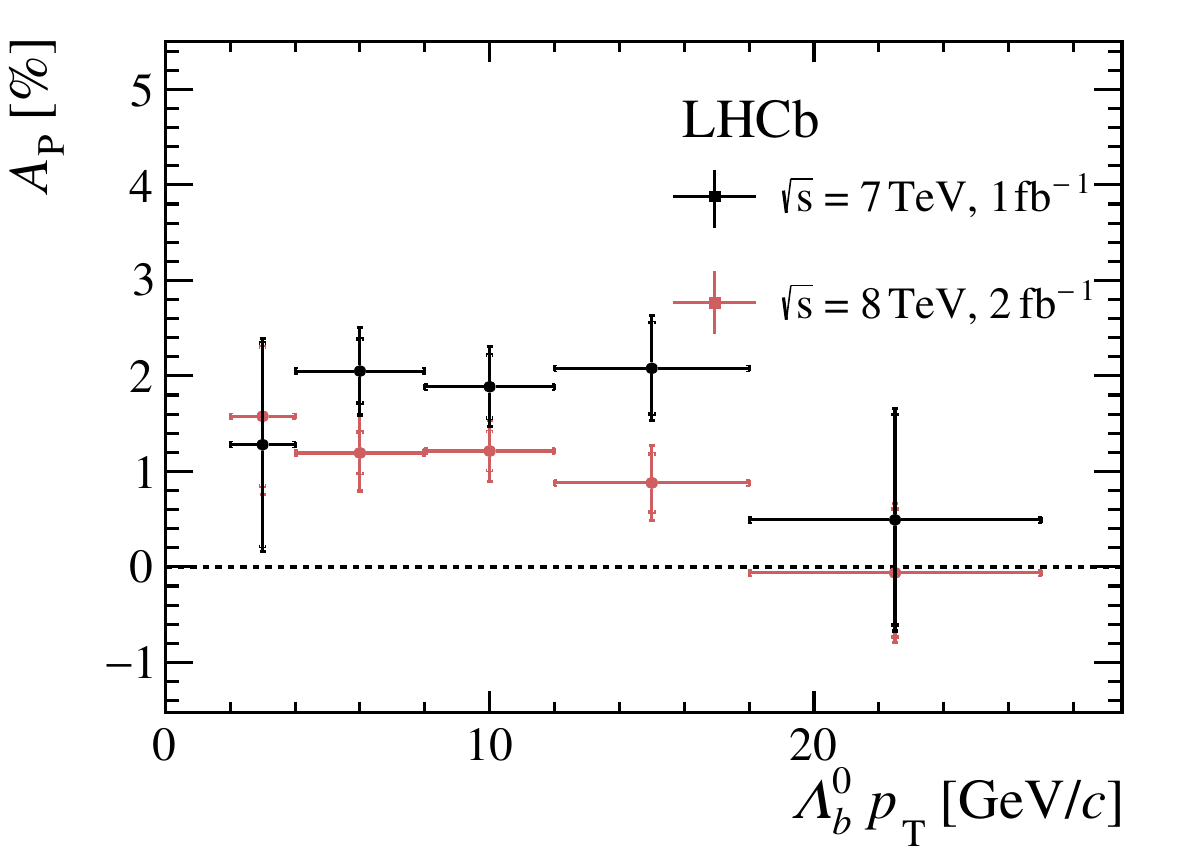}
		\caption{Measured $\Lb$ production asymmetry versus (left) rapidity and (right) transverse momentum. The uncertainties are the quadratic sums of statistical and systematic uncertainties. The results in neighbouring intervals are correlated.}
		\label{fig:results_total_result_plot}
	\end{center}
\end{figure}

The measured production asymmetries as a function of \Lb rapidity and transverse momentum are shown in \cref{fig:results_total_result_plot} for centre-of-mass energies of 7 and 8\tev. The values per kinematic interval including statistical and systematic uncertainties are given in \mbox{Tables \ref{tab:results_total_rapidity} and~\ref{tab:results_total_pt}}. The results for neighbouring intervals are correlated as the data used to determine corrections of detection asymmetries overlap. The correlation matrices including statistical and systematic uncertainties are given in \cref{app:correlation_matrix}. As a consistency check, the measurement is performed independently for magnet-up and magnet-down samples and the $\chi^2$ for the compatibility of equal results is calculated. The obtained $\chi^2$ value when performing the measurement in $y$ (\pt) intervals is 12.9 (5.2) with 14 (10) degrees of freedom, corresponding to a $p$-value of 54\% (88\%). The results split by samples with different magnet polarity and centre-of-mass energy are given in \cref{app:magnet_polarity}.

The production asymmetries are found to be incompatible with zero for both 7 and 8\tev data, with $\chi^{2}=51.1$ and $\chi^{2}=51.7$ with 7 degrees of freedom each, corresponding to $p$-values of $8.9\times10^{-9}$ and $6.6\times10^{-9}$.
To test whether the production asymmetry is independent of rapidity, the asymmetries observed in \(\sqs=7\tev\) and \(8\tev\) data are fitted separately with a constant; the minimized $\chi^2$ values are 13.3 and 30.4, respectively, with 6 degrees of freedom each, corresponding to $p$-values of $3.8\times10^{-2}$ and $3.3\times10^{-5}$. As a difference between the production asymmetries at \(\sqs=7\tev\) and \(\sqs=8\tev\) is expected, the compatibility between the measurements at both energies is calculated. The $\chi^{2}$ is 18.6 with 7 degrees of freedom corresponding to a $p$-value of \(9\times10^{-3}\).

In summary, an asymmetry between $\Lb$ and $\Lbbar$ production is observed at the $5.8$ standard deviations (\(\sigma\)) level for both \(\sqs=7\tev\) and \(\sqs=8\tev\) data.\footnote{The compatibility in units of Gaussian standard deviations, $n\,\sigma$,  is derived from the $p$-value via \mbox{$n(p)=\sqrt{2}\ \text{erf}^{-1}(1-p)$} where $\text{erf}$ is the error function.} There is an evidence for a dependence of this asymmetry on the \Lb{} rapidity at the $4.1\,\sigma$ level  at \(\sqs=8\tev\), $2.1\,\sigma$ at \(\sqs=7\tev\). The asymmetries at \(\sqs=7\tev\) and \(\sqs=8\tev\) are found compatible at the $2.6\,\sigma$ level. No significant variation in the production asymmetry is observed as a function of \pt{}. To illustrate typical production asymmetries for \Lb{} decays reconstructed at the \lhcb experiment, the asymmetries observed across \pt{} intervals are averaged, which implicitly takes the observed \(y\) and \pt{} distributions into account. The results are \((1.92 \pm 0.35)\%\) for \(\sqs=7\tev\) and \((1.09 \pm 0.29)\%\) for \(\sqs=8\tev\).

\begin{table}[ht]
        \setlength{\tabcolsep}{1em}\centering\caption{Measured $\Lb$ production asymmetry in rapidity intervals. Values are given in percent. The first uncertainty is statistical, the second systematic. The uncertainties are partially correlated between rapidity intervals.} 
	\label{tab:results_total_rapidity}\bgroup
	\def\arraystretch{1.25}
    \noindent \begin{tabular}{lcc}
	\toprule
	 & \multicolumn{2}{c}{\Lb{} production asymmetry [\%]}   \\  
	 \cmidrule{2-3} 
	  &	$\sqrt{s}=7\tev$ & $\sqrt{s}=8\tev$\\ \midrule
	 
	$2.15 < y < 2.58$ &  $\phantom{-}1.58 \pm 0.70 \pm 0.27$ & $-0.16 \pm 0.38 \pm 0.30$          \\
	$2.58 < y < 2.80$ &  $\phantom{-}1.11 \pm 0.52 \pm 0.26$ & $\phantom{-}0.60 \pm 0.32 \pm 0.26$\\
	$2.80 < y < 3.00$ &  $\phantom{-}1.11 \pm 0.50 \pm 0.24$ & $\phantom{-}0.68 \pm 0.31 \pm 0.25$\\
	$3.00 < y < 3.20$ &  $\phantom{-}1.64 \pm 0.50 \pm 0.23$ & $\phantom{-}0.78 \pm 0.31 \pm 0.25$\\
	$3.20 < y < 3.43$ &  $\phantom{-}2.73 \pm 0.46 \pm 0.23$ & $\phantom{-}1.14 \pm 0.31 \pm 0.24$\\
	$3.43 < y < 3.70$ &  $\phantom{-}2.06 \pm 0.48 \pm 0.24$ & $\phantom{-}2.70 \pm 0.34 \pm 0.23$ \\
	$3.70 < y < 4.10$ &  $\phantom{-}3.75 \pm 0.68 \pm 0.30$ & $\phantom{-}2.43 \pm 0.44 \pm 0.35$\\
	\bottomrule
	\end{tabular}
	\egroup
\end{table}

\begin{table}[ht]
	\setlength{\tabcolsep}{1em}\centering\caption{Measured $\Lb$ production asymmetry in intervals of transverse momentum. Asymmetry values are given in percent. The first uncertainty is statistical, the second systematic. The uncertainties are partially correlated between transverse-momentum intervals.} 
\label{tab:results_total_pt}\bgroup
\def\arraystretch{1.25}
\noindent \begin{tabular}{lcc}
\toprule
 & \multicolumn{2}{c}{\Lb{} production asymmetry [\%]}   \\  
 \cmidrule{2-3} 
       &	$\sqrt{s}=7\tev$ & $\sqrt{s}=8\tev$\\ \midrule
 $\phantom{0}2 < \pt < 4\gevc$   & $\phantom{-}1.28 \pm 1.08 \pm 0.29$ & $\phantom{-}1.58 \pm 0.74 \pm 0.35$ \\  
 $\phantom{0}4 < \pt < 8\gevc$  & $\phantom{-}2.05 \pm 0.35 \pm 0.30$ & $\phantom{-}1.19 \pm 0.23 \pm 0.33$ \\ 
 $\phantom{0}8 < \pt < 12\gevc$& $\phantom{-}1.89 \pm 0.34 \pm 0.24$ & $\phantom{-}1.21 \pm 0.22 \pm 0.24$ \\ 
 $12 < \pt < 18\gevc$& $\phantom{-}2.08 \pm 0.49 \pm 0.25$ & $\phantom{-}0.88 \pm 0.32 \pm 0.23$ \\
 $18 < \pt < 27\gevc$& $\phantom{-}0.49 \pm 1.11 \pm 0.34$ & $-0.06 \pm 0.68 \pm 0.26$           \\
\bottomrule
\end{tabular}
\egroup
\end{table}

\subsection{Comparison with theory}
In this section, theoretical predictions for the \Lb{} production asymmetry as a function of \(y\) and \pt{} are compared to the measurements. Different models of colour reconnection implemented in \pythia are considered~\cite{Sjostrand:2007gs}, along with results from the heavy-quark recombination model~\cite{Braaten:2001bf} applied to \Lb{} production in $pp$ collisions\cite{LbprodLai}.
The predictions for the models implemented in \pythia are generated using version 8.303. The different settings for \pythia include the standard Monash settings~\cite{Skands:2014pea} and two newer models of colour reconnection: one based on a QCD-inspired scheme (CR1) introduced in Refs.~\cite{Christiansen:2015yqa, Argyropoulos:2014zoa}, and the so-called ``gluon-move'' scheme (CR2) introduced in Refs.~\cite{Christiansen:2015yca, Argyropoulos:2014zoa}. The detailed settings for the \pythia productions are given in \cref{app:pythia}. The predictions for the heavy-quark recombination model are updated with respect to Ref.~\cite{LbprodLai} to have predictions at 7 and 8\tev, and restricted to the same rapidity range, $2.15<y<4.1$, as this measurement. Predictions for heavy-quark recombination are provided for \pt smaller than 4\gevc but are unreliable on energy scales below the $\bquark$-quark mass~\cite{Braaten:2001bf,LbprodLai}.

The predicted asymmetries are sampled with the \(y\) and \pt{} distributions observed in data to correct for efficiency variations within a kinematic interval.
Distributions of the reconstructed $\Lb$ kinematics, along with their one-dimensional projections, are shown in \cref{app:kinematics}.

\begin{figure}[ht]
    \begin{center}
        \includegraphics[width=0.48\textwidth]{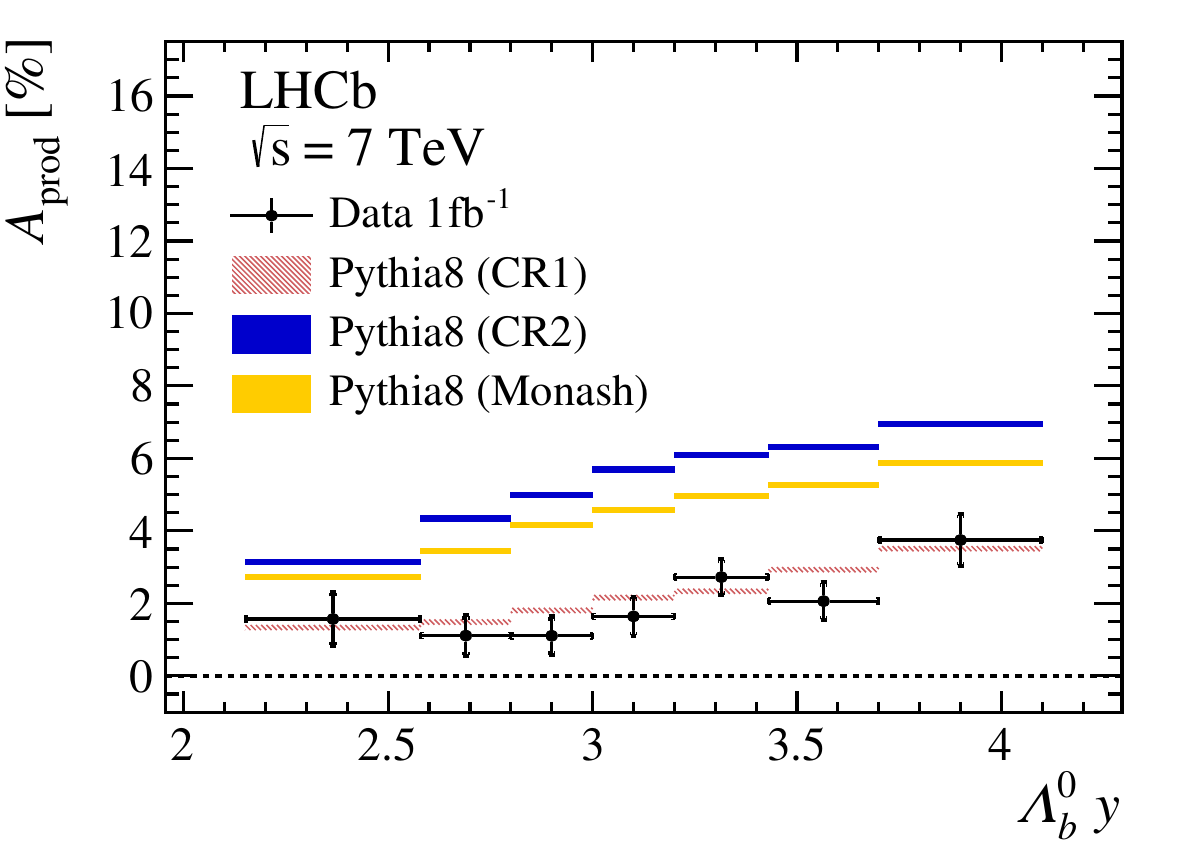}
        \includegraphics[width=0.48\textwidth]{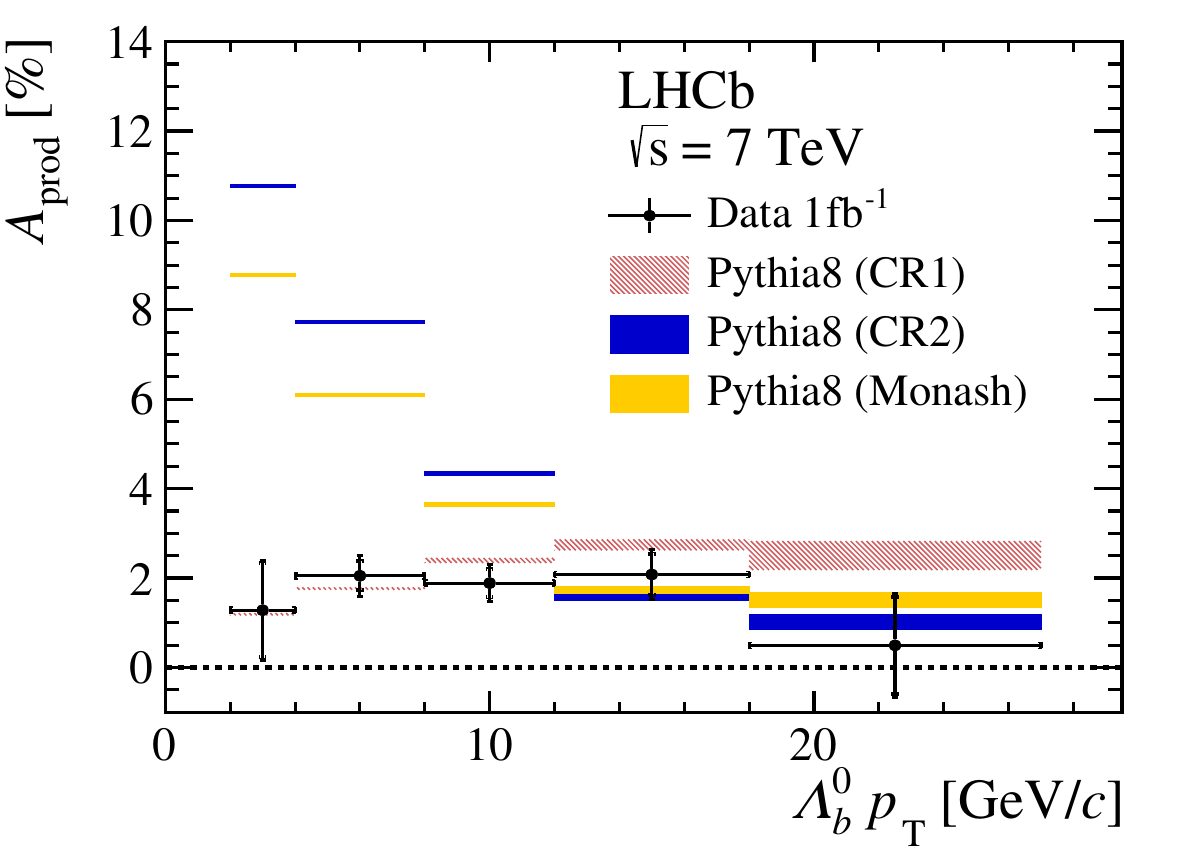} \\
        \includegraphics[width=0.48\textwidth]{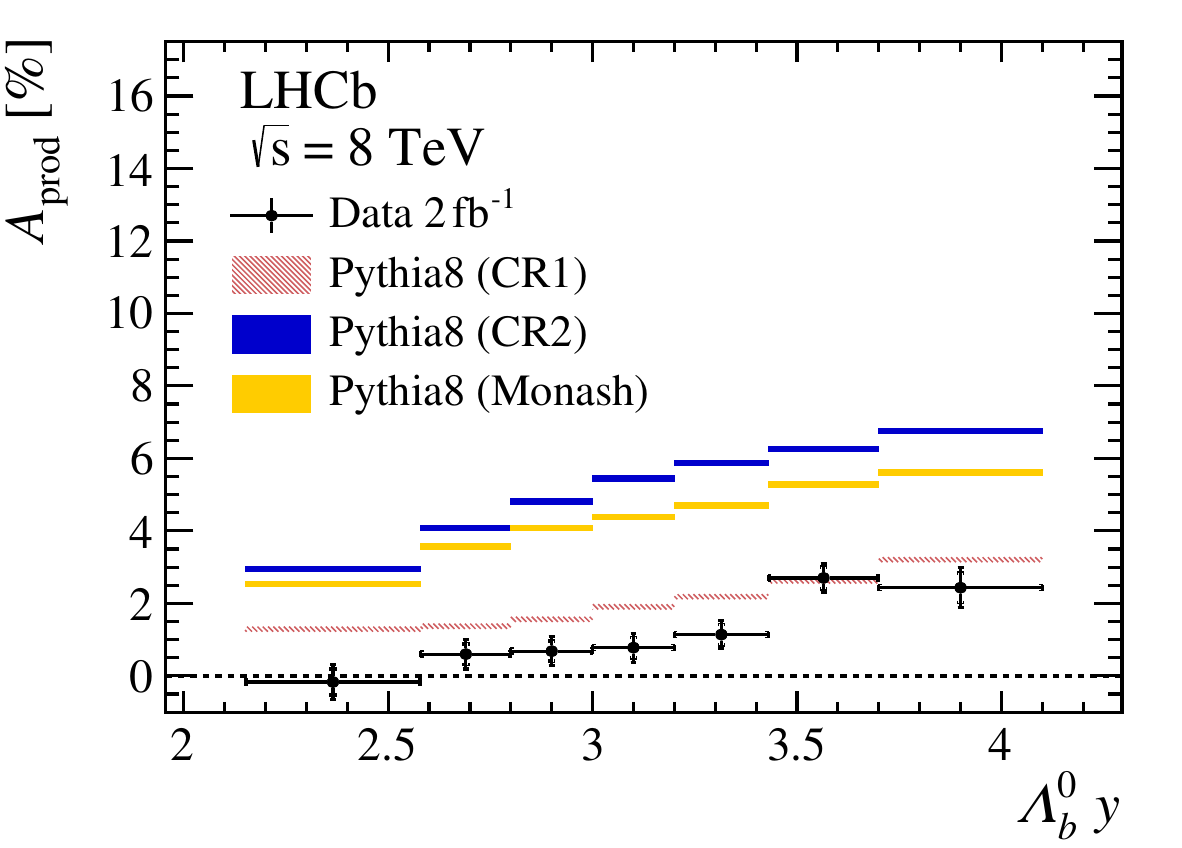} 
        \includegraphics[width=0.48\textwidth]{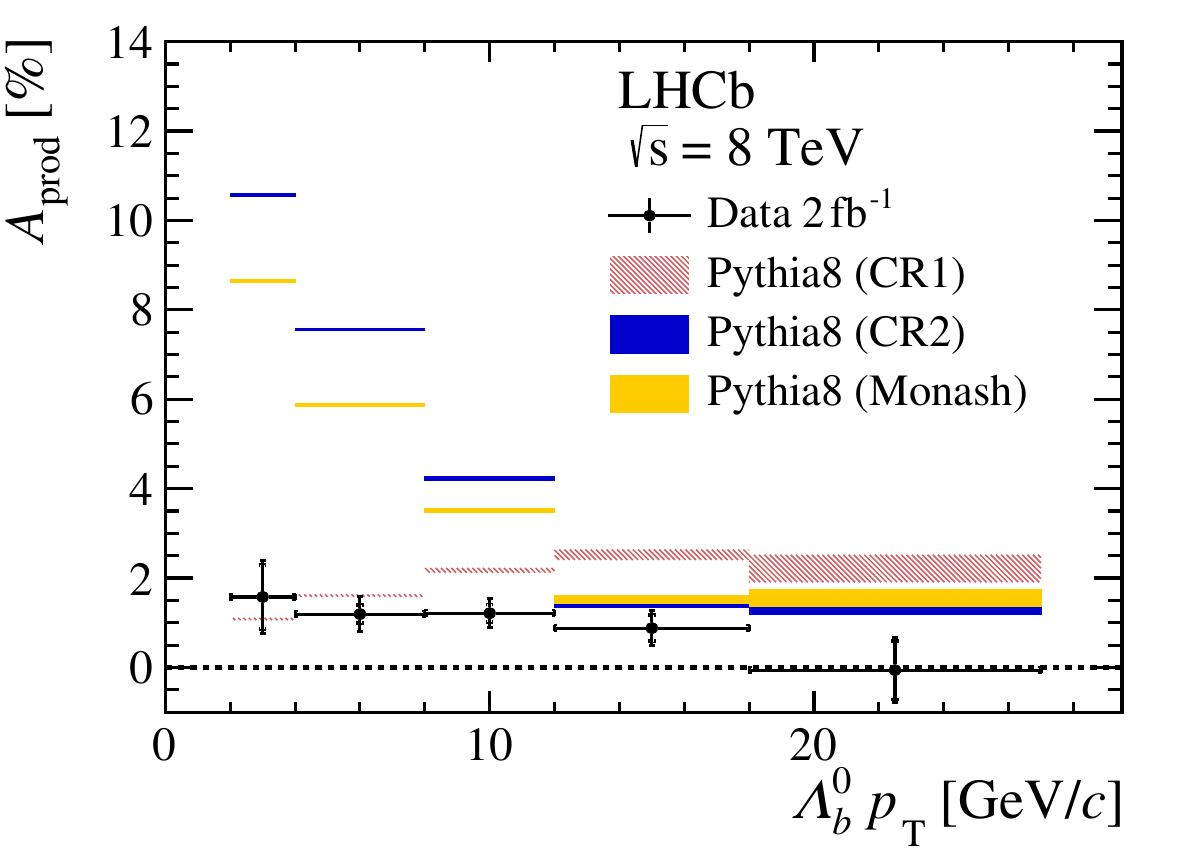} 
    \caption{Comparison of the $\Lb$ production asymmetry predicted by the various \pythia models, where CR1 refers to the QCD-inspired model and CR2 refers to the gluon-move model, and the measured production asymmetries. Results versus $\Lb$ (left) rapidity $y$ and (right) $\pt$ are shown for centre-of-mass energies of (top) $\sqs=7\tev$ and (bottom) $\sqs=8\tev$. Uncertainties on the predictions are due to limited simulation sample sizes.}
    \label{fig:asymmetry_data_pythia}
\end{center}
\end{figure}

\begin{figure}[ht]
    \begin{center}
        \includegraphics[width=0.49\textwidth]{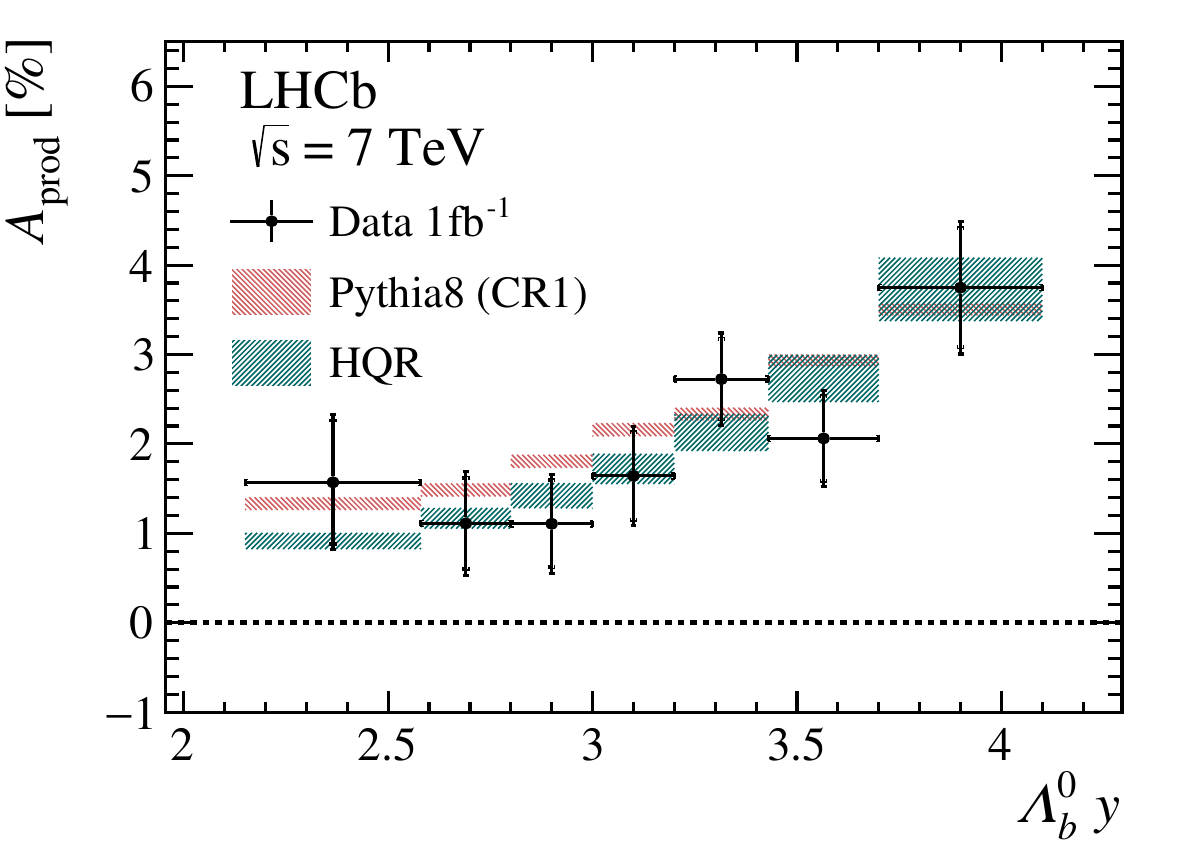} 
		\includegraphics[width=0.49\textwidth]{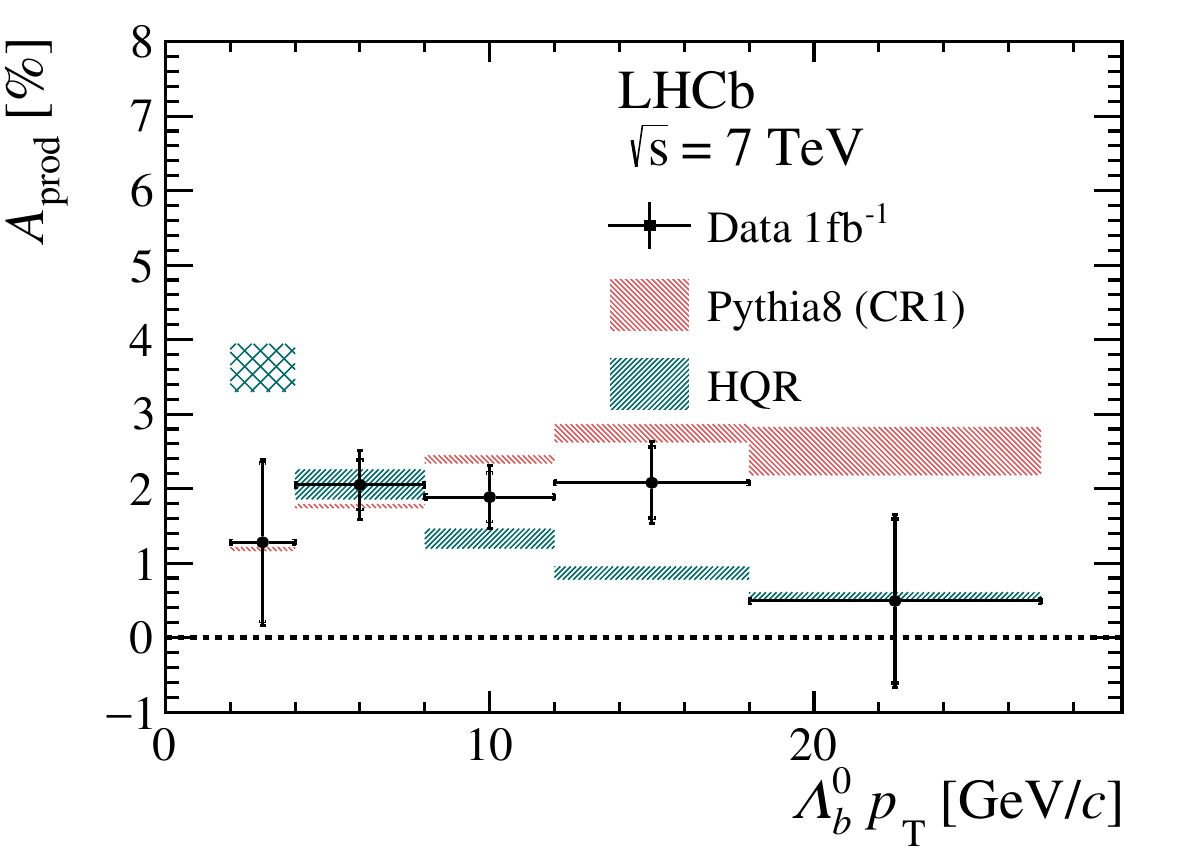}
        \includegraphics[width=0.49\textwidth]{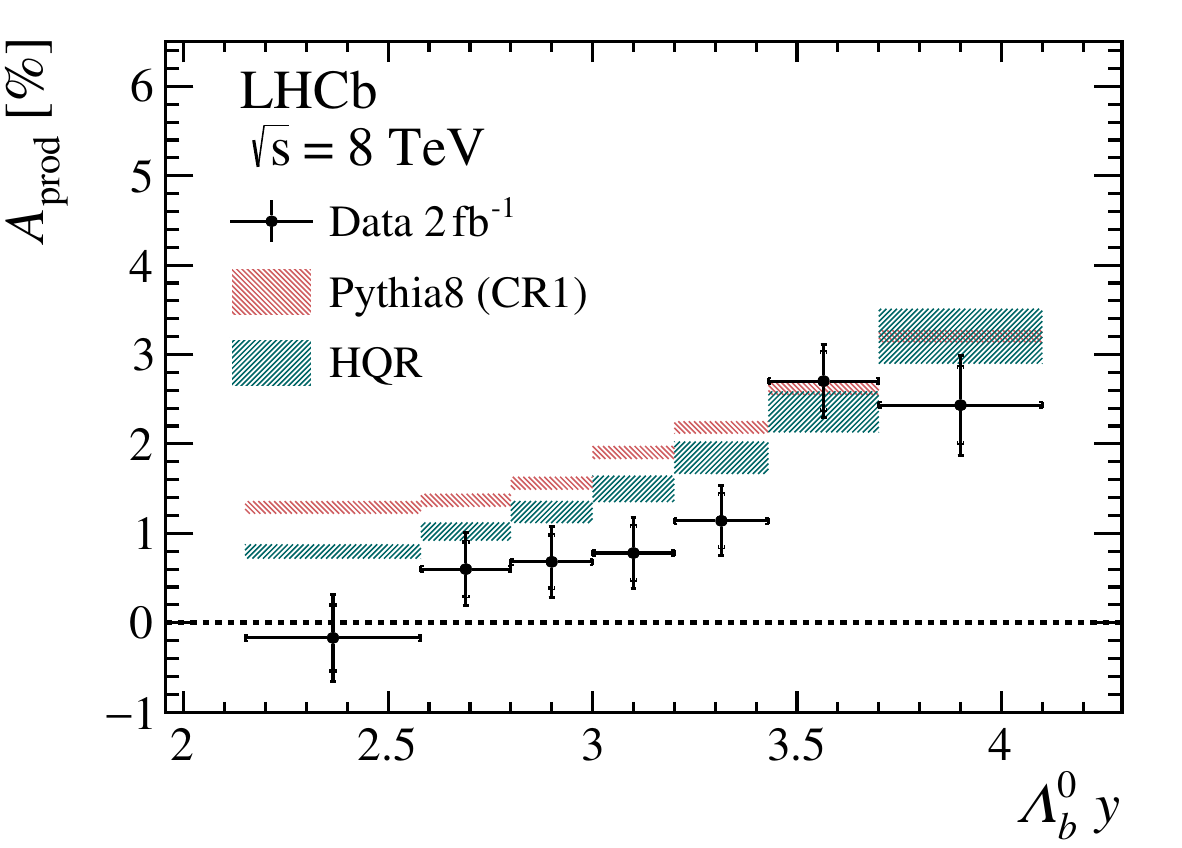}
		\includegraphics[width=0.49\textwidth]{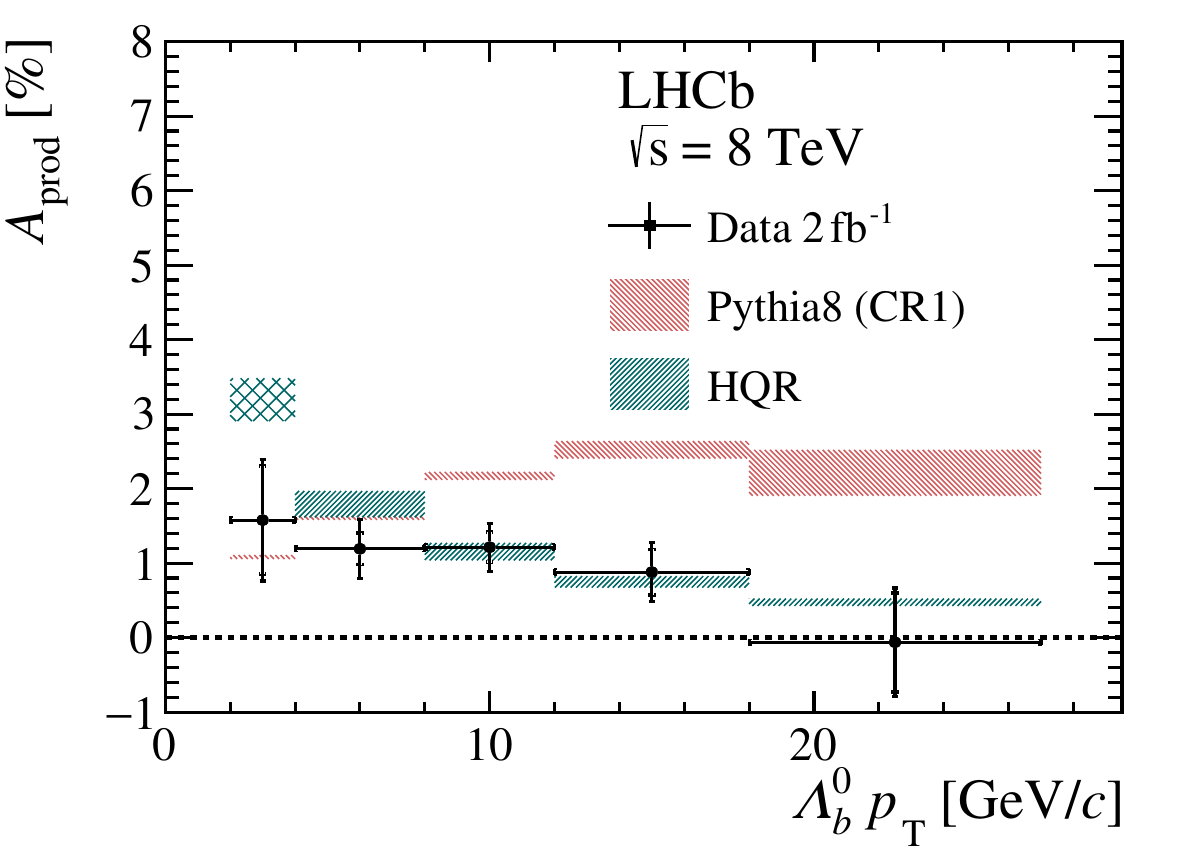} \\      
    \caption{Comparison of the measured $\Lb$ production asymmetry (points with error bars) with the predictions by the most compatible \pythia model and the heavy-quark recombination model (HQR). Results versus $\Lb$ (left) rapidity $y$ and (right) \pt are shown for centre-of-mass energies of (top) $\sqs=7\tev$ and (bottom) $\sqs=8\tev$. Uncertainties for the CR1 \pythia model are due to limited simulation sample sizes. Those from the heavy-quark recombination model are systematic as provided by the authors, and have the results in the lowest \pt interval hatched as the authors claim their results are reliable only for $\pt > 5\gevc$~\cite{Braaten:2001bf,LbprodLai}.}
    \label{fig:asymmetry_data_pythia_lai}
\end{center}
\end{figure}

The comparison between data and the various \pythia models is shown in \cref{fig:asymmetry_data_pythia}. The uncertainties on the \pythia models shown here are only due to the limited sample size of about 12.5 million events. The results of the \pythia hadronisation model describing the data best, along with the predictions of the heavy-quark recombination model are presented in \cref{fig:asymmetry_data_pythia_lai}. The uncertainties on the heavy-quark recombination model are the systematic uncertainties given in Ref.~\cite{LbprodLai}. Overall, the predictions from the heavy-quark recombination model are consistently higher than the $8\tev$ measurements, but remain within uncertainties. For \pythia, only the model CR1 shows a good agreement with the $\sqs=7\tev$ measurements but it is also consistently higher at $8\tev$. The two other tested settings predict asymmetries that are too large, exhibiting the strongest deviation at low transverse momentum.

\section{Conclusions}
The most precise measurements of the \Lb{} production asymmetry in $\sqs=7\tev$ and $8\tev$ proton-proton collisions have been presented. A new method to estimate asymmetries in the interaction of protons and antiprotons with the detector material has been developed. The \Lb{} production asymmetry has been measured in intervals of rapidity and transverse momentum, covering the ranges \(2.15< y <4.10\) and \(2<\pt<27\gevc\). A significant asymmetry in \bquark-hadron production has been observed for the first time with strong evidence for a dependence on \Lb{} rapidity. The results for $\sqs=7\tev$ and $\sqs=8\tev$ proton-proton collisions are compatible at the $2.6\,\sigma$ level, with asymmetries on average being lower at $\sqs=8\tev$. The measured values are consistent with the less precise indirect determinations presented in Ref.~\cite{LHCb-PAPER-2016-062}. A comparison of the obtained results with several theoretical predictions has been performed. The measured asymmetries as a function of rapidity and \pt disfavour the \pythia Monash and CR2 tunes, but are compatible with other colour-reconnection models implemented in \pythia and predictions from heavy-quark recombination.

%% file: acknowledgements.tex
\section*{Acknowledgements}
%
%
\noindent We thank W. K. Lai and A. K. Leibovich for providing us with theory predictions of the production asymmetry for the centre-of-mass energies of this measurement.
We express our gratitude to our colleagues in the CERN
accelerator departments for the excellent performance of the LHC. We
thank the technical and administrative staff at the LHCb
institutes.
We acknowledge support from CERN and from the national agencies:
CAPES, CNPq, FAPERJ and FINEP (Brazil); 
MOST and NSFC (China); 
CNRS/IN2P3 (France); 
BMBF, DFG and MPG (Germany); 
INFN (Italy); 
NWO (Netherlands); 
MNiSW and NCN (Poland); 
MEN/IFA (Romania); 
MSHE (Russia); 
MICINN (Spain); 
SNSF and SER (Switzerland); 
NASU (Ukraine); 
STFC (United Kingdom); 
DOE NP and NSF (USA).
We acknowledge the computing resources that are provided by CERN, IN2P3
(France), KIT and DESY (Germany), INFN (Italy), SURF (Netherlands),
PIC (Spain), GridPP (United Kingdom), RRCKI and Yandex
LLC (Russia), CSCS (Switzerland), IFIN-HH (Romania), CBPF (Brazil),
PL-GRID (Poland) and NERSC (USA).
We are indebted to the communities behind the multiple open-source
software packages on which we depend.
Individual groups or members have received support from
ARC and ARDC (Australia);
AvH Foundation (Germany);
EPLANET, Marie Sk\l{}odowska-Curie Actions and ERC (European Union);
A*MIDEX, ANR, IPhU and Labex P2IO, and R\'{e}gion Auvergne-Rh\^{o}ne-Alpes (France);
Key Research Program of Frontier Sciences of CAS, CAS PIFI, CAS CCEPP, 
Fundamental Research Funds for the Central Universities, 
and Sci. \& Tech. Program of Guangzhou (China);
RFBR, RSF and Yandex LLC (Russia);
GVA, XuntaGal and GENCAT (Spain);
the Leverhulme Trust, the Royal Society
 and UKRI (United Kingdom).

%% file: appendix.tex
\section*{Appendices}
\appendix
\section{Correlation matrices}\label{app:correlation_matrix}
\begin{figure}[ht]
	\begin{center}
		\includegraphics[width=0.79\textwidth]{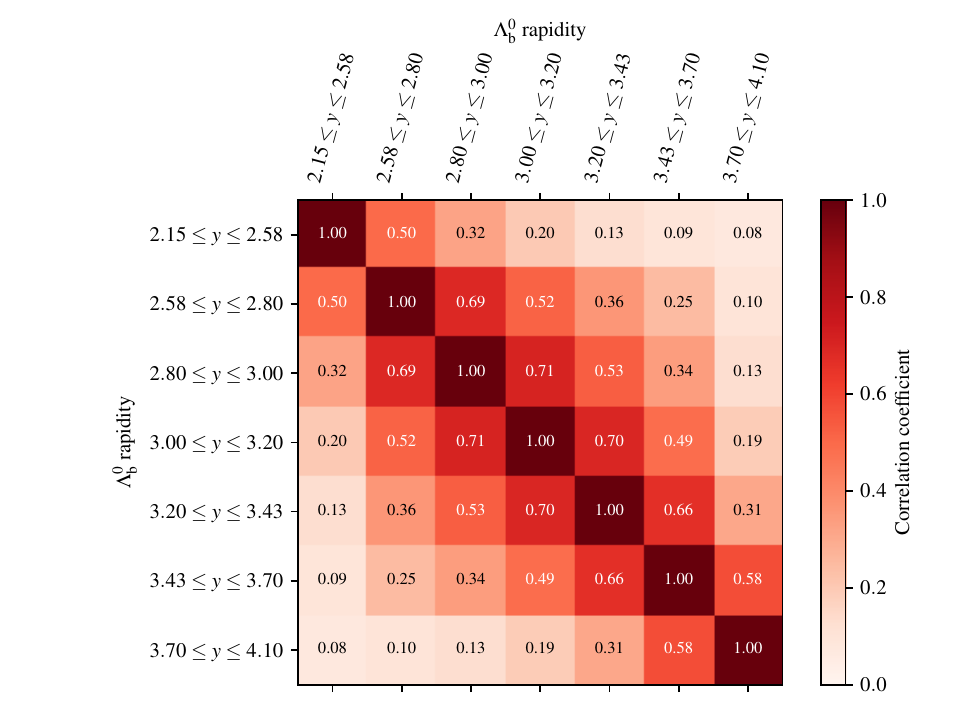}
		\includegraphics[width=0.71\textwidth]{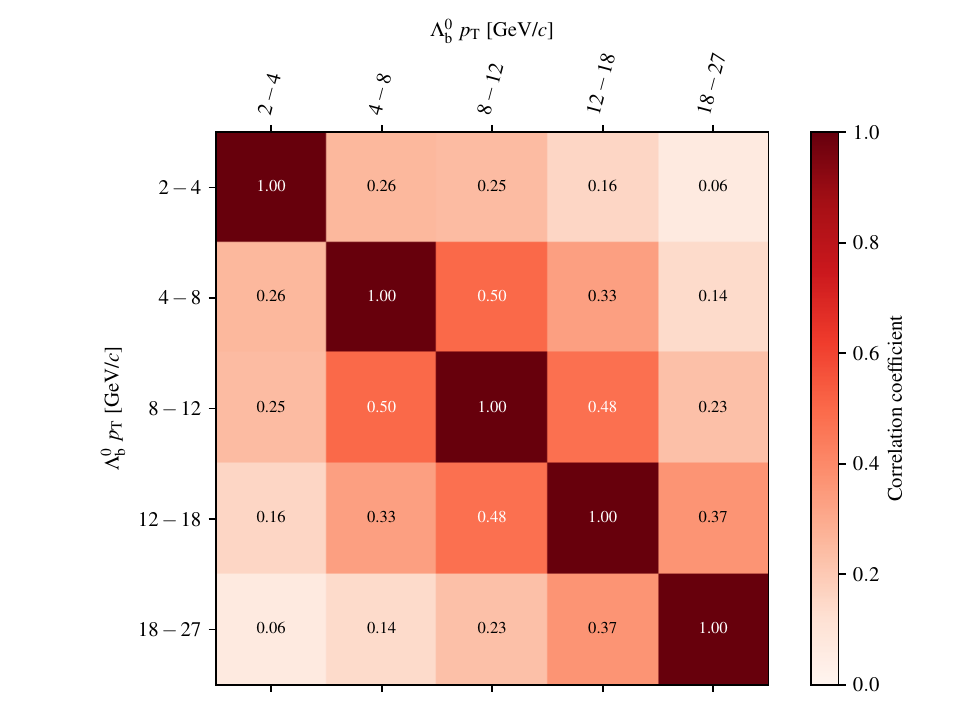}
		\caption{Correlation coefficients of the total (statistical and systematic) uncertainty on the production asymmetry versus $\Lb$ (top) rapidity and (bottom) transverse momentum.}
		\label{fig:correlation_matrix}
	\end{center}
\end{figure}

The correlation matrices for the measurements of \Lb{} production asymmetry in intervals of rapidity and \pt{}, including statistical and systematic uncertainties, are shown in \cref{fig:correlation_matrix}.

\section{\pythia settings}\label{app:pythia}
The settings used in the \pythia generation are given in \cref{tab:pythia_production_overview}.
\begin{table}[htb]
	\caption{\label{tab:pythia_production_overview}Settings for the different \pythia 8 productions. CR stands for Colour Reconnection. \pythia version 8.303 is used.}
   \centering
   \footnotesize
	\begin{tabular}{l l l }
	  Monash  & QCD inspired (CR1) & Gluon move (CR1) \\ 
	  \toprule
	  {\tt PhaseSpace:pTHatMin} 0.4 &  {\tt PhaseSpace:pTHatMin} 0.4 &  {\tt PhaseSpace:pTHatMin} 0.4 \\
	  {\tt CR:mode} 0 &  {\tt CR:mode} 1 &  {\tt CR:mode} 2 \\
	   &  {\tt BeamRemnants:remnantMode} 1 &  \\
	   &  {\tt BeamRemnants:saturation} 5 &  \\
	   &  {\tt CR:allowDoubleJunRem} off &  \\
	   &  {\tt CR:allowJunctions} on &  \\
	   &  {\tt StringZ:aLund} 0.36 &  \\
	   &  {\tt StringZ:bund} 0.56 &  \\
	   \bottomrule
	  \end{tabular}
  \end{table}

\section{Kinematic distributions}\label{app:kinematics}
\begin{figure}[tb]
  \begin{center}
  \includegraphics[width=0.48\textwidth]{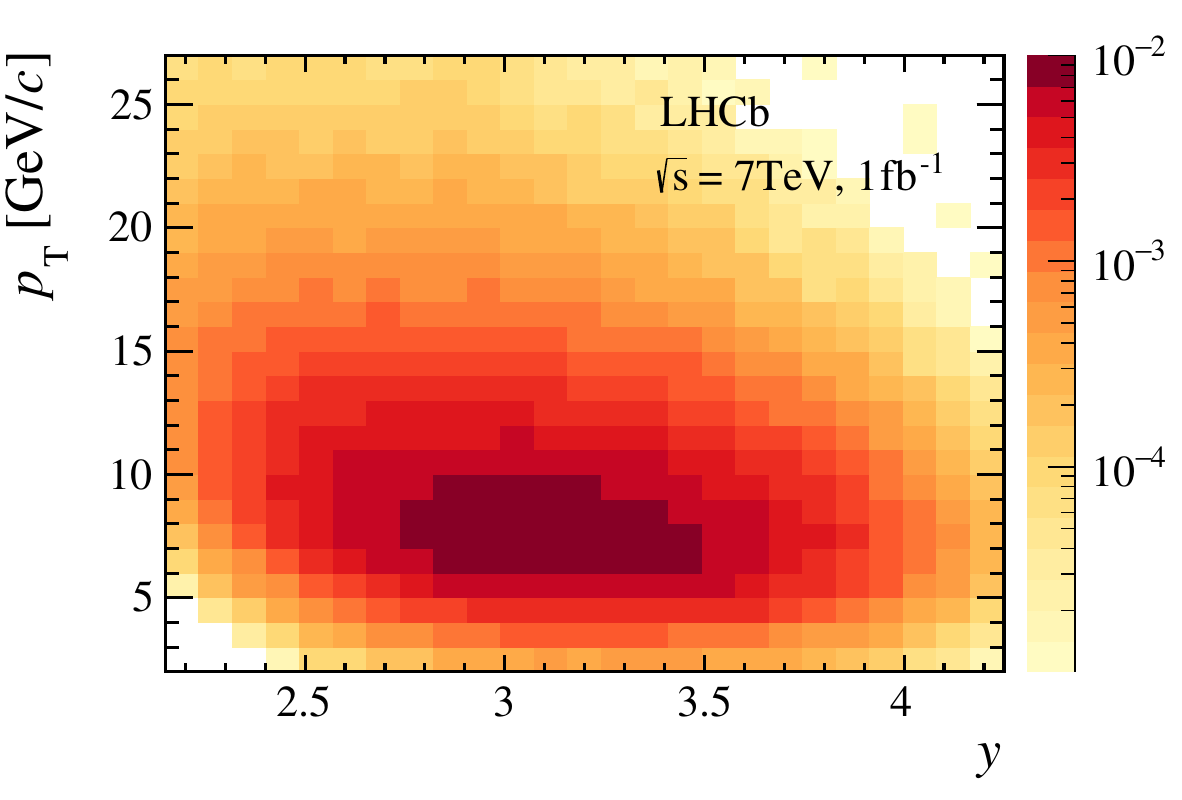}
  \includegraphics[width=0.48\textwidth]{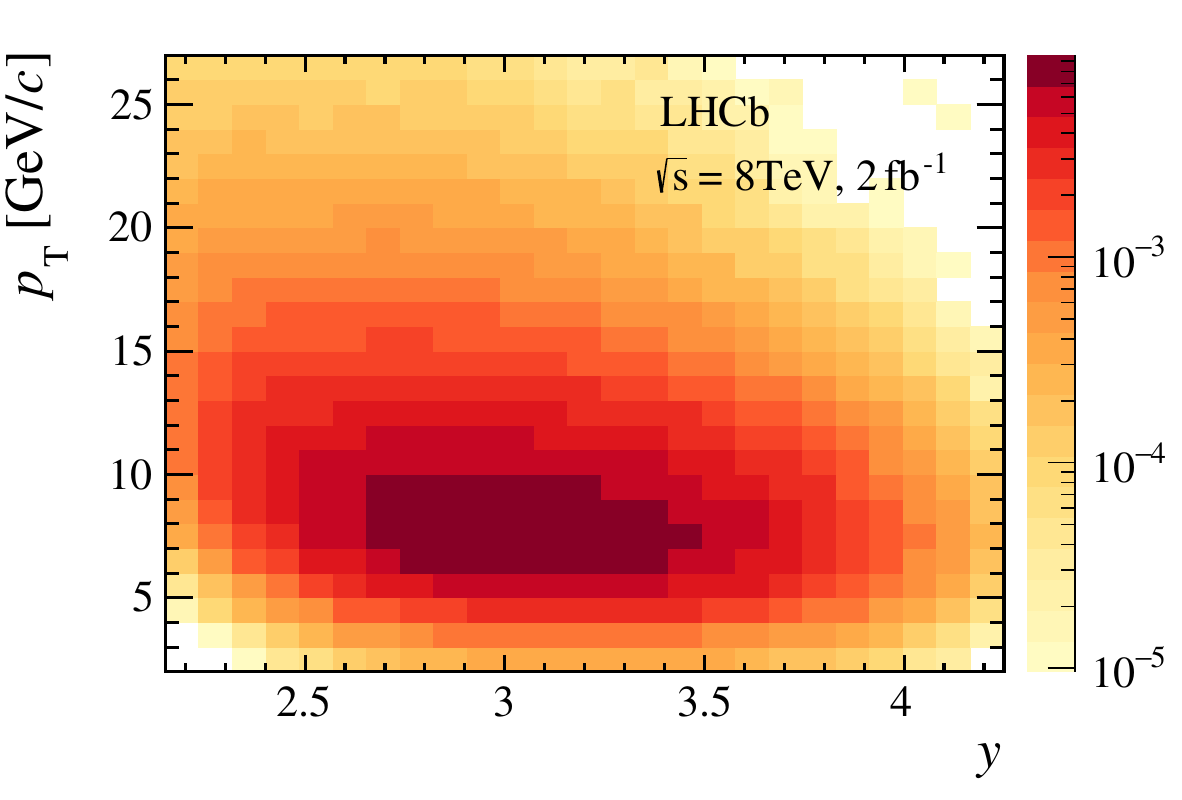}\\
  \includegraphics[width=0.48\textwidth]{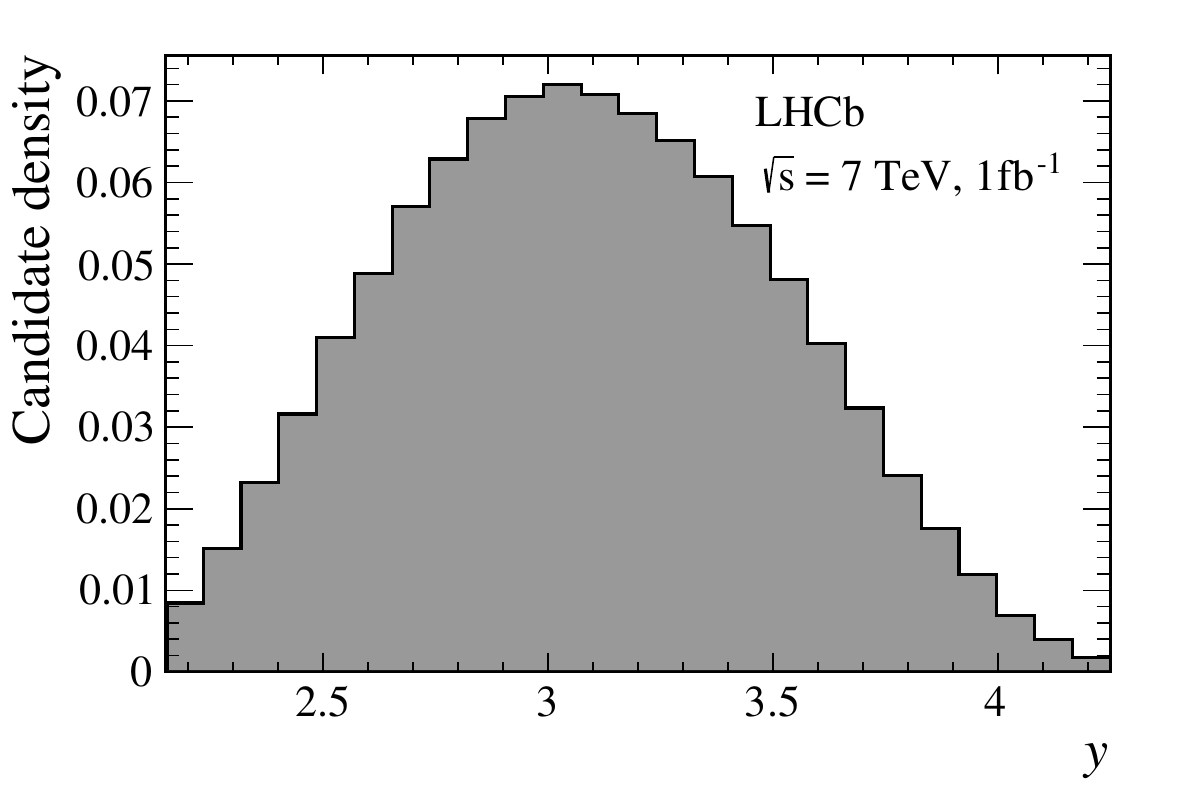}
  \includegraphics[width=0.48\textwidth]{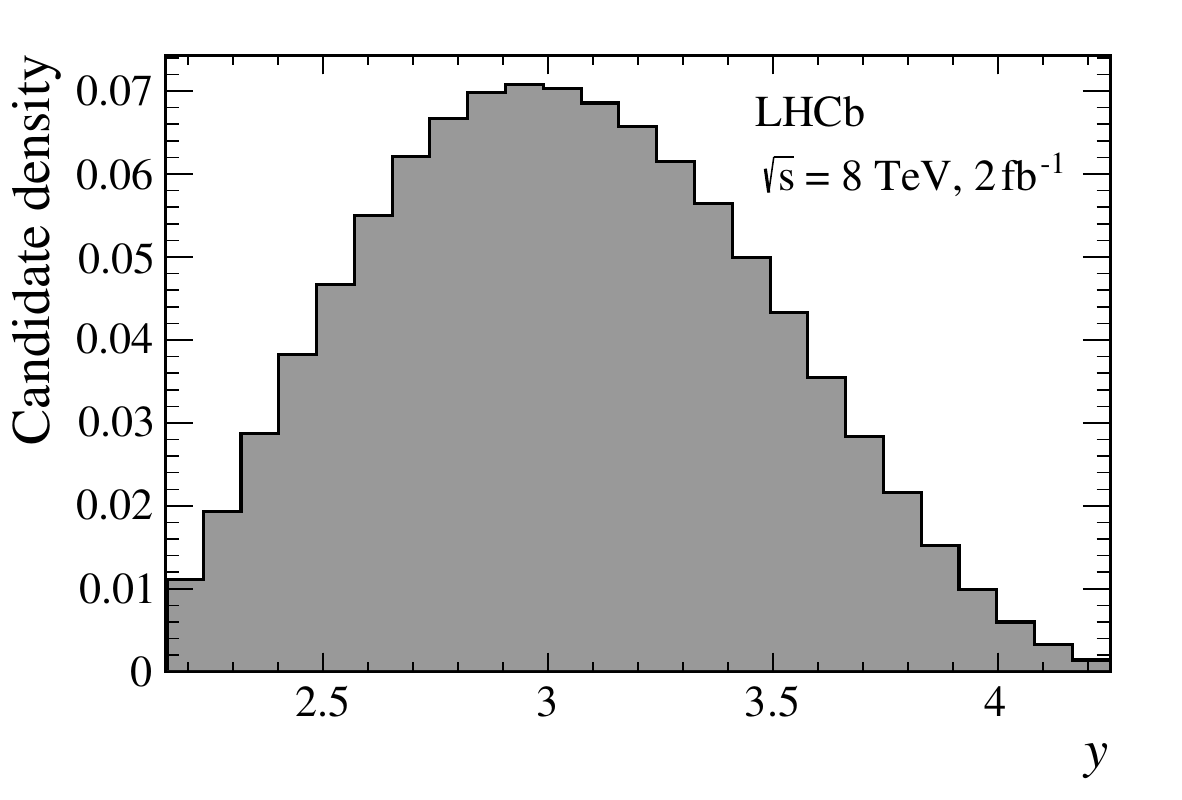}\\
  \includegraphics[width=0.48\textwidth]{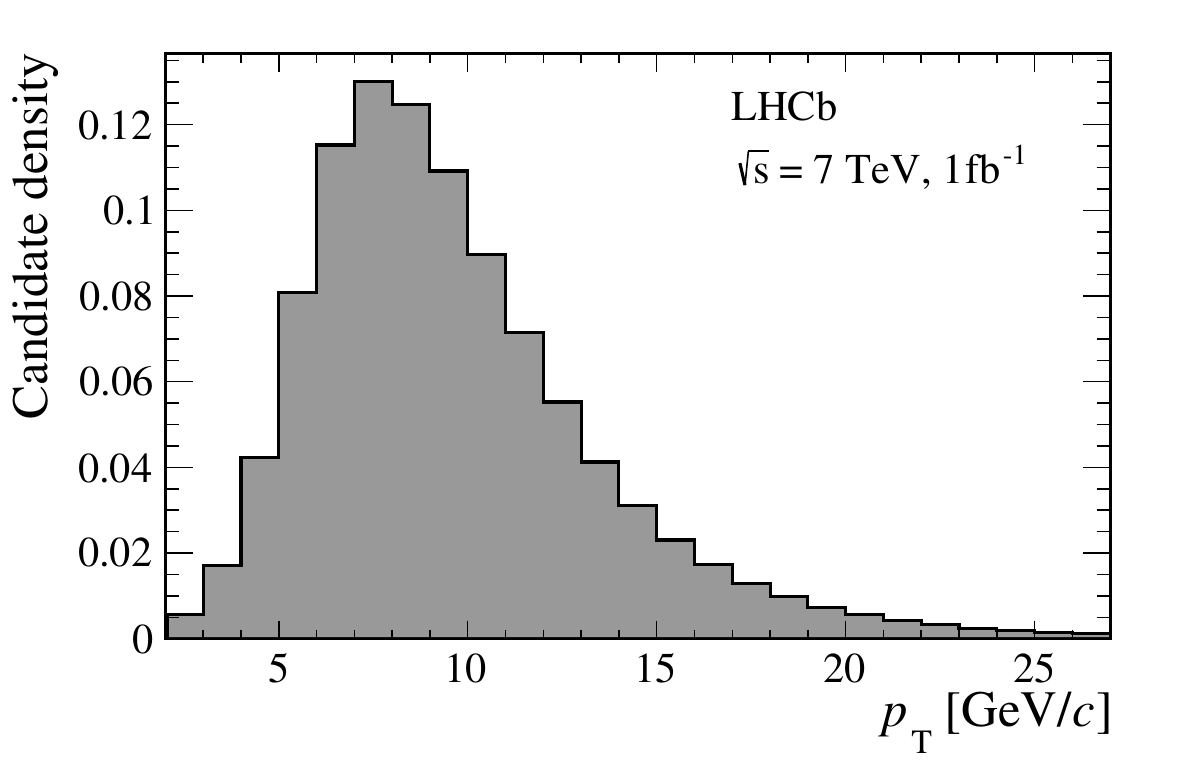}
  \includegraphics[width=0.48\textwidth]{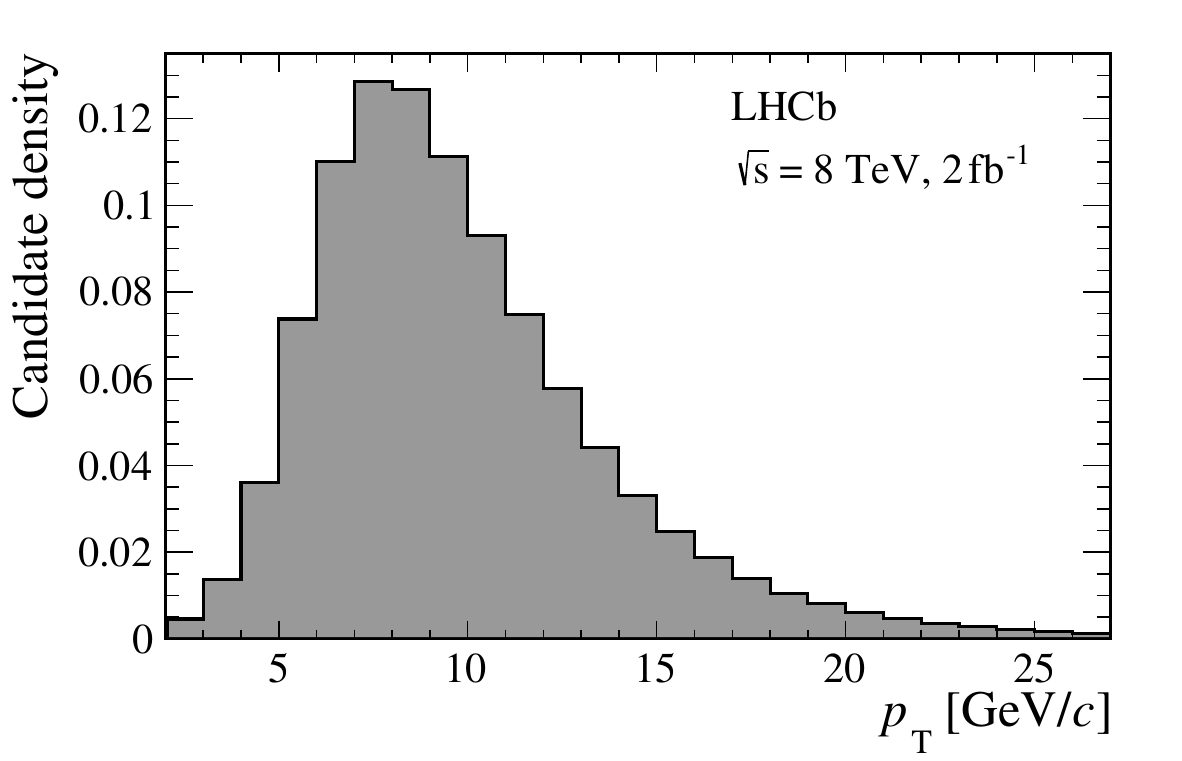}
  \caption{Measured, normalised distributions of $\pt, y$ from background-subtracted $\Lb$ candidates for (left) $\sqs=7\tev$ and (right)  $\sqs=8\tev$ data. The distributions are not corrected for the detector efficiency. Darker areas correspond to more densely populated regions. Below the two-dimensional histogram, the projections in $y$ and $\pt$ are shown. }
  \label{fig:theory_comp_measured_pt_y}
\end{center}
\end{figure}

The kinematic distributions of \Lb{} candidates as observed in data are shown in \cref{fig:theory_comp_measured_pt_y}. The data are not corrected for efficiencies. Background contributions are subtracted statistically. 
\clearpage

\section{Results split by magnet polarity}\label{app:magnet_polarity}
\begin{figure}[ht]
	\begin{center}
		\includegraphics[width=0.49\textwidth]{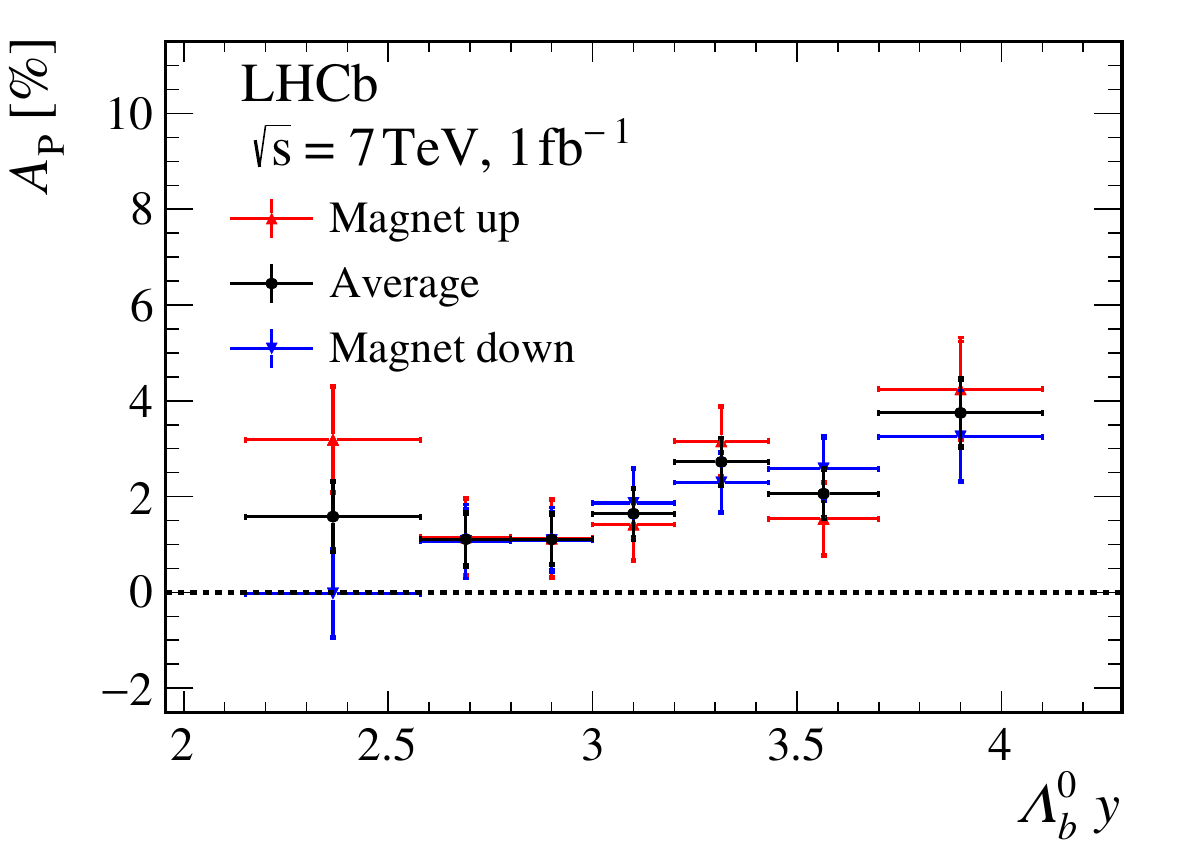}
		\includegraphics[width=0.49\textwidth]{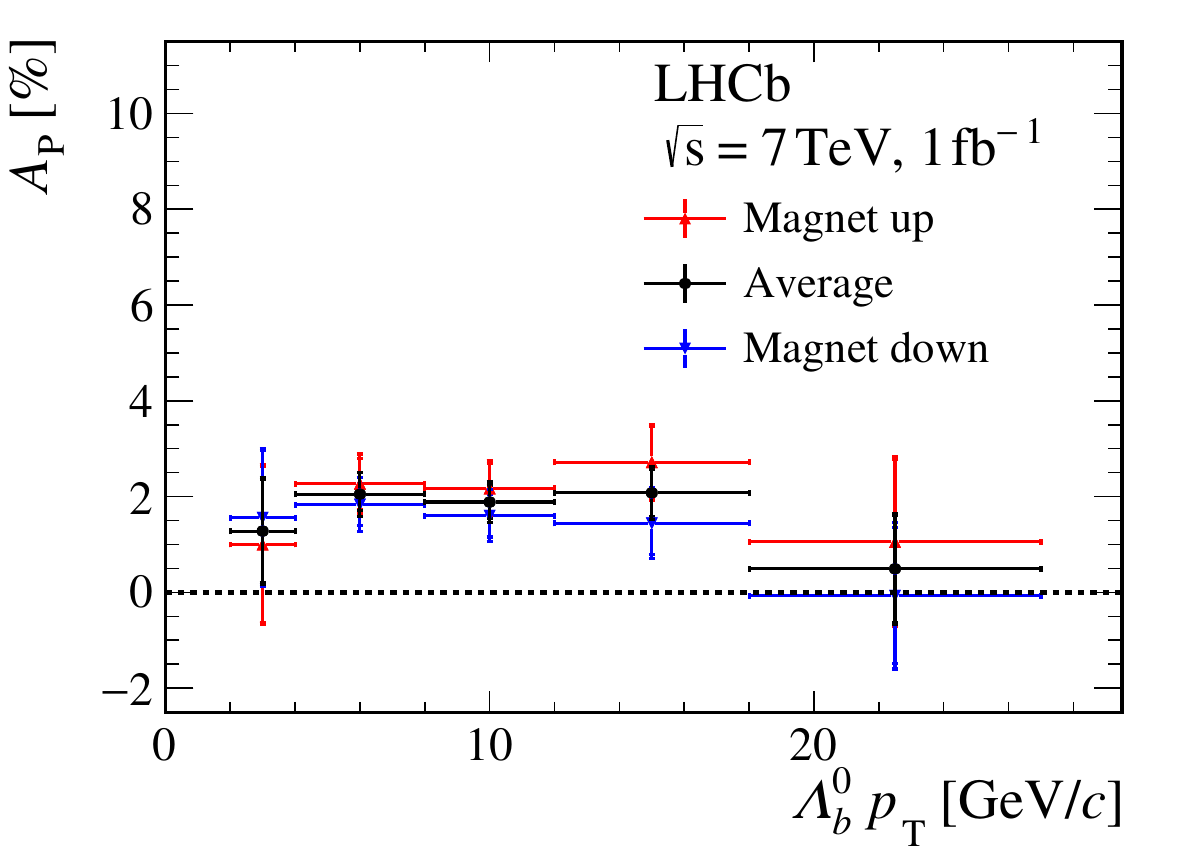}\\
		\includegraphics[width=0.49\textwidth]{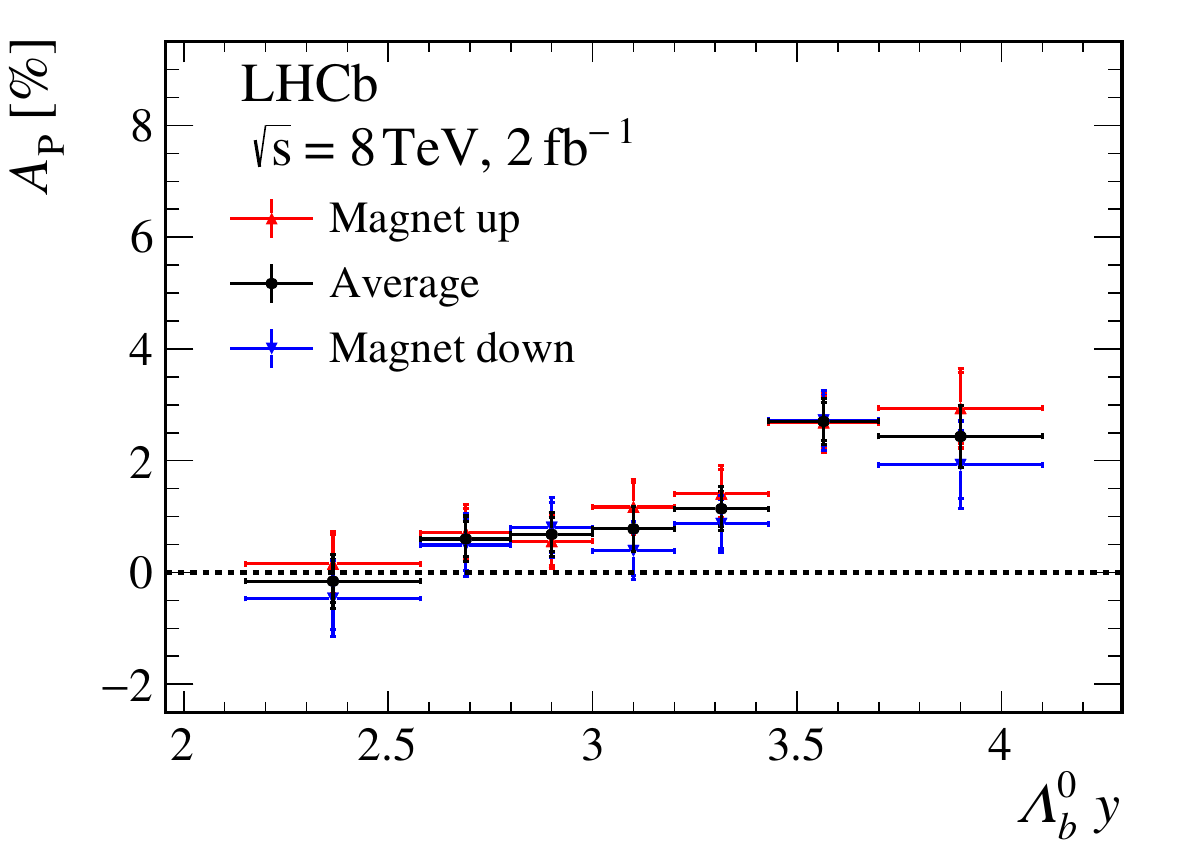}
		\includegraphics[width=0.49\textwidth]{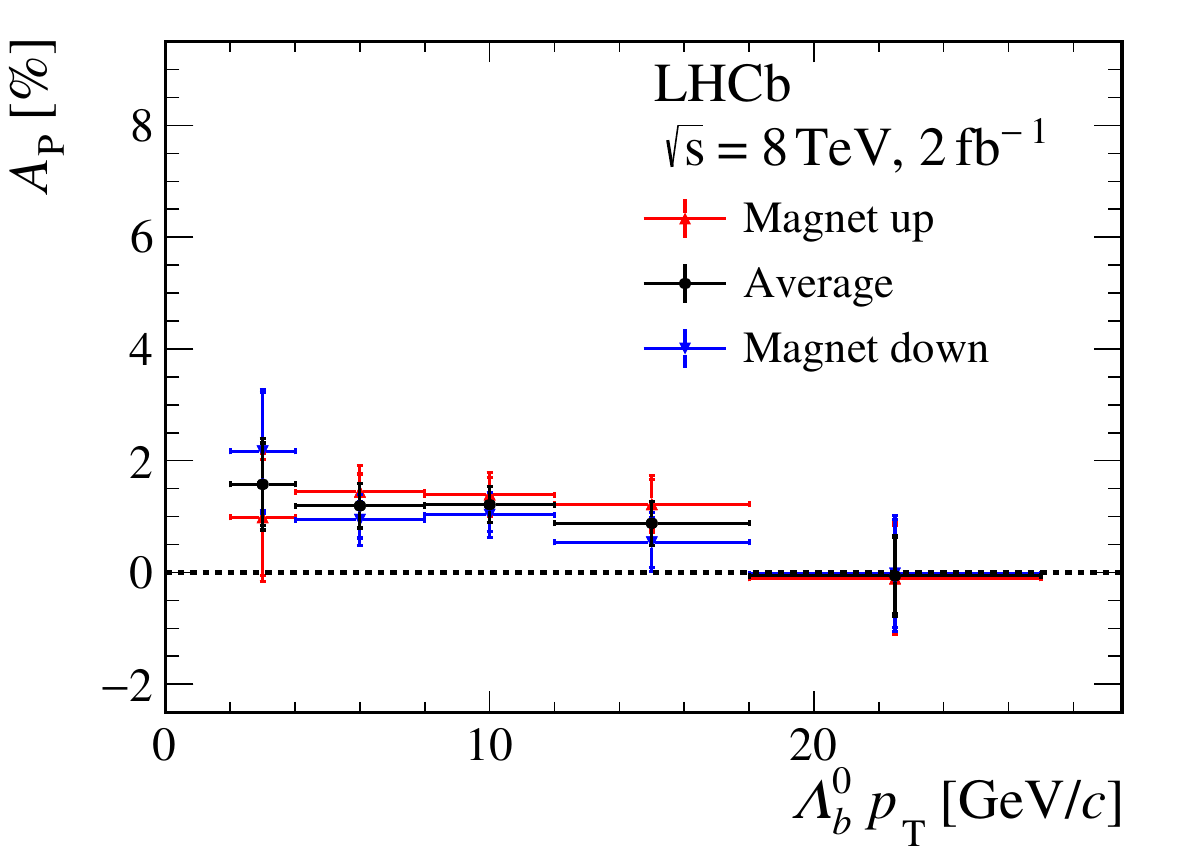}
		\caption{Measured $\Lb$ production asymmetry versus (left) rapidity and (right) transverse momentum separately for data recorded at centre-of-mass energies of (top) 7 and (bottom) 8\tev. The measured production asymmetries are shown separately for (red) magnet up, (blue) magnet down and their (black) average. The uncertainties are the quadratic sums of statistical and systematic uncertainties. }
		\label{fig:production_asymmetry_magnet_dependence}
	\end{center}
\end{figure}

The results as functions of $y$ and $\pt$ separated by centre-of-mass energy and magnet polarity are shown in \cref{fig:production_asymmetry_magnet_dependence}. When performing the measurement in intervals of rapidity, the $\chi^2$ for the compatibility of equal results between the samples taken with magnet up and down polarity is 8.7 for $\sqrt{s}=7\tev$ and 4.3 for $8\tev$. With 7 degrees of freedom in each sample, the corresponding $p$-values are 27.8\% and 75.0\%, respectively. When performing the measurement in intervals of transverse momentum, the obtained $\chi^2$ values are 2.0 and 3.2 for $\sqrt{s}=7\tev$ and 8\tev. With 5 degrees of freedom per sample, the corresponding $p$-values are 84.3\% and 67.6\%.

%% file: Authorship_LHCb-PAPER-2021-016.tex
\centerline
{\large\bf LHCb collaboration}
\begin
{flushleft}
\small
R.~Aaij$^{32}$,
A.S.W.~Abdelmotteleb$^{56}$,
C.~Abell{\'a}n~Beteta$^{50}$,
T.~Ackernley$^{60}$,
B.~Adeva$^{46}$,
M.~Adinolfi$^{54}$,
H.~Afsharnia$^{9}$,
C.A.~Aidala$^{86}$,
S.~Aiola$^{25}$,
Z.~Ajaltouni$^{9}$
% ...

{\footnotesize \it

% ...
$^{9}$Universit{\'e} Clermont Auvergne, CNRS/IN2P3, LPC, Clermont-Ferrand, France\\
% ...
$^{25}$INFN Sezione di Milano, Milano, Italy\\
% ...
$^{32}$Nikhef National Institute for Subatomic Physics, Amsterdam, Netherlands\\
% ...
$^{46}$Instituto Galego de F{\'\i}sica de Altas Enerx{\'\i}as (IGFAE), Universidade de Santiago de Compostela, Santiago de Compostela, Spain\\
% ...
$^{50}$Physik-Institut, Universit{\"a}t Z{\"u}rich, Z{\"u}rich, Switzerland\\
% ...
$^{54}$H.H. Wills Physics Laboratory, University of Bristol, Bristol, United Kingdom\\
% ...
$^{56}$Department of Physics, University of Warwick, Coventry, United Kingdom\\
% ...
$^{60}$Oliver Lodge Laboratory, University of Liverpool, Liverpool, United Kingdom\\
% ...
$^{68}$Syracuse University, Syracuse, NY, United States\\
% ...
$^{86}$University of Michigan, Ann Arbor, United States, associated to $^{68}$\\
\bigskip
% ...
\medskip
}
\end{flushleft}